\documentclass[final,3p,times]{elsarticle}
\usepackage[cp850]{inputenc}
\usepackage{amsmath,amssymb,amsfonts,color}
\usepackage{graphicx}
\usepackage{latexsym}
\usepackage{psfrag}
\usepackage{array}
\usepackage{wasysym}
\usepackage{epsfig}
\biboptions{sort&compress}
\begin{document}
\begin{frontmatter}
\title{Interference traps waves in open system: Bound states in the continuum}
\author{Almas F. Sadreev}
\ead{almas@tnp.krasn.ru}
\address{Kirensky Institute of Physics, Federal Research Center KSC SB RAS,
660036 Krasnoyarsk, Russia}
\date{\today}
\begin{abstract}
I review the four mechanisms of bound states in the continuum
(BICs) in application to microwave and acoustic cavities open to
directional waveguides. The most simple are the symmetry protected
BICs which are localized inside the cavity because of the
orthogonality of the eigenmodes to the propagating modes of
waveguides. However, the most general and interesting is the
Friedrich-Wintgen mechanism when the BICs are result of full
destructive interference of outgoing resonant modes. The third
type of the BICs, the Fabry-Perot BICs, occur in a double
resonator system when each resonator can serve as an ideal mirror.
At last, the accidental BICs can be realized in the open cavities
with no symmetry like the open Sinai billiard in which the
eigenmode of the resonator can become orthogonal to the continuum
of the waveguide accidentally by a smooth deformation of the
eigenmode. We also review the one-dimensional systems in which the
BICs occur owing to full destructive interference of two waves
separated by spin or polarization or by paths in the Aharonov-Bohm
rings. We widely use the method of effective non-Hermitian
Hamiltonian equivalent to the coupled mode theory which detects
bound states in the continuum (BICs) by finding zero widths
resonances.
\end{abstract}

\end{frontmatter}
\tableofcontents
\section{Introduction}
\label{Sect:introd}
More than two centuries has passed since Thomas Young presented
his eminent double-slit experiment which unambiguously proved wave
nature of light but still wave interference offers new phenomena
in physics. Among the last ones to have attracted close attention
of researches are  bound states in the continuum (BICs). In 1929,
von Neumann and Wigner \cite{Neumann} claimed that the
single-particle Schr\"odinger equation could possess localized
solutions that correspond to isolated discrete eigenvalues
embedded in the continuum of positive energy states for some
artificial oscillating bounded potential. Extension and some
correction of this work was done by Stillinger and Herrick
\cite{Stillinger1975} who presented a few examples of spherically
symmetric attractive local potentials with BICs of scattering
states in the context of possible BICs in atoms and molecules
(e.g., \cite{Noeckel92,Pursey1994,Pursey1995,Cederbaum2003}). For
a long time the phenomenon was considered as mathematical
curiosity although physical mechanism is very similar to the
mechanism of the Anderson localization. The BIC as a localized
state is a result of precise destructive  interference of waves
scattered by the bounded potential in such a way that after enough
distance we have no outgoing wave.

The decisive breakthrough came with paper by Friedrich and Wintgen
\cite{Friedrich1985} who formulated a general method to find BICs
in quantum systems. The method based on the effective
non-Hermitian Hamiltonian  originates from Feshbach unified theory
of nuclear reactions \cite{Feshbach,Feshbach1962} and uses the
fact that the occurrence of BICs is directly related to the
phenomenon of avoided level crossing. When two resonance states
approach each other as a function of a certain continuous
parameter, interference causes an avoided crossing of the two
states in their energy positions and, for a certain value of the
parameter, the width of one of the resonance states may vanish
exactly. Since it remains above threshold for decay in to the
continuum, this state becomes a BIC although each resonant state
has a finite width. After numerous model considerations of
different physical systems were presented
\cite{Shahbazyan1994,Magunov1999,Volya,guevara2003,Wunsch2003,
Fedorov2004,Rotter2005,SBR,guevara2006,Solis2008}.

\begin{figure}[ht]
\centering{\resizebox{0.5\textwidth}{!}{\includegraphics{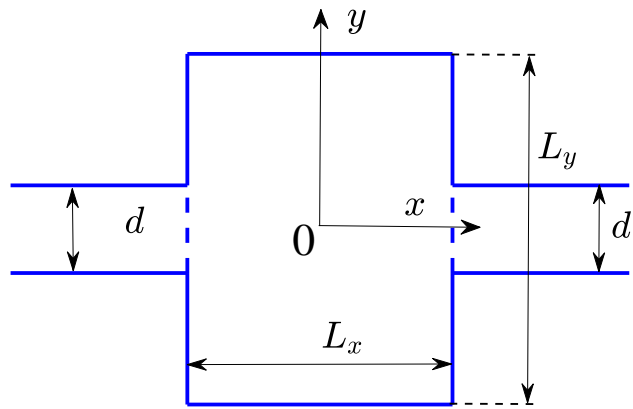}}}
\caption{Two-dimensional plane resonator with two attached plane
waveguides.}\label{fig1}
\end{figure}

The Fridrich-Wintgen (FW) approach of the effective Hamiltonian
was first readily applied to planar metallic integrable billiard
(cavity) open by attachment of two uniform  plane waveguides
\cite{SBR} as shown in Fig. \ref{fig1}. The reader can find the
description of the system in textbooks on electromagnetic (EM)
fields (see, for example, \cite{Jin2010}).
It was shown that for variation of resonator width $W$ numerous
events of degeneracy of the eigenmodes, say $\psi_1$ and $\psi_2$
occur. Then, at the points of degeneracy one can consider the
superposed function $a\psi_1+b\psi_2$. If each eigenmode is
coupled with waveguide first channel by means $W_1$ and $W_2$ for
the superposed function we have obviously the coupling $aW_1+bW_2$
which can be tuned to zero by a proper choice of the superposition
coefficients $a$ and $b$. That is an alternative interpretation of
the BIC occurring at the degeneracy points in the integrable open
resonators. The FW BICs were  first experimentally observed by
Lepetit and Kant\'{e} in metallic waveguide with two ceramic disks
\cite{Lepetit2014}. Similarly, Olendski and Mikhailovska  have
shown that in curved 2d waveguide a quasi-bound state formed as a
result of the bend, at some critical parameters of the curve
becomes a true bound state with in the continuum
\cite{Olendski2002}. Catapan {\it et al} have revealed BICs in 2d
straightforward stubbed quantum waveguide with impurities
\cite{Cattapan2007} and 2d serial structures \cite{Cattapan2007a}.
Thus, it is turned out that going beyond 1d crucially increases
opportunities for BICs.

The question of whether a wave can be perfectly confined (that is,
whether a 'bound state' can exist) in an open system related to a
simple frequency criterion. If the frequency of wave is outside
the continuous spectral range spanned by the propagating waves, it
can exist as a bound state because there is no pathway for it to
radiate away. Conversely, a wave state with the frequency inside
the continuous spectrum can only be a 'resonance' that leaks and
radiates out to infinity. This is the conventional wisdom
described in many books. A bound state in the continuum (BIC) is
an exception to this conventional wisdom: it lies inside the
continuum and coexists with extended waves, but it remains
perfectly confined without any radiation.

Besides the Friedrich-Wintgen mechanism of full destructive
interference  other mechanisms for BICs exist. The most simple
mechanism is the symmetry protection. Bolsterli has treated the
special case in which there occur discrete states in the continuum
in separable potentials \cite{Bolsterli}. In such a system a
symmetry incompatibility decouples the square-integrable
eigenmodes from the  propagating modes of the waveguides
\cite{Robnik,Schult,Moiseyev2009}. It is accepted to determine
such BICs as the symmetry protected ones. Less obvious but similar
to the symmetry protected BICs are the accidental BICs when in
spite of absence of symmetry arguments the coupling between the
cavity eigenmode and the mode of the continuum can turn to zero
accidentally by variation of the shape of the cavity as it was
demonstrated in the open Sinai billiard \cite{Pilipchuk2017}.
Firstly, such a possibility was mentioned by Friedrich and Wintgen
in the paper of physical realization of BICs in hydrogen atom in a
magnetic field \cite{Friedrich1985a}. After accidental BICs were
demonstrated in photonic systems \cite{Hsu2013,Bulgakov2014}.

More sophisticated but transparent mechanism of BICs is the
Fabry-Perot one. Assume, we have two ideal metallic mirrors
parallel each other and separated by the  distance $L$ between
them. All states are bounded in this system with eigenfrequencies
$\omega_n=\pi n/L, n=1, 2, 3, \ldots$. If the mirrors have finite
transmission probability all bound states become resonant states
with finite line widths because of leakage through the mirrors
\cite{Stratton}. Such a system is analogous to the simplest
quantum mechanical problem of single particle in double barrier
potential. In this one-dimensional system there are no BICs.
However, in 1999 Kim and Satanin \cite{Kim1999} put forward the
idea to go beyond the one-dimensional case applying a temporally
periodically driven  barriers. Then, the effective dimensionality
of the one-dimensional double barrier potential becomes two
\cite{Sambe,SadreevPRE86} allowing for transmission zeros  even
for a finite height of the potential barriers. A possibility to
localize quantum particle in a tight-binding chain with an
off-channel impurity driven by an ac field was later considered by
Longhi and Della Valle in a series of papers on Floquet BICs
\cite{Longhi2011,Longhi2013,Longhi2014}.

More straightforward Fabry-Perot models which supports BICs was
considered by Shanhui Fan {\it et al} \cite{Fan1999} in the
framework of coupled mode theory \cite{Suh2004}. In a series of
papers \cite{Rotter2004,Rotter2005,Sadreev2005a,Sadreev2005b}
two-dimensional identical quantum dots were used as Fabry-Perot
mirrors. Then, the BICs are engineered by tuning the distance
between the resonators coupled by wire. A similar approach was
also used by Ordonez \cite{Ordonez2006}. The same mechanism of
BICs was exploited in photonic crystal systems
\cite{Marinica,BS2008,Ndangali2010,Li16}. The occurrence of BICs
in these systems is accompanied by the collapse Fano resonances
when transmission zero coalesces with the transmission unit
\cite{SBR,Kim1999,Lepetit2014}. Another variant of waveguide which
supports BICs is double bend waveguide \cite{Sadreev2015} due to
transmission zeros in the bend  \cite{Olendski2002}.

Up to now we briefly discussed bound states with discrete
frequencies embedded into the waveguide continua which are given
by quantized due to a finite width of the waveguide (see Fig.
\ref{fig1}). It is easy to realize the FW BIC embedded into the
first continuum which is separated from the next continua by
finite gap, say, by variation of length of resonator
\cite{SBR,Hein2008,Hein2012,Vargiamidis2009} or obstacle size  in
the waveguide \cite{Rowe2005,Cattapan2007a}. The state of art is
BICs embedded into a few continua of the waveguide
\cite{Bulgakov2011}. At the first glance it seems impossible to
support the BICs in the radiation continuum of free space which is
given by continuous spectrum of light line (cone) $\omega=ck$
where $c$ is the light velocity. The closed metallic resonator in
free space is exceptional case because of its equivalence to
quantum mechanical well potential with infinitely high walls.
Similarly there might be the BICs in plasmonic nanostructures
\cite{Monticone14,Silveirinha14}. It agrees with theorem that
there are no BICs in the bounded domain which is complement of an
unbounded domain \cite{Silveirinha14,Colton}. However, in infinite
periodic arrays of dielectric particles light can leak only into a
discrete number of diffraction orders allowing to find BICs
embedded into a finite number of diffraction continua
\cite{Yang,Bulgakov2015,Bulgakov2017}. Therefore the infinite
periodic dielectric structures can support BICs that attracts
growing interest of the optical community because of possibility
to confine light. The extremely large quality factor of the BICs,
or better to say, quasi-BICs and possibility to manipulate the
BICs above light line has become of extreme importance in modern
science and opens up many applications. In what follows we skip
photonic BICs since they have already been a subject of recent
reviews \cite{Hsu16,Krasnok2019,Koshelev2019,Peng2020}.

\section{The effective non Hermitian Hamiltonian}
\label{Sect:Heff}
 One of the powerful and unambiguous means
to diagnose BICs is the method of effective non Hermitian
Hamiltonian
\cite{Feshbach,Rotter1991,Dittes2000,Rotter2003,Savin2003,SR2003}
which is equivalent to the coupled mode theory (CMT)
\cite{Suh2004,Maksimov2015}. An important advantage of the
effective non Hermitian Hamiltonian approach is a possibility to
calculate the coupling matrix between closed system and continuum
when the eigenmodes of subsystems are known
\cite{Pichugin2001,Rotter2003}. The approach of the effective
non-Hermitian Hamiltonian
\cite{Rotter1991,Dittes2000,Okolowicz2003} have found numerous
applications in various branches of physics including atomic
nuclei \cite{Weidenmuller,Savin2003}, chaotic billiards
\cite{Pichugin2001,Stockmann,Alhassid,Stockmann2002,Akguc,Auerbach},
tight-binding models
\cite{Datta,SR2003,Sadreev,Hatano,Hatano2014}, potential
scattering \cite{Savin2003}, photonic crystals
\cite{Bulgakov2011a}, etc.

The  objective  of  the  present  paper is to revisit the concept
of the effective non-Hermitian Hamiltonian in application to open
resonators with the Dirichlet or Neumann boundary conditions. The
problem of resonant scattering typically involves a cavity (which
could be quantum dot, microwave or acoustic cavity  {\it etc}) and
scattering channels coupled to the cavity. The mainstream idea is
to split the full Hilbert space into subspaces: subspace $B$
formed by the eigenfunctions of discrete spectrum localized within
the scattering center, and subspaces $C$ which spans the extended
eigenfunctions of the scattering channels. Therefore, the exact
description of open system meets a problem of matching the wave
functions of discrete and continuous spectra. In 1958 Feshbach
\cite{Feshbach} introduced the idea to project the total Hilbert
space onto the discrete states of subspace $B$. Given the Hamilton
operator of the whole system as
\begin{equation}\label{Htot}
\widehat{H}=\widehat{H}_B+\sum_C(\widehat{H}_C+\widehat{V}_{BC}+\widehat{V}_{CB})
\end{equation}
the projection onto the discrete subspace leads to the concept of the effective non-Hermitian
Hamiltonian \cite{Feshbach,Rotter1991,Dittes2000,Rotter2003}
\begin{equation}
\widehat{H}_{eff} = \mathcal{\widehat{H}}_B + \sum_C
\widehat{V}_{BC} \frac{1}{E^+ - \widehat{H}_C}\widehat{V}_{CB}.
\label{Heff}
\end{equation}
Here $\widehat{H}_B$ is the Hamiltonian of the closed system,
$\widehat{H}_C$ is the Hamiltonian of the scattering channel $C$,
 $\widehat{V}_{BC},~\widehat{V}_{CB}$ stand for the coupling matrix elements between the
eigenstates of closed cavity and the eigenstates of the scattering
channels, and $E$ is the energy of scattered particle (wave). For
EM wave or acoustic transmission $E=\omega^2$ where $\omega$ is
the frequency. The term $E^{+} =E+i0$ ensures that only outgoing
waves will be present in the solution after the scattering occurs.
As a result the effective Hamiltonian (\ref{Heff}) is a
non-Hermitian matrix with complex eigenvalues $z_{\lambda}$ which
determine the positions and lifetimes of the resonant states as
$Re(z_{\lambda})$, and $-2Im(z_{\lambda})$
\cite{Rotter1991,Rotter2003}. If to assume that the propagation
band of the continuum is not bounded then the effective
non-Hermitian Hamiltonian takes the most simple form widely used
in the scattering theory
\cite{Dittes2000,Weidenmuller,Stockmann2002}
\begin{equation}\label{Heffphen}
    \widehat{H}_{eff}=\widehat{H}_B-i\sum\limits_{C}\widehat{W}_C
    \widehat{W}_C^{\dagger},
\end{equation}
where $\widehat{W}_C$ is a column matrix whose elements account
for the coupling of each individual inner state to the scattering
channel $C$. The scattering matrix $\mathcal{S}_{CC'}$ is then
given by the inverse of $E-\widehat{H}_{eff}$
 \cite{Dittes2000,Stockmann2002}
\begin{equation}\label{S}
    \widehat{S}=\delta_{CC'}-2i\widehat{W}^{+}\frac{1}{\widehat{H}_{eff}-E+i0}\widehat{W},
\end{equation}
where $C=L,R$. Therefore for the case of energy/frequency
independent coupling matrix the complex eigenvalues coincide with
the poles of the S-matrix.

However this formulation of the effective Hamiltonian is
oversimplified because of unbounded spectrum of the continuum.
Commonly the spectrum is bounded, at least below. For example, the
electron has the spectrum $E=\frac{\hbar^2k^2}{2m}$ and
electromagnetic (EM) waves has the spectrum $\omega=ck$. Although
the form for the effective Hamiltonian (\ref{Heffphen}) is
preserved the coupling matrix elements become dependent on the
energy or frequency \cite{Pichugin2001,SR2003,Maksimov2015}. In
what follows we apply the method of the effective non Hermitian
Hamiltonian to several physical systems: 1) One-dimensional wires
with off-channel cavities the Aharonov-Bohm rings, 2)
Two-dimensional microwave planar metallic waveguide consisted of
the cavity  and two attached waveguides and microelectronic
waveguides (Fig. \ref{fig1}), and 3) Three-dimensional acoustic
cylindrical and spherical resonators with attached cylindrical
waveguides.

\section{Friedrich-Wintgen concept of BIC}
\label{Sect:FW}
One can see that the effective non Hermitian Hamiltonian
(\ref{Heffphen}) consists of Hermitian part $\widehat{H}_B$ whose
eigenvalues are the eigenfrequencies of the closed cavity and the
second anti-symmetric imaginary part. This part is a result of
coupling of the cavity with the continua $C$ of waveguides. The
complex eigenvalues of the effective Hamiltonian have clear
physical meaning. Their real parts respond for position of
resonances while their imaginary parts respond for half resonant
widths \cite{Rotter1991,Dittes2000}. Other words, if to prepare
some field as the eigenmode of the closed cavity it will decay
because of leakage of the mode into waveguides. Therefore the BIC
is easily found out by turning to zero one of the imaginary parts
of the complex eigenvalues of the non Hermitian effective
Hamiltonian that was first established by Friedrich and Wintgen
\cite{Friedrich1985} in generic two-level Hamiltonian. When two
resonance states approach each other as a function of a certain
continuous parameter, interferences cause an avoided crossing of
the two states in their energy positions and, for a certain value
of the parameter, the width of one of the resonance states
vanishes exactly. Since it remains above threshold for decay into
the continuum, this state becomes a BIC. The Friedrich and Wintgen
(FW) approach is significant by that it can be applicable to any
waveguide system, in particular to microelectronic, microwave or
acoustic resonators opened by attachment of waveguides
\cite{SBR,Datta,Lyapina2015}.

Let the cavity undergoes degeneracy, for example,  variation of
shape. In the neighborhood of this degeneracy it is reasonable to
truncate the Hamiltonian of the cavity by only those eigenvalues,
say  $E_1$ and $E_2$, which are crossing. Moreover we assume that
there is the only continuum with which the cavity modes are
coupled. That gives the following two-level effective Hamiltonian
\begin{equation}\label{Heff2}
\widehat{H}_{eff}=\left(\begin{array}{cc} \epsilon-i\gamma_1 &
u-i\sqrt{\gamma_1\gamma_2}\cr u-i\sqrt{\gamma_1\gamma_2}
&-\epsilon-i\gamma_2\end{array}\right)
\end{equation}
where without lose of generality we take $E_{1,2}=\pm \epsilon$.
Also we introduce the $\gamma_1=W_1^2, ~\gamma_2=W_2^2$ which
could define the resonant widths of the levels $E_{1,2}$ if the
effective Hamiltonian (\ref{Heff2}) was diagonal. $W_n, n=1,2$ are
the coupling constants of the cavity modes with waveguide
propagating mode. Parameter $u$ is responsible for repulsion of
the eigenfrequencies of the closed cavity due to, for example,
inner perturbation which removes the integrability of the cavity.
For example, in Section \ref{Sect:Accidental} we consider a hole
inside the cavity transforming into the Sinai billiard where the
eigenlevels are avoided.

The advantage of the two-level approximation is that the BIC can
be considered analytically \cite{Volya,SBR}. Let us write the
transmission amplitude in the biorthogonal basis of the
eigenstates of the effective non Hermitian Hamiltonian
(\ref{Heff2})
\begin{equation}\label{bio}
\widehat{H}_{eff}|\lambda)=z_{\lambda}|\lambda), ~~(\lambda|\lambda')=
\delta_{\lambda,\lambda'}, ~|\lambda)=|\lambda\rangle, ~(\lambda|=|\lambda)^c=
\langle \lambda|^*,
\end{equation}
i,e., the left states are related to the right states via transposing.
Then using the condition of completeness
$$\sum_{\lambda}|\lambda)(\lambda)|=1$$
we can rewrite the transmission amplitude as sum of the resonant
terms \cite{SR2003}
\begin{equation}\label{transz}
    T=-2i\sum_{\lambda}\frac{V_{\lambda}^{L}V_{\lambda}^{R}}{E-z_{\lambda}}
=-2i\sum_{\lambda}\frac{V_{\lambda}^2}{E-z_{\lambda}},
\end{equation}
where $V_{\lambda}^{C}=V_{\lambda}, C=L, R$ are the coupling
constants of resonant states with the continuum or the propagating
mode of waveguides. The expression (\ref{transz}) immediately
shows us that the complex eigenvalues $z_{\lambda}$ are the poles
of the S-matrix provided that the matrix elements of the effective
Hamiltonian are energy independent. Otherwise, we are to use the
complex scaling method \cite{Moiseyev1998} or to solve nonlinear
fix point equations for real and imaginary parts of the complex
eigenvalues $z_{\lambda}$ which define the resonant positions and
the resonant widths \cite{Rotter2003}. Relation of $V_{\lambda}$
with the coupling constants $W_n$ of the states of closed cavity
with waveguides will be given below. Let us first consider the
integrable resonator with $u=0$ shown in Fig. \ref{fig1}. Then
\begin{equation}\label{eigszu0}
  z_{1,2}=-i\Gamma\pm \sqrt{(\epsilon-i\Delta\Gamma)^2-\gamma_1\gamma_2},\\
\end{equation}
where $$\Gamma=\frac{\gamma_1+\gamma_2}{2},
\Delta\Gamma=\frac{\gamma_1-\gamma_2}{2}.$$

For simplicity we take the coupling constants of the cavity
eigenmodes with the propagating mode of waveguide equal
$\gamma_1=\gamma_2$. Such an simplification substantially shortens
algebra of the eigenstates of the effective non Hermitian
Hamiltonian. Then the right eigenstates are
\begin{equation}
\label{eigfunab}
|1)=\frac{1}{\sqrt{2\eta(\eta+i\epsilon)}}\left(\begin{array}{c} -\gamma \cr
     \eta+i\epsilon \cr
    \end{array}\right),\quad
     |2)=\frac{1}{\sqrt{2\eta(\eta-i\epsilon)}}\left(\begin{array}{c} \gamma \cr
      \eta-i\epsilon \cr \end{array}\right)
\end{equation}
with corresponding eigenvalues
\begin{equation}\label{eigsz}
  z_{1,2}=-i\gamma \pm \eta,
\end{equation}
where $\eta=\sqrt{\gamma^2-\epsilon^2}$. Let us write the
following identity
\begin{equation}\label{ll}
    \widehat{W}=\widehat{W}\sum_{\lambda}|\lambda)(\lambda|=\sum_{\lambda}V_{\lambda}|\lambda)
\end{equation}
where $V_{\lambda}$ are the coupling constants between the
resonant states and the continuum.  Therefore from Eq.
(\ref{eigfunab}) we  obtain the link between coupling constants
$W_n$ where $n$ enumerates the closed resonator states and
$V_{\lambda}$ where $\lambda$ enumerates the resonant states:
\begin{equation}
\label{link}
V_1=\frac{W}{\sqrt{2\eta(\eta+i\epsilon)}}(\eta+i\epsilon-\gamma),
~~V_2=\frac{W}{\sqrt{2\eta(\eta-i\epsilon)}}(\eta-i\epsilon+\gamma)
\end{equation}
where $W=W_1=W_2$.

The BIC occurs when $\epsilon=0$. The eigenstates limit to
\begin{equation}
\label{eigfunBIC}
|1)=\frac{1}{\sqrt{2}}\left(\begin{array}{c} -1 \cr
     1 \cr     \end{array}\right),\quad
     |2)=\frac{1}{\sqrt{2}}\left(\begin{array}{c} 1 \cr
     1 \cr     \end{array}\right).
\end{equation}
From Eq. (\ref{link}) one can see that the resonant state $|1)$
decouples from the continuum at $\epsilon=0$ while the state $|2)$
acquires maximal coupling with the continuum (superradiant state).
Therefore the state $|1)$ can be qualified as the FW BIC decoupled
from the continuum owing to exact destructive interference of
leaking eigenmodes of the closed cavity $|1\rangle$ and
$|2\rangle$. For such a simplified case of equal coupling
constants and $u=0$ we see the difference between the FW BIC which
has $V_1(\epsilon=0)=0$ in respect to the eigenstates $|\lambda)$
of $\widehat{H}_{eff}$ and the symmetry protected BIC which has
$W_1=0$ in respect to the eigenstate of $\widehat{H}_B$ of the
closed cavity. General case of $N$ levels was considered in Ref.
\cite{Bulgakov2007} where it is proved that decoupling from all
channels of the continuum described is a necessary and sufficient
condition for a resonance state to be the BIC, i.e., the state
with vanishing decay width.

\begin{figure}[hbt]
\centering{\resizebox{0.5\textwidth}{!}{\includegraphics{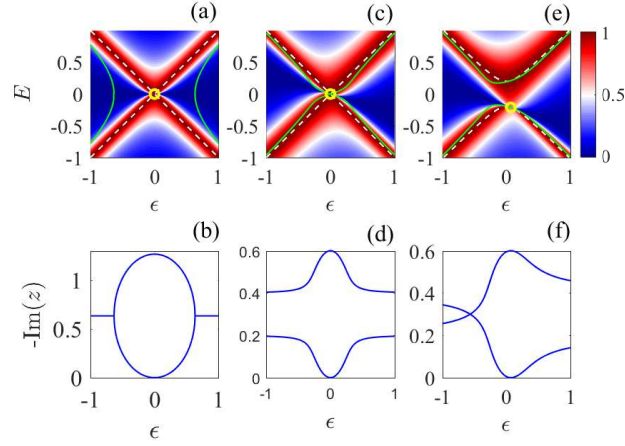}}}
\caption{The transmittance, eigenlevels of closed system (dash
white lines), resonant positions (solid green lines), resonant
widths (solid lines, below) and BICs (open circles) in two-level
description of the effective Hamiltonian (\ref{Heff2}). (a) and
(b) $\gamma_1=\gamma_2=0.1, u=0$; (c) and (d) $\gamma_1=0.1,
\gamma_2=0.2, u=0$; (e) and (f) $\gamma_1=0.1, \gamma_2=0.2,
u=25$.} \label{fig2}
\end{figure}

The transmittance is plotted in Fig. \ref{fig2} (a) which
demonstrates that at the BIC point $E=0, \epsilon=0$ the maximal
transmittance coalesces with the maximal reflectance (collapse of
Fano resonance \cite{Kim1999}).
Simultaneously at the BIC point we observe in Fig. \ref{fig2} (b)
that the resonant width turns to zero.

Let consider the transmittance in the vicinity of the BIC's point
$\epsilon=0, E=0$. The eigenvalues of $\widehat{H}_{eff}$ can be
approximated as $z_1\approx -i\epsilon^2/2\Gamma, ~z_2\approx
-2i\Gamma$. Then the transmission amplitude (\ref{transz}) takes
the simple form
\begin{equation}\label{TBIC}
T(E,\epsilon)\approx -\frac{2iE\Gamma}{2E\Gamma+i\epsilon^2}.
\end{equation}
It follows $|T|=0$  for $E=0, ~\epsilon\neq 0$, and $|T|=1$ for
$\epsilon=0, ~E\neq 0$. Therefore, the BIC is a singular point in
the sense that the value of the transmission amplitude depends on
the way to approach this point. If $\Delta\Gamma\neq 0$ the
transmission zero follows $E=\epsilon\Delta\Gamma/\Gamma$.
In general case of different coupling constants $\gamma_1, ~\gamma_2$
and $u\neq 0$ we will follow Kikkawa {\it et al} \cite{Kikkawa2019}.
We have for the eigenvalues of the effective Hamiltonian (\ref{Heff2})
\begin{equation}\label{z12}
    (z+\epsilon+i\gamma_1)(z-\epsilon+i\gamma_2)-(u-i\sqrt{\gamma_1\gamma_2})^2=0.
\end{equation}
Then we have for the roots of this equation according to the Vietta's formula
\begin{eqnarray}\label{Vietta}
    z_1+z_2=-i(\gamma_1+\gamma_2),&\nonumber\\
   z_1z_2=-(\epsilon+i\gamma_1)(\epsilon-i\gamma_2)-(u-i\sqrt{\gamma_1\gamma_2})^2=
   -\epsilon^2-u^2+2i(u\sqrt{\gamma_1\gamma_2}-\epsilon(\gamma_1-\gamma_2)).&
\end{eqnarray}
At the BIC's point one of the roots, say $z_1$,  is real. In that case, the roots can
be expressed using real quantities $A$ and $B$ as
\begin{eqnarray}\label{z1real}
    &z_1=A,&\nonumber\\
    &z_2=B-i(\gamma_1+\gamma_2).&
\end{eqnarray}
Substitution of Eq. (\ref{z1real}) into Eq. (\ref{Vietta}) gives
\begin{eqnarray}\label{Vietta1}
   & A+B=0,&\nonumber\\
   & AB=-\epsilon^2-u^2.&
\end{eqnarray}
On the other hand, by comparing the imaginary parts of both sides of Eq. (\ref{Vietta})
after substitution we obtain
\begin{equation}\label{A}
    A=-\frac{\epsilon(\gamma_1-\gamma_2)+2u\sqrt{\gamma_1\gamma_2}}{\gamma_1+\gamma_2}.
\end{equation}
Finally from Eqs. (\ref{Vietta1}) and (\ref{A}) we obtain the following equation
for the BIC's point
\begin{equation}\label{BICpoint}
u(\gamma_1-\gamma_2)=2\epsilon\sqrt{\gamma_1\gamma_2}.
\end{equation}
First, this equation for the BIC point in two-level approximation
was obtained by Volya and Zelevinsky \cite{Volya}, and solution is
shown in Fig. \ref{fig2} (e) and (f).

\section{Application to one-dimensional structures}
\label{Sect:1d}
\subsection{Potential well}
\label{PW}
 Let us consider the textbook problem of quantum
particle propagation in one-dimensional potential relief like
shown in Figure 1 in a review by Hsu {\it et al} \cite{Hsu16}. The
wave functions in the segments of the structure are the follows
\begin{eqnarray}\label{1dpot}
&\psi_L(x)=\exp(ikx)+r\exp(-ikx),&\nonumber\\
&\psi(x)=a\exp(iqx)+b\exp(-iqx),&\\
 &\psi_R(x)=t\exp(ikx).&\nonumber
\end{eqnarray}
By use of the boundary conditions we can write the following equation for the solution
\begin{equation}\label{ASpot}
\hat{L}\vec{\psi}= \vec{g},
\end{equation}
where $\hat{L}(k)$ is the following matrix
\begin{equation}\label{L}
  \left(\begin{array}{cccc}
    -1 & 1 &  1 & 0 \cr
     k & q & -q & 0 \cr
     0 & e^{iqL} & e^{-iqL} & -e^{ikL} \cr
     0 & qe^{iqL} & -qe^{-iqL} & -0e^{ikL}
      \end{array}\right),
\end{equation}
$\vec{g}^{T}=(1~ k ~0 ~0 ), ~~\vec{\psi}^{T}=(r ~a ~b ~t)$, and
$L$ is the width of the potential well. The determinant of matrix
$\hat{L}(k)$ equals:
\begin{equation}\label{DetL}
2i(k^2+q^2)\sin(kL)+4kq\cos(kL)
\end{equation}
is the denominator of the S-matrix \cite{Markos} zeros of which
define its poles. The BIC is the solution of the inhomogeneous
part of Eq. (\ref{ASpot}) when $\vec{g}=0$. In order there were a
BIC the determinant (\ref{DetL}) is to be turn to zero that can
not be fulfill for that case of one-dimensional potential well.
Therefore the one-dimensional potential can not support localized
states with energy embedded into the continuum of extended states
with $E>0$. This is the conventional wisdom described in many
books. A bound state in the continuum (BIC) is an exception to
this conventional wisdom: it lies inside the continuum and
coexists with extended waves, but it remains perfectly confined
without any leakage. In 1929, von Neumann and Wigner
\cite{Neumann} discovered that the long-range oscillating
attractive one-dimensional potential can support BICs. The BIC is
a classical paradox of a quantum particle with the energy enough
to leak from the potential well and nevertheless remaining
spatially confined. The Neumann-Wigner  BIC emerges due to precise
destructive interference of waves scattered by a bound potential
in such a way that, after enough distance, we obtain localized
state. The physics of localization is similar to Anderson
localization in random potential \cite{Anderson1958}. For a long
time the phenomenon was considered as mathematical curiosity
because hardly such potentials invented by von Neumann and Wigner
(corrections of the potentials were done by Stillinger
\cite{Stillinger1975}) can be realized experimentally.
\subsection{BICs in Aharonov-Bohm rings}
\label{AB}
The Aharonov-Bohm oscillations of conductance are another bright
example of wave interference when electron encircling upper or
down arms of ring acquires additional magnetic flux phases $\pm
\gamma/2$ where $\gamma=2\pi\Phi/\Phi_0, ~\Phi=B\pi R^2$ is the
magnetic flux, $ \Phi_0=2\pi\hbar c/e$ \cite{Aharonov1959}. In
this subsection we show that particular case of full destructive
interference gives rise to localization of electron inside the
ring, i.e., BICs \cite{Bulgakov2006}.

\begin{figure}[h]
\centering
\includegraphics[width=7cm,clip=]{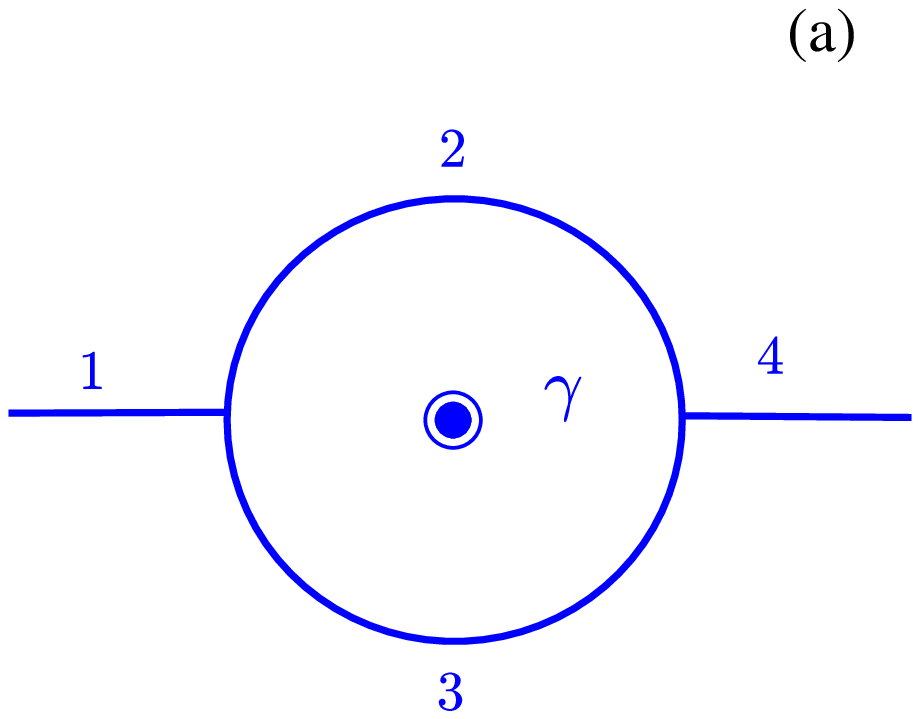}
\includegraphics[width=8cm,clip=]{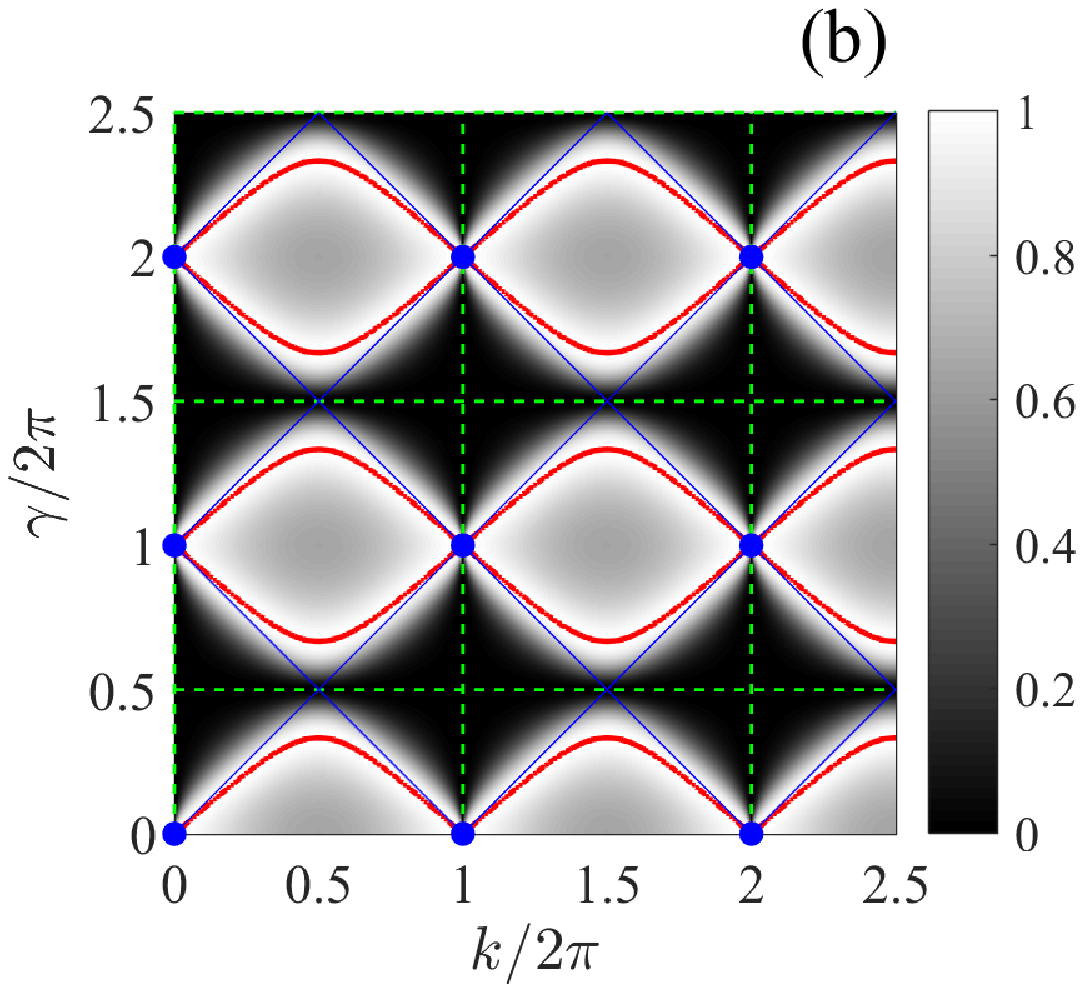}
\caption{(a) One-dimensional ring thread by the magnetic flux
$\gamma$ and opened by attachment of two leads. (b) Transmission
through the ring vs the magnetic flux and wave number
$k=\sqrt{E}$. Green dash lines show transmission zeros $|t|^2=0$
and solid red lines show transmission ones $|t|^2=1$. The BICs are
marked by blue closed circles. Blue thin solid lines show wave
numbers as dependent on the flux $k=(m-\gamma)$ where $m$ are
integers.} \label{fig3}
\end{figure}

Following Xia \cite{Xia1992} we write the wave functions in the
segments of the structure shown in Fig. \ref{fig3} (a) as
\begin{equation}\label{psi1d}
\begin{array}{rcl}
\psi_1(x)&=&\exp(ikx)+r\exp(-ikx),\\
\psi_2(x)&=&a_1\exp(ik^-x)+a_2\exp(-ik^+x),\\
\psi_3(x)&=&b_1\exp(ik^+x)+b_2\exp(-ik^-x),\\
\psi_4(x)&=&t\exp(ikx),
\end{array}
\end{equation}
where $k^-=k-\gamma, ~k^+=k+\gamma$. All variables are
dimensionless via the ring length $2\pi R$. The boundary
conditions (the continuity of the wave functions and the
conservation of the current density) allow to find all
coefficients in (\ref{psi1d}). We write the corresponding equation
in matrix form
\begin{equation}\label{AS}
\hat{F}\vec{\psi}= \vec{g},
\end{equation}
where $\hat{F}(k,\gamma)$ is the following matrix
\begin{equation}\label{matrixF}
  \left(\begin{array}{cccccc}
    -1 & 0 & 1 & 1 & 0 & 0 \cr
    -1 & 0 & 0 & 0 & 1 & 1 \cr
    0 & -1 & e^{ik^-/2} & e^{-ik^+/2} & 0 & 0 \cr
    0 & -1 & 0 & 0 & e^{ik^+/2} & e^{-ik^-/2} \cr
    1 & 0 & \frac{k^-}{k}& -\frac{k^+}{k}& \frac{k^+}{k} &
    -\frac{k^-}{k}\cr
    0 & -1 &\frac{k^-}{k}e^{i\frac{k^-}{2}} & -\frac{k^+}{k}e^{-i\frac{k^+}{2}} &
    \frac{k^+}{k}e^{i\frac{k^+}{2}} &
    -\frac{k^-}{k}e^{-i\frac{k^-}{2}} \end{array}\right),
\end{equation}
$\vec{g}^{T}=(1~ 1 ~0 ~0 ~1 ~0)$. The vector $\vec{\psi}^{T}=(r ~t
~a_1 ~a_2 ~b_1 ~b_2)$ is the solution for the scattering wave
function:
\begin{equation}\label{rta}
\begin{array}{rcl}
r&=&2(3\cos{k}-4\cos{\gamma}+1)/Z,\\
t&=&16i(\sin{\frac{k}{2}}\cos{\frac{\gamma}{2}})/Z\\
a_1&=&2(2e^{i\gamma}-3e^{-ik}+1)/Z,\\
a_2&=&2(e^{ik}+1-2e^{i\gamma})/Z,\\
Z&=&8\cos{\gamma}-9e^{-ik}-e^{ik}+2,\\
\end{array}
\end{equation}
$b_{1,2}(k,\gamma)=a_{1,2}(k,-\gamma)$. In Fig. \ref{fig3} we show
lines of the transmission zeros ($|t(k,\gamma)|=0$, dashed lines)
which cross the lines of the transmission ones ($|t(k,\gamma)|=1$,
solid lines) at points
\begin{equation}\label{BIC_points}
\begin{array}{rcl}
  k_m&=&2\pi m, m=\pm 1, \pm2, \ldots,\\
  \gamma_n&=&2\pi n, n=0, \pm 1, \pm2, \ldots.
\end{array}
\end{equation}
As can be seen from the expression for the denominator $Z$ in Eq.
(\ref{rta}), the imaginary part of the  poles vanishes at these
points. Simultaneously, a degeneracy of eigen energies of closed
ring $(k_m-\gamma)^2$ occurs at these points. Here $m$ is the
azimuthal index (magnetic quantum number) that defines the
eigenfunctions of the closed ring $\psi_m(x)=\exp(ik_m x)$. The
point $k=0$ is excluded from the consideration since it gives zero
conductance. An existence of the peculiar points
(\ref{BIC_points}) were shown in Ref. \cite{Wunsch2003} as a
points where the density of states shows similar collapses as
collapses of the Fano resonance in the transmission. To show that
the BICs appear at the points (\ref{BIC_points}), let us consider
one of the points, say, ${\bf s}_0= (k_1,\gamma_1)=2\pi(1,~1)$.
All the other points are equivalent because of the periodical
dependence of the system on $k$ and $\gamma$. In the vicinity of
the point ${\bf s}_0$ we write Eq. (\ref{rta})   in the following
approximated form
\begin{eqnarray}\label{Sappr}
t\approx \frac{\Delta k}{\Delta k+i(\Delta\gamma)^2/2},~ r\approx
\frac{i(3\Delta k^2-4\Delta\gamma^2)}{4(2\Delta
k+i\Delta\gamma^2)},\nonumber\\
 a_1\approx
\frac{3\Delta k+2\Delta\gamma}{4\Delta k+2i\Delta\gamma^2},~
a_2\approx \frac{\Delta k-2\Delta \gamma}{4\Delta
k+2i\Delta\gamma^2},
\end{eqnarray}
where $\Delta k=k-k_1, ~\Delta\gamma=\gamma-\gamma_1$. The
transmission amplitude in the vicinity of the BIC point ${\bf
s}_0$ in (\ref{Sappr}) is similar to the expressions obtained for
a shifted von Neumann-Wigner potential \cite{Pursey1994} or
two-level approximated approach (see Eq. (\ref{TBIC})). One can
see that all amplitudes $a_{1,2}, ~b_{1,2}$ of the inner wave
functions are singular at the point ${\bf s}_0$. Such a result for
the BIC points was firstly found by Pursey and Weber
\cite{Pursey1994}. At this point the matrix (\ref{matrixF}) takes
the following form
\begin{equation}\label{matrixa}
  \hat{F}({\bf s}_0)=\left(\begin{array}{cccccc}
    -1 & 0 & 1 & 1 & 0 & 0 \cr
    -1 & 0 & 0 & 0 & 1 & 1 \cr
    0 & -1 & 1 & 1 & 0 & 0 \cr
    0 & -1 & 0 & 0 & 1 & 1 \cr
    1 & 0 & 0& -2 & 2 & 0 \cr
   0 & -1 & 0 & -2 & 2 & 0 \end{array}\right).
\end{equation}
The determinant of the matrix $\hat{F}({\bf s}_0)$ equals zero.
Therefore, $\hat{F}\vec{f_0}=0$. By direct substitution of the
vector
 $\vec{f_0}^{T}=\frac{1}{2}(0 ~0 ~1 ~-1 ~-1 ~1)$
one can verify that $\vec{f}_0$ is the right eigenvector which is
the null vector. The corresponding left null eigenvector is
$\tilde{\vec{f}}_0=\frac{1}{2}(-1 ~1 ~1 ~-1 ~0 ~0)$. It is well
known from linear algebra, that if the determinant of matrix
$\hat{F}$ is equaled to zero, then the necessary and sufficient
condition for existence of solution of the equation (\ref{AS}) is
that the vector $\tilde{\vec{f}}_0$ is orthogonal to vector
$\vec{g}$ \cite{Smirnov}. In holds, indeed,
$\tilde{\vec{f}}_0\cdot\vec{g}=0$. Therefore, the null vector
$\tilde{\vec{f}}_0$ is proven to be the BIC. The general solution
of Eq. (\ref{AS}) at the point ${\bf s}_0$ can therefore be
presented as
\begin{equation}\label{BIC_sol}
\vec{\psi}({\bf s}_0)=\alpha\vec{f}_0+ \vec{\psi_p},
\end{equation}
where $\alpha$ is an arbitrary coefficient and $\vec{\psi_p}$ is
particular transport solution of Eq. (\ref{AS}). By direct
substitution one can verify that $\vec{\psi_p}^T=\left(0 ~1
~\frac{3}{4} ~\frac{1}{4} ~\frac{3}{4} ~\frac{1}{4}\right)$ is the
particular solution of Eq. (\ref{AS}). It is worthwhile to note
that this result completely agrees with the scattering theory on
graphs \cite{Texier2002,Texier2003}. Texier  has shown that for
certain graphs the stationary scattering state gives the solution
of the Schr\"odinger equation for the continuum spectrum apart for
discrete set of energies where some additional states are
localized in the graph and thus are not probing by scattering,
leading to the failure of the state counting method from the
scattering.
\subsection{Zeeman localization}
\label{Zeeman}
 Although open the Aharonov-Bohm ring consists of 1d
wires, the ring is two-dimensional in order electron could
encircle the flux. In this subsection we present the model which
is indeed one-dimensional but capable to localize electron. We go
beyond the scalar Helmhotz equation and employ the interference of
spin polarized resonant states of the one-dimensional electron
transmission \cite{Pankin2020}.
\begin{figure}[ht]
\centering{\resizebox{0.5\textwidth}{!}{\includegraphics{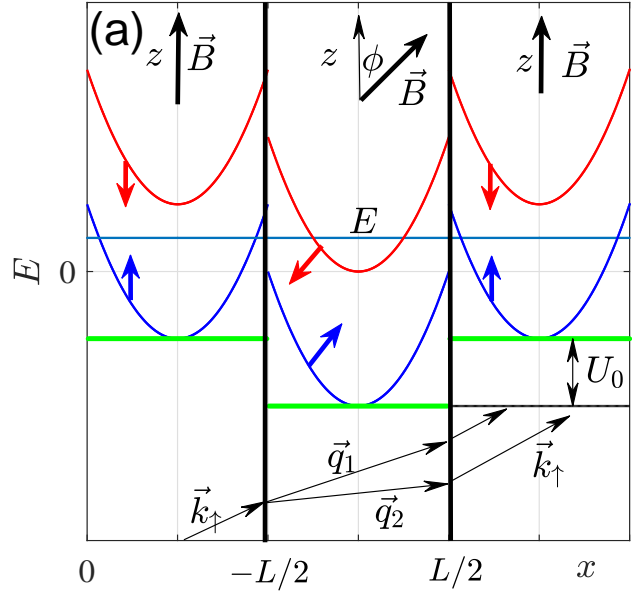}}}
\caption{An one-dimensional spin model for illustration of BICs
due to full destructive interference of spin polarized resonant
states. Beyond the central layer, the magnetic field $\vec{B}$ is
directed along the $z$-axis, inside the central layer $\vec{B}$ is
tilted by angle $\phi$. The spin-up electron falls by angle
$\theta$ with the energy below the spectrum of spin-down and
splits into two states specified by $\vec{k}_1$ and $\vec{k}_2$. }
\label{fig4}
\end{figure}
Let us consider three domains in which external stationary
magnetic field is applied as sketched in Fig. \ref{fig4}. Assume
the external magnetic field $\vec{B}$ inside the central layer is
tilted relative to the outer magnetic field oriented along
$z$-axis. We also assume that the inner layer has the potential
shifted relative to the outer layers by a value $U_0$. Outside of
the central layer electron has two split energy spectra
$E=k_{\sigma}^2 \mp B, \; \sigma= \uparrow, \downarrow$ which
specify the continua by the wave vector
$\overrightarrow{k}_{\sigma}$. In the central layer the spin
dependent spectra have the following form $E=q_{s}^2+U_0 \mp B, \;
s=1,2$ which specify spin dependent channels by the vector
$\overrightarrow{q}_s$. Owing to choice of the potential step
($U_0=-20$) as depicted in Fig.~\ref{fig4}  by green both spin
channels are open in the central layer while outside only the spin
up continuum is open for $E<B$. Therefore only the electron with
spin up participates in electron transmission and reflection.

Let us write the Schr\"odinger equation for the toy model of
electron in magnetic field (see Fig. \ref{fig4}):
\begin{equation}
[\frac{1}{2m}(i\hbar\nabla+\frac{e}{c}{\bf A})^2 + U_0(z) -{\bf
\sigma B}(z) -  E ] \Psi = 0. \label{Schrod}
\end{equation}
The orbital motion has characteristic length $a_B^2=\frac{\hbar c}{eB}$ which in the
magnetic field of $10^3 Oe$ equals 100 nm. Then for layer of thickness $L\ll a_B$
we can disregard the orbital contribution in Eq. (\ref{Schrod}) and rewrite
as follows
\begin{equation}
\left[\nabla^2 - U_0(z) + {\bf \sigma B}(z)+ E \right]\Psi = 0.
\label{Schrodinger}
\end{equation}
Next we substitute the step-wise magnetic filed as shown in Fig.
\ref{fig4}. Then Hamiltonian (\ref{Schrodinger}) will take the
following form
\begin{equation}
\label{H}
\widehat{H}= \left\{\begin{array}{cc}
-\frac{d^2}{dz^2}-\sigma_xB & \mbox{if $|z|>L/2$};
\\ -\frac{d^2}{dz^2}+U_0
-\sigma_x B\cos{(\phi)}-\sigma_zB\sin{(\phi)} & \mbox{if $|z|<L/2$}.
\end{array}\right.
\end{equation}
In the outer layers which form the radiation continua with the following propagating
solutions
\begin{equation}
\label{psi_out}
    \Psi_{\vec{k}_{\sigma}}(\vec{x})=
    \exp(i\vec{k}_{\sigma}\vec{x})|\sigma\rangle,
\end{equation}
where $\sigma=\uparrow,\downarrow$,
\begin{equation}\label{sigma}
|\uparrow\rangle=\left(%
\begin{array}{c}
  1 \\   0 \\
\end{array}%
\right), ~~
|\downarrow\rangle=\left(%
\begin{array}{c}
  0 \\  1 \\
\end{array}%
\right),
\end{equation}
and
\begin{equation}\label{E_sigma}
    E=k_{\sigma}^2\mp B.
\end{equation}

Respectively for the inner layer we have
\begin{equation}
\label{psi_in}
    \Psi_{\vec{q}_s}(\vec{x})=
    \exp(i\vec{q}_s\vec{x})|s\rangle,
\end{equation}
where $s =1,2$,
\begin{equation}\label{inside1}
|1\rangle=\left(%
\begin{array}{c}
\cos{(\phi/2)} \\  \sin{(\phi/2)} \\
\end{array}%
\right), ~~
|2\rangle=\left(%
\begin{array}{c}
  -\sin{(\phi/2)} \\ \cos{(\phi/2)} \\
\end{array}%
\right),
\end{equation}
and
\begin{equation}\label{E_s_inside}
    E=q_s^2+U_0\mp B.
\end{equation}

\begin{figure}[hbt]
\centering{\resizebox{0.8\textwidth}{!}{\includegraphics{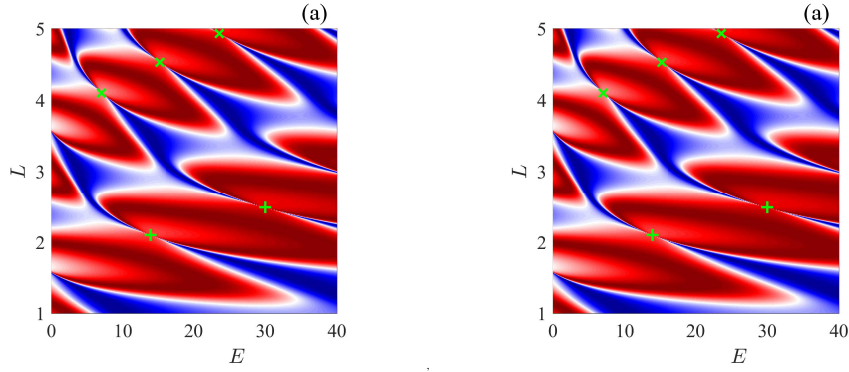},\includegraphics{Fig5a}}}
\caption{Reflection probability of the spin-up electron for $B=10$
tilted by angle $\phi=\pi/3$ and $U_0=-20$ vs (a) incident energy
$E$ and central layer thickness $L$ at angle of incidence
$\theta=\pi/4$, and (b) vs angle of incidence $\theta$ and central
layer thickness $L$ at incident energy $E = 30$ and $\phi=\pi/4$.
Magenta pluses mark the points of the BICs symmetric with respect
to the center of the layer, while crosses mark the points of
antisymmetric BICs.}
\label{fig5}
\end{figure}

\begin{figure}[ht]
\centering{\resizebox{0.35\textwidth}{!}{\includegraphics{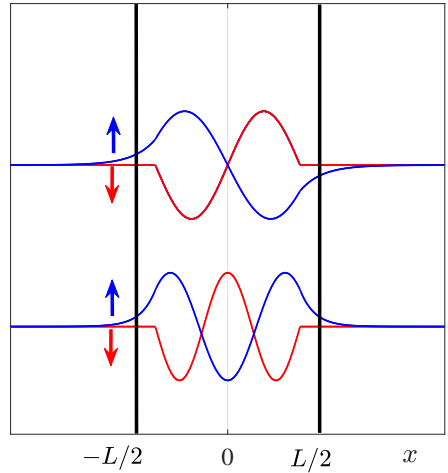}}}
\caption{The BIC solutions,  symmetric and antisymmetric, in the
layered structure.} \label{fig6}
\end{figure}

Let us choose the energy of incident electron that only spin up
channel is open. Then at the left ($z<-L/2$) we have
\begin{equation}\label{left}
\Psi_L(\vec{x})=(e^{i\vec{k}_{\uparrow}\vec{x}}
+r_{\uparrow}e^{-i\vec{k}_{\uparrow}\vec{x}})|\uparrow\rangle.
\end{equation}
Inside the defect layer ($|z|<L/2$) both channels are open due to
proper choice of the potential $U_0=-20$ and therefore one can
present the solutions as follows
\begin{equation}\label{inside}
\Psi_B(\vec{x})=\sum_{s=1,2}(a_se^{i\vec{q}_s
\vec{x}}+b_se^{-i\vec{q}_s\vec{x}})|s\rangle.
\end{equation}
At last at the right side ($z>L/2$) we write
\begin{equation}\label{right}
\Psi_R(\vec{x})=t_{\uparrow}e^{i\vec{k}_{\uparrow}
\vec{x}}|\uparrow\rangle.
\end{equation}
Here $r_{\uparrow}$ and $t_{\uparrow}$ are the reflection and
transmission amplitudes. Next, assume electron with spin $\sigma$
incidents with wave vector $\vec{k}_{\sigma}=(k_{x \sigma},k_{z
\sigma})$ and reflecting with the reflection amplitude
$r_{\sigma}$. Because of preservation of transverse component of
moment $k_{x \sigma}=k_{x s}=k_x$ we obtain the following
equations:
\begin{eqnarray}\label{sew}
&1+r_{\uparrow}=(a_1+b_1)\cos{(\phi/2)}-(a_2+b_2)\sin{(\phi/2)}, &\nonumber\\
&k_{z\uparrow}(1-r_{\uparrow})=q_{z1}(a_1-b_1)\cos{(\phi/2)}-q_{z2}(a_2-b_2)\sin{(\phi/2)}, &\nonumber\\
&r_{\downarrow}=(a_1+b_1)\sin{(\phi/2)}+(a_2+b_2)\cos{(\phi/2)}, &\nonumber\\
&-k_{z\downarrow}r_{\downarrow}=q_{z1}(a_1-b_1)\sin{(\phi/2)}+q_{z2}(a_2-b_2)\cos{(\phi/2)}, &\nonumber\\
&t_{\uparrow}e^{ik_{z\uparrow}L}=(a_1e^{iq_{z1}L}+b_1e^{-iq_{z1}L})\cos{(\phi/2)}
-(a_2e^{iq_{z2}L}+b_2e^{-iq_{z2}L})\sin{(\phi/2)}, &\nonumber\\
&k_{z\uparrow}t_{\uparrow}e^{ik_{z\uparrow}L}=q_{z1}(a_1e^{iq_{z1}L}-b_1e^{-iq_{z1}L})\cos{(\phi/2)}
-q_{z2}(a_2e^{iq_{z2}L}-b_2e^{-iq_{z2}L})\sin{(\phi/2)}, &\nonumber\\
&t_{\downarrow}e^{ik_{z\downarrow}L}=(a_1e^{iq_{z1}L}+b_1e^{-iq_{z1}L})\sin{(\phi/2)}
+(a_2e^{iq_{z2}L}+b_2e^{-iq_{z2}L})\cos{(\phi/2)}, &\nonumber\\
&k_{z\downarrow}t_{\downarrow}e^{ik_{z\downarrow}L}=q_{z1}(a_1e^{iq_{z1}L}-b_1e^{-iq_{z1}L})\sin{(\phi/2)}&\nonumber\\
&+q_{z2}(a_2e^{iq_{z2}L}-b_2e^{-iq_{z2}L})\cos{(\phi/2)} .&
\end{eqnarray}
The transmission probability versus the thickness of potential
well $L$ and incident energy  or angle of incidence $\theta$  is
plotted in Fig. \ref{fig5} (a) and (b) respectively where one can
see typical points for BICs with collapse of Fano resonance is
observed.
These points unambiguously indicate the BIC points. Indeed, the
BIC as localized mode inside the layer can be found from equations
of continuity at the interfaces. These equations can be simplified
with account of symmetry relative to $z\rightarrow -z$. Then the
symmetric BIC can be written as
\begin{equation}\label{BIC_s}
    \psi_{BIC,sym}(z)=\left\{\begin{array}{cc}
    a\cos{(q_{z1}z)} |1\rangle +b\cos{(q_{z2}z)} |2\rangle & \mbox{if $|z|< L/2$}\\
    c e^{(-|k_{z\downarrow}|z)} |\downarrow\rangle   & \mbox{if $|z| > L/2$}.
 \end{array}\right.
    \end{equation}
where the last contribution is the result of evanescent mode with
spin down and asymmetric BIC
\begin{equation}\label{BIC_a}
    \psi_{BIC,asym}(z)=\left\{\begin{array}{cc}
    a\sin{(q_{z1}z)} |1\rangle +b\sin{(q_{z2}z)} |2\rangle & \mbox{if $|z|< L/2$}\\
    sign(z) c e^{(-|k_{z\downarrow}|z)}|\downarrow\rangle   & \mbox{if $|z| > L/2$}.
    \end{array}\right.
    \end{equation}
We imply that the modes equal zero at the spin up continuum and obey the continuity equations
for the evanescent mode spin down. As the result we obtain the
following equations for the symmetric BIC
\begin{eqnarray}\label{BICsym}
&a\cos{(\phi/2)}\cos{(q_{z1}L/2)} - b\sin{(\phi/2)}\cos{(q_{z2}L/2)}=0,&\nonumber\\
&a\sin{(\phi/2)}\cos{(q_{z1}L/2)} + b\cos{(\phi/2)}\cos{(q_{z2}L/2)}=c e^{(-|k_{z\downarrow}|L/2)},&\nonumber\\
&aq_{z1}\sin{(\phi/2)}\sin{(q_{z1}L/2)} + bq_{z2}\cos{(\phi/2)} \sin{(k_{z2}L/2)} =
c|k_{z\downarrow}| e^{(-|k_{z\downarrow}|L/2)}.&
\end{eqnarray}
Thus, we obtain the following equation for the symmetric BIC
points
\begin{equation}\label{BICasym}
    -\tan^2{(\phi/2)}=\frac{q_{z2}\tan{(q_{z2}L/2)}-|q_{z\downarrow}|}
    {k_{z1}\tan{(q_{z1}L/2)}-|k_{z\downarrow}|},
\end{equation}
and respectively for the asymmetric BIC points
\begin{equation}\label{BIC_p}
    -\tan^2{(\phi/2)}=\frac{q_{z2}\cot{(q_{z2}L/2)}+|k_{z\downarrow}|}
    {q_{z1}\cot{(q_{z1}L/2)}+|k_{z\downarrow}|}.
\end{equation}
The solutions of Eqs. (\ref{BICsym}) and (\ref{BICasym}) are
marked in Fig. \ref{fig5} by pluses and crosses respectively which
exactly coincide with points of Fano resonance collapse. The
lowest symmetric and antisymmetric BIC solutions are shown in Fig.
\ref{fig6}.

In Table \ref{Tab1} we establish the one-by-one correspondence
between the spin of the electron and the polarization state of
light. Owing that these BICs were verified experimentally by full
destructive interference of light paths with TM and TE
polarizations in anisotropic layer \cite{Pankin2020}.
\begin{table}
        \centering
                \caption{Quantum/Optical correspondence.} \label{Tab1}
\begin{tabular}{c|c}
\hline\hline
Quantum mechanics & Optics \\
\hline
electron & photon \\
$\psi$, $\frac{\partial \psi}{\partial z}$ & ${\bf E}$, ${\bf B}$ \\
spin & polarization \\
energy & frequency\\
$|\downarrow\rangle$ & TE-wave \\
$|\uparrow\rangle$ & TM-wave \\
magnetic field & anisotropy axis \\
\hline \hline
\end{tabular}
\end{table}

\section{BICs in two-dimensional planar open cavities}
\label{Sect:2dres}
Two- and three-dimensional wave transmission through cavities is
distinct of one-dimensional transmission. First, by change of
shape of the 2d or 3d cavity we can achieve a degeneracy in 1d
resonator and therefore avoided crossing of resonances. Second, 2d
and 3d waveguides attached to the 2d and 3d cavities can support
finite number of open channels as dependent on wave frequency. The
other channels are closed forming evanescent modes whose role is
crucially important for BICs. The evanescent modes of waveguide
shift the BIC points and "blow out" the BIC modes from the open
resonator. Moreover in 3d resonators the evanescent modes play
principal role to give rise to the BICs.

\begin{figure}
\centering{\resizebox{0.4\textwidth}{!}{\includegraphics{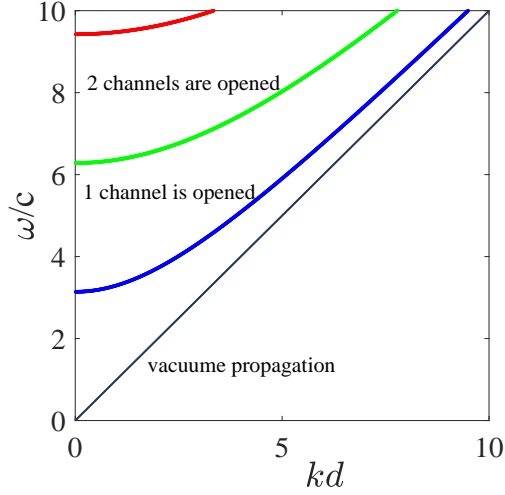}}}
\caption{Dispersion curves of open channels in planar waveguide
with rectangular cross-section.}\label{fig7}
\end{figure}

\begin{figure}[ht]
\centering{\resizebox{0.4\textwidth}{!}{\includegraphics{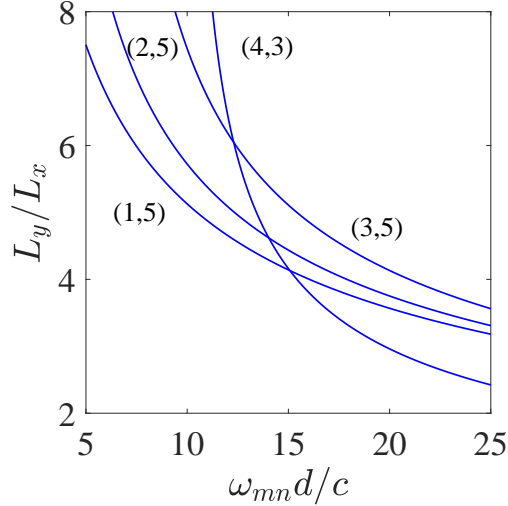}}}
\caption{Selected frequencies (\ref{eigfreqrect}) of the closed
rectangular resonator vs width $L_y$. $E_0=c^2/d^2$.}\label{fig8}
\end{figure}

\begin{figure}[ht]
\centering{\resizebox{0.4\textwidth}{!}{\includegraphics{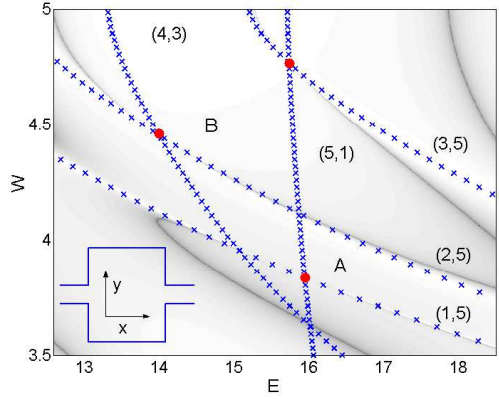}}}
\caption{The transmittance in Log scale through the rectangular
resonator shown in the inset versus energy  $E=\omega^2$ and width
$W$ of the resonator (in terms of the width of the lead). The dark
areas correspond to low transmittance.  The length of the
resonator along the transport axis equals 4. The eigenfrequencies
of the closed billiard are marked by crosses. The positions of the
BICs are shown by closed red circles. The patterns of the two BICs
A and B are shown in Fig. \ref{fig11}.}\label{fig9}
\end{figure}

In order to illustrate these statements we start with the planar
microwave metallic cavity or resonator with the Dirihclet boundary
conditions at the walls. Such a system is convenient by that the
solutions with different polarizations, TE and TM, are separated
\cite{Jackson}. The total system can be viewed as consisted of
three subsystems: two semi infinite planar waveguides and
rectangular plane resonator. In each subsystem the solution obeys
the Helmgoltz  equation \cite{Stockmann}
\begin{equation}\label{Helm}
    -\nabla^2\psi(x,y)=\frac{\omega^2}{c^2}\psi(x,y).
\end{equation}
In what follows all quantities are measyred in terms of the light
velocity $c$. This equation is completely equivalent to the case
of electron transmission in microwave waveguides
$$ -\nabla^2\psi(x,y)=\frac{2m^{*}E}{\hbar^2}\psi(x,y)$$ where
$m^*$ is the effective electron mass with energy $E$. In the plane
waveguides the solutions are given by TE propagating waves
\cite{Jackson}
\begin{equation}\label{Ez}
    \psi_p(x,y)=\sqrt{\frac{1}{2\pi k_p}}\exp(ik_px)\phi_p(y)
\end{equation}
\begin{equation}\label{phip}
    \phi_p(y)=\sqrt{2}\sin(\pi py)
\end{equation}
 with the eigenfrequency spectra
\begin{equation}\label{disp}
    \omega^2=k_p^2+\pi^2 p^2, p=1,2,3,\dots.
\end{equation}
Here $\psi_p(x,y)=E_z(x,y)$ responses for the electric field
component of EM field. The integer $p$ numerates channels which
are opened for increasing of the frequency as shown in Fig.
\ref{fig7}.
Other components of EM field can be easily expressed through
$\psi(x,y)$ by use of the Maxwell equations \cite{Jackson}. The
solutions inside the closed rectangular resonator are the
following
\begin{equation}\label{eigfuncrect}
    \psi_{mn}(x,y)=2\sqrt{\frac{1}{L_xL_y}}\sin\left(\frac{\pi mx}{L_x}\right)
\sin\left(\frac{\pi ny}{L_y}\right)
\end{equation}
with the discrete eigenfrequencies
\begin{equation}\label{eigfreqrect}
    \omega_{m,n}^2=\frac{\pi^2 m^2}{L_x^2}+\frac{\pi^2 n^2}{L_y^2}
\end{equation}
where $m$ and $n$ are integers. Here and further all dimensional
quantities are measured in the terms of the waveguide's width $d$,
i.e., $d=1$. These eigenfrequencies as dependent on the resonator
width $W$ are shown in Fig. \ref{fig8}.
In respect to the non Hermitian effective Hamiltonian approach it
is important to note that the Helmgoltz equation (\ref{Helm}) one
by one is equivalent to the quantum mechanical description of the
electron transmission through quantum dots with attached quantum
wires. The squared frequency can be expressed as the quantum
energy $E=\omega^2$ and the electric field directed perpendicular
to metallic planes is equivalent to the quantum wave function
\cite{Stockmann} $E_z=\psi(x,y)$.
However the effective non Hermitian Hamiltonian (\ref{Heffphen})
is to be modified with account of dispersion properties of
microwave waveguides (\ref{disp}) as follows
\cite{Pichugin2001,SR2003}:
\begin{equation}\label{Heffplan}
    \mathcal{\widehat{H}}_{eff}=\mathcal{\widehat{H}}_B-i\sum\limits_{C=L,R}\sum_p W_{Cp}W_{Cp}^{\dagger},
\end{equation}
where the matrix elements of the coupling matrix elements between the $m,n$-th eigenmode of the closed resonator
and the $p$-th propagation channel of the $C$-th waveguide equal
\begin{equation}\label{WD}
    W_{mn;pC}=\sqrt{\frac{1}{\pi k_p}}\left.\int_{-1/2}^{1/2}dy\sin\left(\frac{\pi py}{d}\right)
    \frac{\partial{\psi_{mn}(x,y)}}{\partial x}\right|_{x=x_C},
\end{equation}
$C=L,R$ enumerates the interfaces between the left and right
waveguides shown in Fig. \ref{fig1} by dashed lines $x_L=-L_x/2,
x_R=L_x/2$. We pay attention that the overlapping is given by
derivatives of the eigenfunctions of closed resonator over the
transmission direction but not the eigenfunctions themselves which
equal zero at the boundaries shown in Fig. \ref{fig1} by dash
lines. In the present case of planar resonator this direction is
the x-direction as shown in Fig. \ref{fig1}. For the case of TM
waves the magnetic field $H_z(x,y)$ serves as the  wave function
$\psi(x,y)$ with the Neumann boundary conditions at the metallic
walls of waveguide that makes the problem fully equivalent to
transmittance of acoustic waves in hard wall resonator. In that
case the form of the effective Hamiltonian remains the same but
the coupling matrix elements takes the following form
\cite{Maksimov2015,Pichugin2001}
\begin{equation}\label{WN}
    W_{mn;pC}=\sqrt{\frac{k_p}{\pi}}\left.\int_{-1/2}^{1/2}dy\sin\left(\frac{\pi py}{d}\right)
    \psi_{mn}(x,y)\right|_{x=x_C}.
\end{equation}
The S-matrix is given \cite{Dittes2000,Stockmann2002}
\begin{equation}\label{trans}
    S_{CC'}=\delta_{CC'}-2i\widehat{W}^C\frac{1}{E-\widehat{H}_{eff}}
    \widehat{W}^{C'}.
\end{equation}
In Fig. \ref{fig9} we show the transmittance in the first open
channel vs incident frequency $E=\omega^2$ and the width of
resonator $W$.

In the framework of this formalism, the positions and decay widths
of the resonance states follow from the complex eigenvalues of the
non-Hermitian effective Hamiltonian
\begin{equation}\label{Heffeig}
    \hat{H}_{eff}|\lambda)=z_{\lambda}|\lambda)
\end{equation}
where $z_{\lambda}=E_{\lambda}-i\gamma_{\lambda}/2$. The
biorthogonal eigenstates are normalized as
$(\lambda|\lambda')=\delta_{\lambda,\lambda'}$, where $(\lambda|$
is given by transpose of $|\lambda)$. Similar to the two-level
approach for description of BICs in Section 3 the BIC of the
present formalism is given by those eigenstate of the effective
Hamiltonian, whose eigenvalue is real. However as distinct of
phenomenological case by Friedric and Wintgen \cite{Friedrich1985}
(see also Ref. \cite{Volya}) the coupling matrix elements
(\ref{WN}) are frequency dependent through Eq. (\ref{disp}). Then
the resonant positions and widths are obtained by solving the
corresponding fixed-point equations \cite{Rotter2003}
\begin{equation}\label{fix point}
  E_{\lambda}=Re({z_\lambda(E_{\lambda})}), ~~ 2\gamma_{\lambda}=
-Im({z_\lambda(\gamma,E_{\lambda})}).
\end{equation}
Moreover the rank of matrix of the effective Hamiltonian is
defined by number of the eigenmodes of closed resonator whose
number rigorously speaking is infinite. In order to solve the
eigenvalue problem one has to decimate the matrix however a
convergence of the matrix of the effective Hamiltonian is
controversial for the Dirichlet BC \cite{Maksimov2015}. In
practice we explore  the tight-binding approach for the effective
Hamiltonian \cite{SR2003} which is equivalent to finite difference
method of solution of the Helmholtz equation (\ref{Helm}).
\begin{figure}[ht]
\centering{\resizebox{0.35\textwidth}{!}{\includegraphics{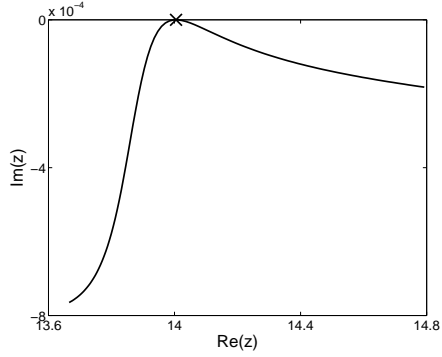}}}
\caption{ The evolution of  resonance width $Im(z)$ and position
$Re(z)$ of one of the two resonance states in the vicinity of the
BIC B in Fig. \ref{fig9}. $Im(z)$ vanishes at $L_y=4.45$ (marked
by a cross).  The widths of the other  resonance states are much
larger and are not shown here.} \label{fig10}
\end{figure}
The one half of eigenvalues of the effective non Hermitian
Hamiltonian (\ref{Heffplan}) are real and correspond to the
symmetry protected BICs because they  are antisymmetric relative
to $y\rightarrow -y$ and therefore have zero couplings (\ref{WD})
with the first channel continuum $p=1$ provided that
$\omega^2<4\pi^2$. The second half of the eigenvalues is complex
and correspond to resonances for wave transmission through the
rectangular resonator. However a very few of these complex
eigenvalues have a tendency to acquire zero imaginary parts for
variation of the width of the resonator at the vicinity of those
points where a degeneracy of the eigenfrequencies
(\ref{eigfreqrect}). One of such events is shown in Fig.
\ref{fig10} where other eigenfrequencies are excluded in order to
avoid obscure picture.

One can see from Fig. \ref{fig9} that these BICs are located in
the very close vicinity to the points of degeneracy of the
eigenmodes of the close resonator. Indeed when the eigenmodes, say
$\psi_1$ become degenerate one can superpose the eigenmodes as
$\psi=a_1\psi_1+a_2\psi_2$. Although each eigenmode is coupled
with the continuum $|C\rangle $ via the coupling constants
$W_1=\langle C|\psi_1\rangle\neq 0$ and $W_2=\langle
C|\psi_2\rangle \neq 0$ the coupling of the superposed state
$W=\langle C|\psi\rangle=a_1W_1+a_2W_2$ can be cancelled by a
proper choice of the superposition coefficients $a_1$ and $a_2$
\cite{SBR}. Then this state $\psi$ becomes the BIC which is
decoupled from the waveguides for the case $E<4\pi^2$.

{\bf The BIC function.}
 In general case the scattering wave function
obeys the following equation \cite{SR2003,Maksimov2015}
\begin{eqnarray}\label{scatfun}
&\psi_L(x,y)=\frac{1}{\sqrt{4\pi k_1}}[e^{ik_1x}\phi_1(y)+
\sum_pS_{1L;pL}e^{-ik_px}\phi_p(y)], \ x<-L_x/2,&\nonumber\\
&\psi_B(x,y)=-i\sum_{m'n'}G_{mn;m'n'}\sqrt{\frac{k_{p=1}}{\pi}}W_{m'n';1L}\psi_{m'n'}(x,y),\ -L_x/2<x<L_x/2,&\\
&\psi_R(x,y)=\sum_p\frac{1}{\sqrt{4\pi
k_p}}S_{1L;pR}e^{ik_px}\phi_p(y), x>L_x/2,&\nonumber
\end{eqnarray}
where $S_{pC,p'C'}$ are components of the S-matrix (\ref{trans})
and the Green function $\widehat{G}$ is the inverse of the matrix
$\omega^2-\widehat{H}_{eff}$. So, inside the resonator the wave
function is given by the Lippmann-Schwinger equation
\cite{SR2003,Maksimov2015}
\begin{equation}
\label{LS}
    (\omega^2-\widehat{H}_{eff})|\psi_B\rangle=\hat{W}_{Lp=1}a^*_{L,p=1}|L,p=1\rangle
    \nonumber
\end{equation}
where the waveguide states are given by incoming wave amplitude
$a^{+}_{L,p=1}$  for the present case of 2d wave transmission
shown in inset of Fig. \ref{fig9}. We imply that the wave incomes
through the left waveguide. Eq. (\ref{LS}) has unambiguous
solution until the operator at the left can be inverted. However,
if
\begin{equation}\label{det}
||\omega^2-\hat{H}_{eff}||=0,
    \end{equation}
the inverse operator does not exist, and the solution becomes
ambiguous.

Such a precedent was revealed  in a periodical structure (grating
slab) \cite{Bonnet-Bendhia1994} and  is a consequence of bound
states in the diffraction continuum \cite{Hsu2013,Bulgakov2014}.
If Eq. (\ref{det}) is fulfilled, then the solution of Eq.
(\ref{LS}) can be presented as superposition \cite{Smirnov}
\begin{equation}\label{BICmode}
|\psi_B\rangle=\alpha|BIC\rangle+ |\psi_p\rangle,
\end{equation}
where the first part is the solution of the homogeneous equation
\begin{equation}
\label{LShom}
    (\omega^2-\hat{H}_{eff})|BIC\rangle=0
\end{equation}
while the second contribution is the particular solution of Eq.
(\ref{LS}). In the presentation of eigenstates (\ref{Heffeig}) Eq.
(\ref{det}) takes the following form
\begin{equation}\label{detlambda}
\prod_{\lambda}(\omega^2-z_{\lambda})=0.
    \end{equation}
Obviously, Eq. (\ref{det}) is fulfilled if some of complex
eigenvalues becomes real, i.e. at the BIC point. Then the
necessary and sufficient condition for existence of solution of
the equation (\ref{LS}) is that the vector \cite{Smirnov}
\begin{equation}\label{Smirnov}
\langle BIC|\psi_p\rangle=0.
\end{equation}
This equation has clear physical meaning of that the BIC solution
is orthogonal to the solution which propagates in waveguide,
therefore can not leakage from the cavity.

It might be seemed that the BIC solution (\ref{inside}) can be
presented by only those eigenfunctions (\ref{eigfuncrect}) which
undergo degeneracy, events of which are shown in Fig. \ref{fig8}.
In particular, let us consider the eigenmodes $\psi_{4,3}$ and
$\psi_{2,5}$ with corresponding eigenfrequencies
\begin{eqnarray}
\label{eigrecd}
&\omega_{4,3}^2=\omega_a^2=\frac{4^2}{L_x^2}+\frac{3^2}{L_y^2},&\nonumber\\
&\omega_{2,5}^2=\omega_b^2=\frac{2^2}{L_x^2}+\frac{5^2}{L_y^2}.&
\end{eqnarray}
All dimensional units are measured in term of the waveguide width
$d$ and frequency is measured in term of $\sqrt{E_0}=\frac{c}{d}$.
The degeneracy point is given by relation
$\frac{L_y}{L_x}=\frac{2}{\sqrt{3}}$ and respectively the BIC
frequency equals
$\omega_c=\frac{4\pi}{L_x}\sqrt{1+\frac{27}{64}}$. In numerics we
have chosen $L_x=4$ that gives $\omega_c=3.746$. Then the coupling
matrix elements (\ref{WD}) equal
$$W_{4,3;1L}=W_{4,3;1R}=W_a=\sqrt{\frac{2}{k_1}}\frac{8\pi}{L_x^{3/2}L_y^{1/2}}
\int_{-1/2}^{1/2}dy\cos(\pi y)\cos(3\pi y/L_y)\approx 0.618,$$
$$W_{2,5,1L}=W_{2,5;1R}=W_b=\sqrt{\frac{2}{k_1}}\frac{4\pi}{L_x^{3/2}L_y^{1/2}}
\int_{-1/2}^{1/2}dy\cos(\pi y)\cos(5\pi y/L_y)\approx 0.4.$$ Thus,
the BIC solution in two-level approximation can be written as the
linear superposition, at least, at the point of degeneracy
$\omega=\omega_c$
\begin{equation}\label{BIC0}
    \psi_{BIC}(x,y)=\psi_0(W_b\psi_a(x,y)-W_a\psi_b(x,y))
\end{equation}
where indices $4,3$ and $2,5$ are absorbed by the indices $a$ and
$b$ respectively. One can easily verify this function is
orthogonal to the first continuum of both waveguides given by
$p=1$ and turns to zero at the boundaries $x=\pm L_x/2$ and
therefore is localized inside the resonator. The matrix of the
effective Hamiltonian (\ref{Heffplan}) takes the following form
\begin{equation}
\label{noevanes}
\left(%
\begin{array}{ccccc}
  \omega_1^2-2iW_1^2 & -2iW_1W_2 & \ldots & -2iW_1W_a & -2iW_1W_b  \\
  -2iW_1W_2 & \omega_2^2-2iW_2^2 & \dots & -2iW_2W_a &  -2iW_2W_b\\
  \vdots & \vdots & \ddots & \vdots & \vdots \\
   -2iW_1W_a& -2iW_2W_a & \ldots & \omega_a^2-2iW_a^2 & -2iW_aW_b  \\
 -2iW_1W_b& -2iW_2W_b & \ldots & -2iW_b^2 & \omega_b^2-2iW_aW_b \\
\end{array}%
\right)=0.
\end{equation}
Equation for the BIC takes the following form
\begin{equation}
\label{detab}
\left|%
\begin{array}{ccccc}
  \frac{\omega_1^2-\omega^2}{2iW_1^2}+1 & 1 & \ldots & 1 & 1  \\
  1 & \frac{\omega_2^2-\omega^2}{2iW_2^2} & \dots & 1 & 1 \\
  \vdots & \vdots & \ddots & \vdots & \vdots \\
   1& 1 & \ldots & \frac{\omega_a^2-\omega^2}{2iW_a^2} & 1  \\
 1& 1 & \ldots & 1 & \frac{\omega_b^2-\omega^2}{2iW_b^2} \\
\end{array}%
\right|.
\end{equation}
One can see that at the point of degeneracy  $\omega_a=\omega_b$
and $\omega=\omega_a$  the determinant (\ref{detab}) turns to zero
to realize the BIC as the linear superposition of degenerate
states (\ref{BIC0}).

Which is role of evanescent modes? First, we show that the
evanescent modes shift the BIC point. The effective Hamiltonian
(\ref{Heffplan}) can be rewritten as follows for $\omega^2<4\pi^2$
\begin{equation}\label{Hefftilda}
    \widehat{H}_{eff}=\widehat{\widetilde{H}}_B-2i\widehat{W}_{1}\widehat{W}_{1}^{\dagger},
\end{equation}
where
\begin{equation}\label{Htilde}
    \widehat{\widetilde{H}}_B=\widehat{H}_B-2\sum_{p>1}\widehat{\widetilde{W}}_{p}\widehat{\widetilde{W}}_{p}^{\dagger},
\end{equation}
where the coupling matrix $\widehat{W}_{p=1}$ is defined by Eq.
(\ref{WD}) or Eq. (\ref{WN}) while the coupling matrix
$\widehat{\tilde{W}}_{p>1}$ originated from the evanescent modes
and equal
\begin{equation}\label{Wtilde}
    \widetilde{W}_{mn;p>1}=\sqrt{\frac{1}{\pi |k_p|}}\left.\int_{-1/2}^{1/2}dy\sin\left(\frac{\pi py}{d}\right)
    \frac{\partial{\psi_{mn}(x,y)}}{\partial x}\right|_{x=\pm L_x/2}.
\end{equation}
The factor 2 in Eqs. (\ref{Hefftilda}) and (\ref{Htilde}) is the
result of equal contribution of both left anf right waveguides.
The matrix $\widehat{\widetilde{H}}_B$ is Hermitian and can be
interpreted as the effective Hamiltonian of the cavity modified by
evanescent modes. Substituting modified eigenvalues into
(\ref{detab}) we obtain that points of degeneracy of them define
the exact BIC points.

Second, the approximate BIC solution (\ref{BIC0}) turns to zero at
boundaries between the resonator and waveguides $x=\pm L_x/2$. The
exact BIC solution defined by Eq. (\ref{LShom}) which can be
expressed in series of the eigenfunctions of the closed resonator
\begin{equation}\label{BICinside}
    \psi_{BIC}(x,y)=\sum_{mn}a_{mn}\psi_{mn}(x,y)
\end{equation}
where the expansion coefficients are given by eigenvector of Eq.
(\ref{LShom}). Although each eigenfunction $\psi_{mn}(x=\pm
L_x/2,|y|<1/2)=0$ the BIC solution (\ref{BICinside}) is to be
sewed with the evanescent modes in the waveguides which
exponentially decay when we move away from the boundary of the
closed resonator
\begin{equation}\label{sewing}
\sum_{mn}a_{mn}\psi_{mn}(x=\pm
L/2,|y|<1/2)=\sum_{p>1}\frac{a_p}{\sqrt{4\pi k_p}}\phi_p(y)\approx
\frac{a_2}{\sqrt{4\pi k_2}}\phi_2(y).
\end{equation}

\begin{figure}[ht]
\centering{\resizebox{0.6\textwidth}{!}{\includegraphics{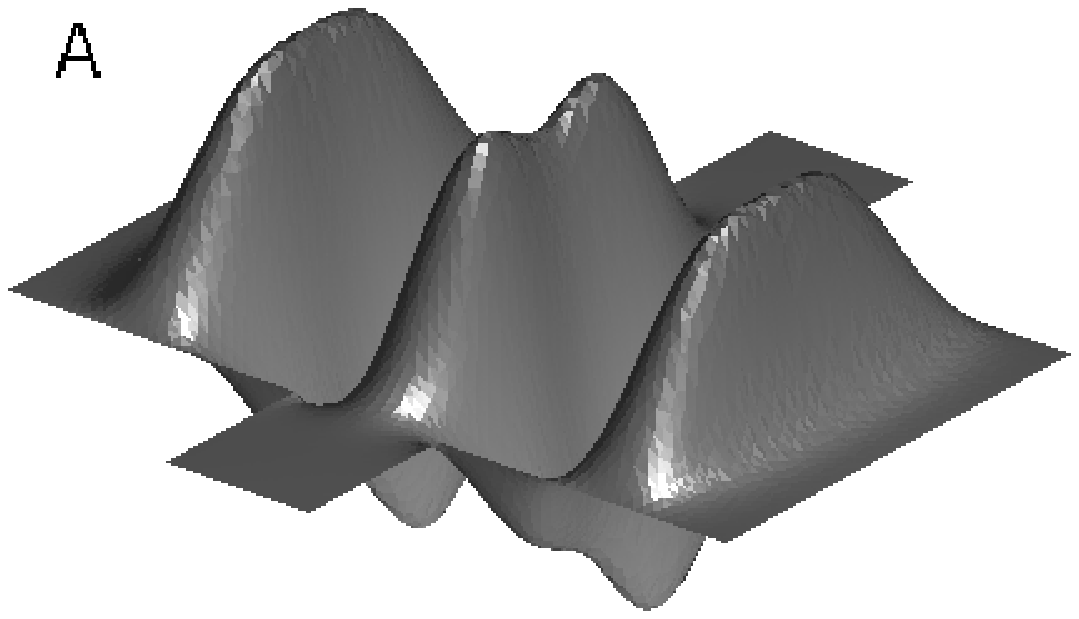},\includegraphics{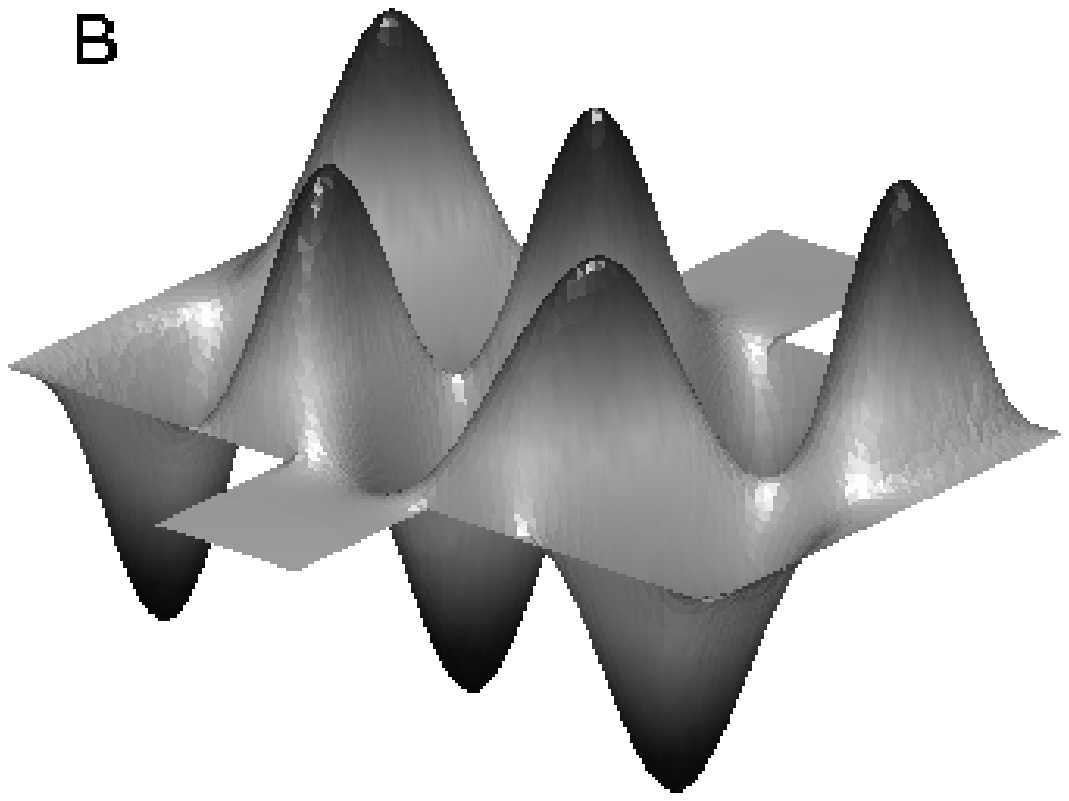}}}
\caption{The patterns of the two BICs A and B marked in Fig.
\ref{fig9} by bold circles.} \label{fig11}
\end{figure}

We pay attention that if we restricted by only two eigenfunctions
which undergo degeneracy in the vicinity of the BIC point the left
hand expression in Eq. (\ref{sewing}) would turn to zero. Only due
to infinite series over the eigenfunctions the left hand
expression (\ref{sewing}) differs from zero. Thus, the second role
of the evanescent modes is in exponential weak blowing of the BIC
solution into waveguides that provides smooth behavior of the BIC
solution as seen from Fig. \ref{fig11}. These BIC solutions are
found numerically from Eq. (\ref{LShom}) with sufficiently large
rank of the effective Hamiltonian.
Thus, there are two important features of the BICs caused by
evanescent modes. First, the BIC solution is overflowed from the
resonator due to coupling to the evanescent modes as seen from
Fig. \ref{fig11}. A degree of the overflowing is given by the
exponential contribution of the first evanescent mode
$\exp(-\sqrt{(2\pi)^2-\omega^2}(x-L_x/2))$ in the right waveguide.
The same holds in the left waveguide however for $x< -L/2$.
Second, the BIC point is shifted relative to points of degeneracy
of eigenmodes of the closed resonator because of contribution in
the effective Hamiltonian (\ref{Htilde}). Details of these effect
will be given below for the 3d resonators where the contribution
of evanescent modes has principal importance for existence of
BICs.

\section{Accidental BICs in the Sinai shaped open cavity}
\label{Sect:Accidental}
\begin{figure}
\centering{\resizebox{0.4\textwidth}{!}{\includegraphics{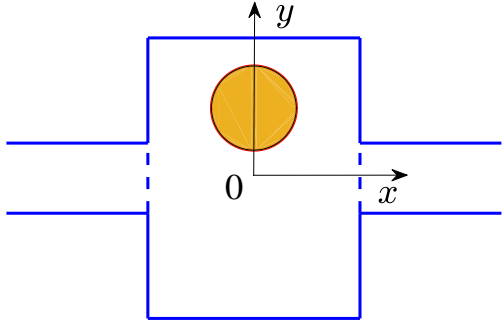}}}
\caption{(Color online) Dielectric or metallic disk inside the
rectangular open resonator.} \label{fig12}
\end{figure}

The rectangular cavity is an example of integrable system when the
variables $x$ and $y$ are separated that reduces the eigenvalue
problem to the one-dimensional one with multiplicative
eigenfunctions (\ref{eigfuncrect}). Because of that for variation
of one of the scales of the resonator, say width $W$, we have
multiple events of degeneracy each of which gives rise to BICs in
the Friedrich-Wintgen scenario as it was described in previous
section. In fact there are only a few integrable resonators,
elliptic and equilateral triangle which are specified by Poisson
distribution of the nearest distances between eigenlevels. All the
rest falls into non integrable whose eigenlevels undergo avoided
crossings for variation of some parameter with the Wigner
distribution and form so called chaotic billiards
\cite{Stockmann}. The Bunimovich and Sinai billiards are the well
known examples of chaotic billiards. Experimentally it is easy to
transform the integrable billiard into the chaotic one by
embedding of dielectric or metallic disk inside the plane
rectangular cavity as sketched in Fig. \ref{fig12}.
Then the FW mechanism of the BIC due to degeneracy of eigenstates
of closed billiard is not applicable.  However there is another
way to realize the BIC by decoupling of an individual eigenmode of
the Sinai billiard from the the first continuum of waveguides
\cite{Bulgakov2011,Sadreev2015}. For that we smoothly deform the
eigenmodes by, for example, variation of radius or position of
disk inserted inside the rectangular cavity. The effect of disk
can be described by a circular potential perturbation
\begin{equation}\label{pot}
    V(x,y)=V_g\exp\left[-\frac{(x-x_0)^2+(y-y_0)^2}{R^2}\right]
\end{equation}
added into the effective Hamiltonian (\ref{Heffplan}). To be
specific we consider the Neumann boundary conditions because of
good convergence of the results with growth of rank of the matrix
$\widehat{H}_{eff}$ for low lying eigenfrequencies
\cite{Maksimov2015}.

In what follows we fix the radius $R=1.5$ and position of circular
potential at $x_0=0, y_0=1$ in terms of the waveguides width $d$
and vary the height $V_g$ of the potential (\ref{pot}) that
effectively varies the radius of the circular potential. Because
of symmetry of full system relative to $x\rightarrow -x$ the
continua of both waveguides are identical. Respectively we have
identical coupling matrix elements of the Sinai resonator with
waveguide continua
\begin{equation}\label{WSinai}
    W_{b,pC}=\sqrt{\frac{1}{k_p}}\left.\int_0^ddy\phi_p(y)\frac{\partial{\psi_b(x,y)}}
    {\partial x}\right|_{x=\pm L/2},
\end{equation}
where $C=L,R$ enumerates the interfaces between the left and right
waveguides shown in Fig. \ref{fig12} by dashed lines, $\psi_b$ are
the eigenfunctions of the closed Sinai billiard.

\begin{figure}
\centering{\resizebox{0.8\textwidth}{!}{\includegraphics{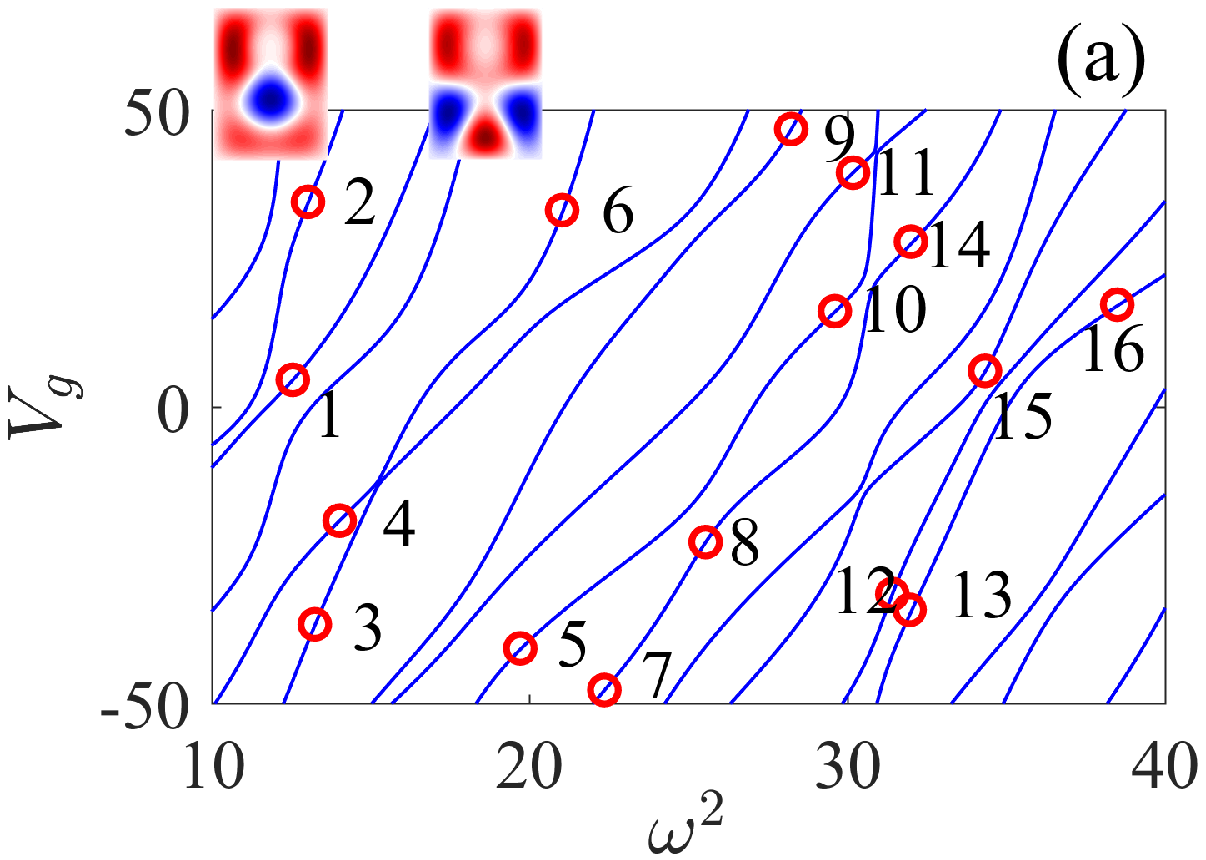},\includegraphics{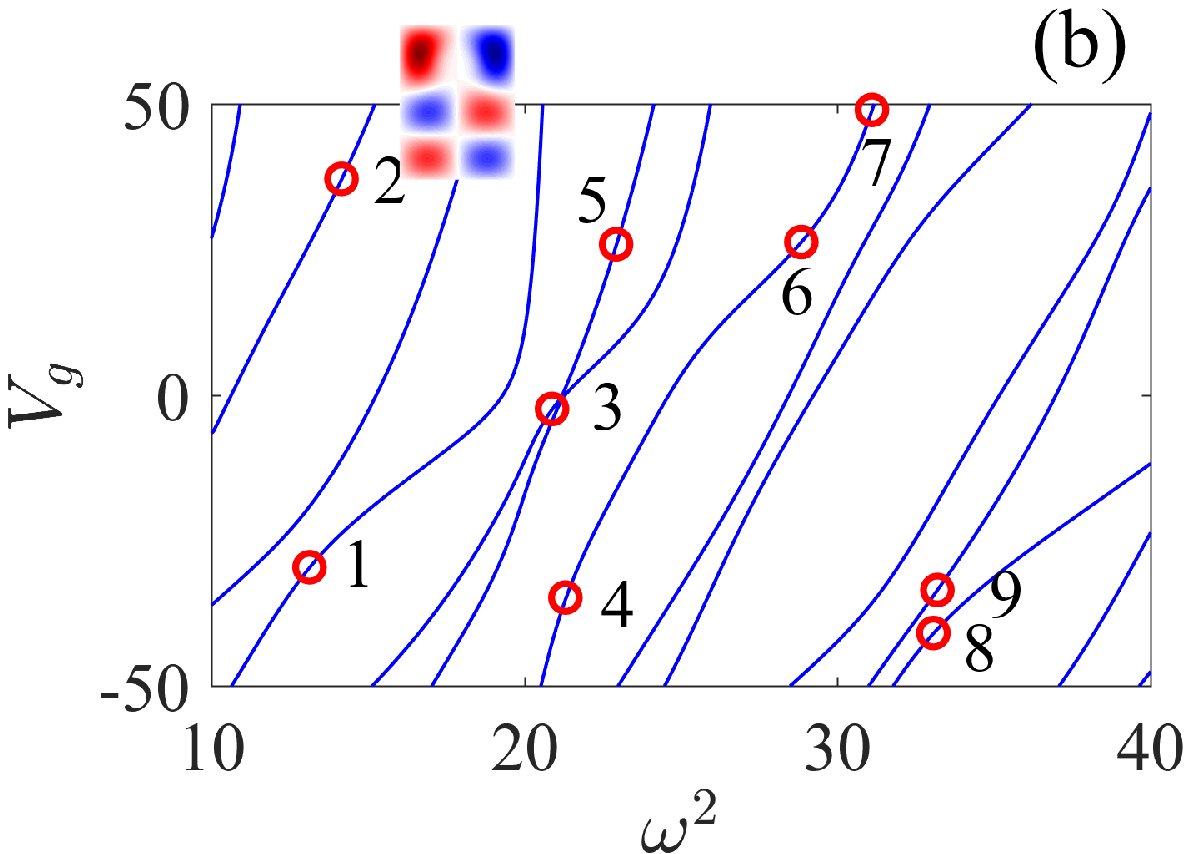}}}
\caption{(Color online) Eigenvalues and eigenfunctions of the soft
Sinai resonator vs height of the potential (\ref{pot}). Insets
show a few patterns of  corresponding eigenfunctions are even (a)
and odd (b) relative to $x\rightarrow -x$. Open circles mark the
BIC points.} \label{fig13}
\end{figure}

The eigenfunctions are classified as even and odd
$\psi(x,y)=\pm\psi(-x,y)$. Respectively, the eigenvalues in each
irreducible representation undergo avoided crossings with
variation of $V_g$ as illustrated in Fig. \ref{fig13}.
For clarity we show some patterns of the eigenfunctions at
$V_g=50$ in Fig. \ref{fig13}. One can see that the eigenfunctions
are depleted inside by the potential (\ref{pot}) at $V_g=50$. A
variation of another parameter of the potential (\ref{pot}), for
example, the radius or position shows a similar result. Thus, we
have no degeneracy of the eigenfunctions of the same irreducible
representation in the chaotic Sinai resonator.

\begin{figure}
\centering{\resizebox{0.4\textwidth}{!}{\includegraphics{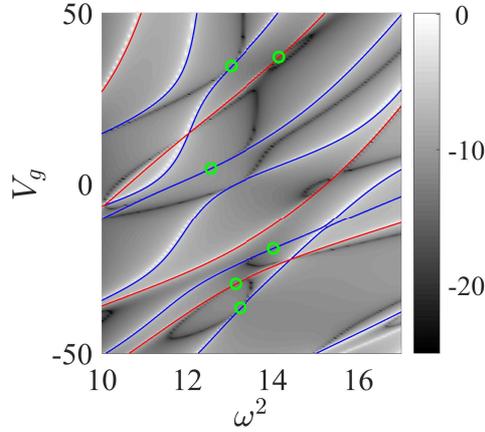}}}
\caption{Transmittance  of the Sinai resonator in Log scale vs
$V_g$ (effective radius of the circular hole shown in Fig.
\ref{fig12}) and incident frequency. The BICs are shown by open
circles.} \label{fig14}
\end{figure}

Fig. \ref{fig14} shows the transmittance calculated via Eq.
(\ref{trans}). In order the reader can observe that peaks of the
transmittance follow the eigenvalues of the closed Sinai resonator
we reduce the coupling between the waveguides and the resonator by
implementation of diaphragms between the waveguides and the
billiard \cite{Rotter2004a} that narrows transmission peaks.
\begin{figure}
\centering{\resizebox{0.4\textwidth}{!}{\includegraphics{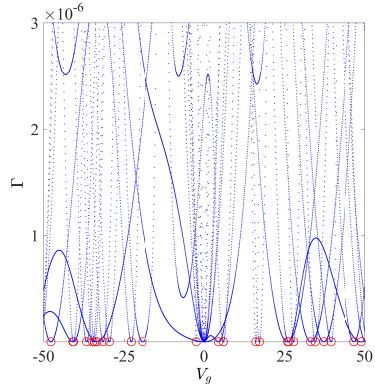}}}
\caption{Evolution of the resonant widths for variation of the
potential. Red open circles mark BICs.} \label{fig15}
\end{figure}
The BIC occurs if the resonance width turns to zero which are
given by the imaginary part of the complex eigenvalues of the
effective non-Hermitian Hamiltonian
\begin{equation}\label{HeffSinai}
    \widehat{H}_{eff}=\widehat{H}_B+V_g-i\sum_{C=L,R}\sum_p
    \widehat{W}_{C,p}\widehat{W}_{C,p}^{\dagger},
\end{equation}
where $\mathcal{\widehat{H}}_B$ is the Hamiltonian of closed
rectangular resonator, and $\widehat{W}_{C,p}$ are columns of
matrix elements (\ref{WSinai}) labelled by the eigenstate indices
$b$. Numerically computed evolution of the resonant widths is
presented in Fig. \ref{fig15} which shows multiple events of the
resonant widths turning to zero, i.e., BICs in the Sinai
resonator. The even BICs sorted by their energies are shown in
Fig. \ref{fig13} (a) by open circles. Respectively the odd BICs
are shown in Fig. \ref{fig13} (b). Besides these BICs one can see
in Fig. \ref{fig15} numerous symmetry protected BICs at the point
$V_g=0$ which are the eigenfunctions of the rectangular resonator
antisymmetric relative to $y\rightarrow -y$ for $\omega<2\pi$. And
therefore they are incompatible with the symmetric propagating
mode in the first channel $p=1$ (\ref{phip}).

\begin{figure}
\centering{\resizebox{0.4\textwidth}{!}{\includegraphics{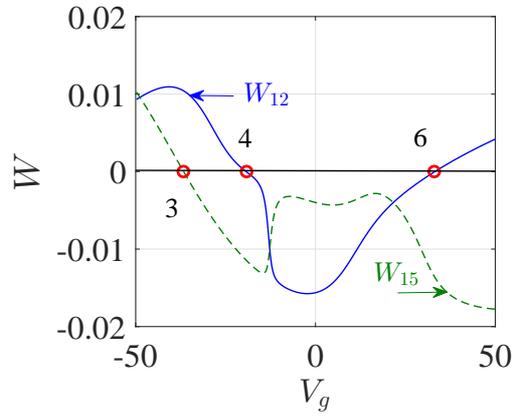}}}
\caption{Evolution of the coupling matrix element (\ref{WSinai})
with $V_g$.} \label{fig16}
\end{figure}

\begin{figure}
\centering{\resizebox{0.8\textwidth}{!}{\includegraphics{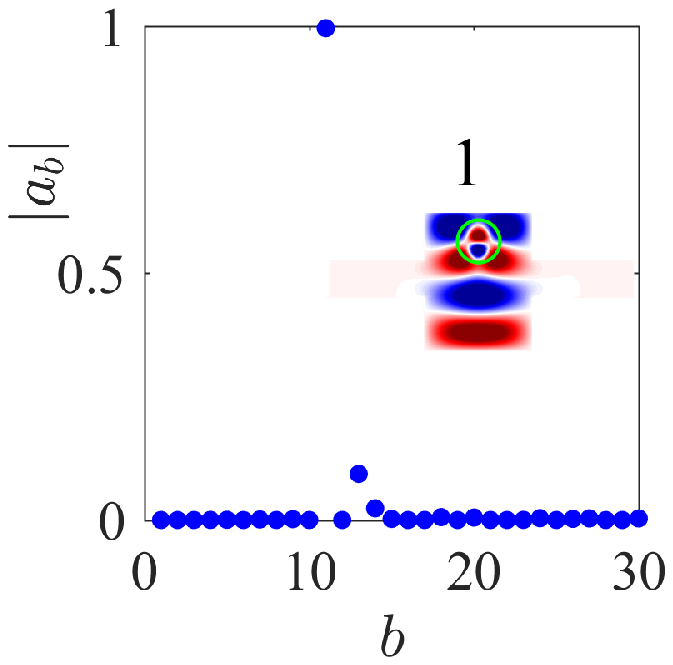},\includegraphics{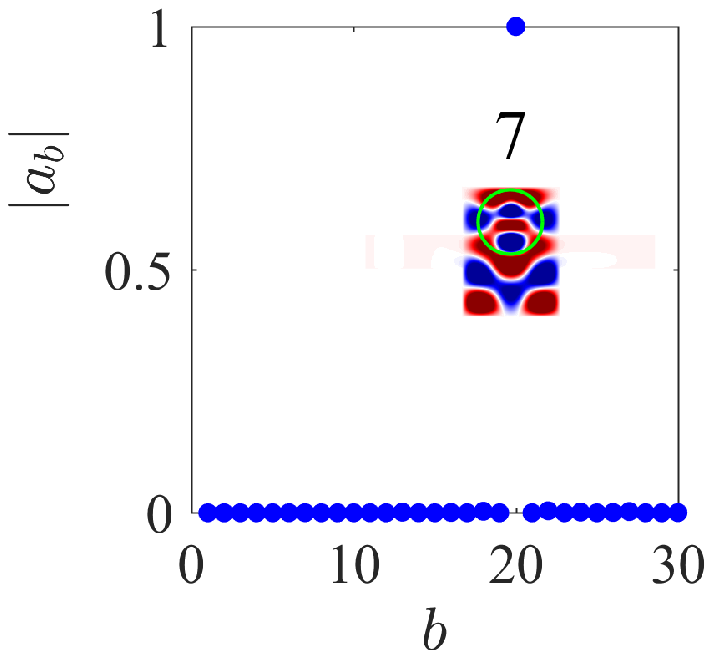}}}
\caption{(Color online) Patterns of even BICs enumerated
according to Table \ref{SinaiTab1} with coefficients of the modal
expansions. Position of potential (\ref{pot}) is shown by green
circle.} \label{fig17}
\end{figure}

\begin{figure}
\centering{\resizebox{0.8\textwidth}{!}{\includegraphics{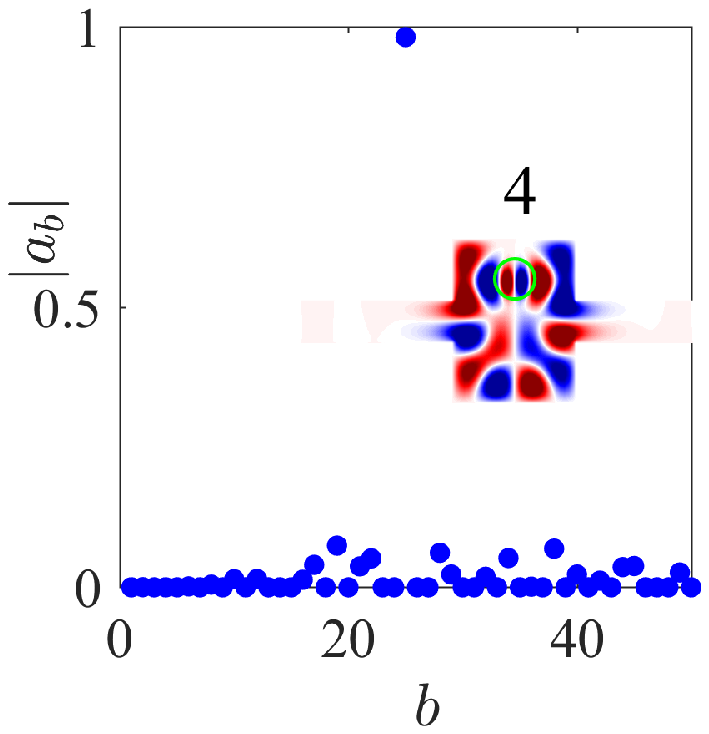},\includegraphics{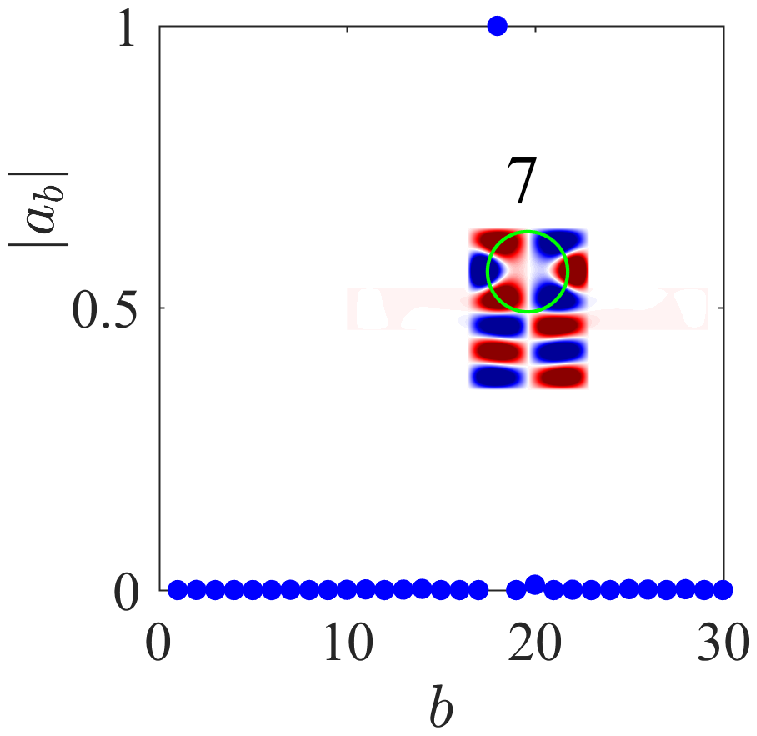}}}
\caption{(Color online) Patterns of odd BICs enumerated according
to Table \ref{SinaiTab2} with coefficients of the modal
expansions.} \label{fig18}
\end{figure}

Fig. \ref{fig14} clearly demonstrates that the BIC points are
positioned at those points in the parametric space of $E$ and
$V_g$ where the transmission zero coalesces with the transmission
unit similar to the FW BICs \cite{SBR} illustrating the collapse
of Fano resonance \cite{Kim1999}. However in the Sinai billiard
the BICs occur accidentally under variation of the circular
potential (\ref{pot}) that changes the eigenfunctions of the
closed Sinai resonator as shown in insets in Fig. \ref{fig13}.
That in turn changes the coupling matrix elements (\ref{WSinai})
so that some of them can turn to zero as illustrated in Fig.
\ref{fig16}.
That is a mechanism of the accidental BICs patterns shown in Figs.
\ref{fig17} and \ref{fig18}.
These Figures also depict the modal expansion coefficients $|a_b|$
of BICs over the eigenmodes of the closed Sinai resonator
\begin{equation}\label{modal expansion}
    \psi_{BIC}(x,y)=\sum_ba_b\psi_b(x,y).
\end{equation}
\begin{table}
        \centering
                \caption{BICs even relative to $x\rightarrow -x$
marked by open circles in Fig.~\ref{fig13} (a).} \label{SinaiTab1}
\begin{tabular}{|c|c|c|c|}
\hline
Number of the even BIC & $E$ &$V_{g}$ \\
\hline
1 & 12.550 & 4.5\\
2 & 13.029 &  34.45\\
3 & 13.244 & $-$36.65\\
4 & 14.026  &  $-$19.2\\
5 & 19.709 &  $-$40.7\\
6 & 21.025 &  33.05\\
7 & 22.355 &  $-$47.7\\
8 & 25.541 &  $-$22.8\\
9 & 28.236  &  46.7\\
10 & 29.608  & 16.05\\
11 & 30.181  & 39.35\\
12 & 31.418  &  $-$31.55\\
13 & 31.960  & $-$34.2\\
14 & 32.002 &  27.75\\
15 & 34.333  & 6.00\\
16 & 38.495 &  17.15\\
\hline
\end{tabular}
\end{table}
\begin{table}
\centering \caption{BICs odd relative to $x\rightarrow -x$ marked
by open circles in Fig.~\ref{fig13} (b).} \label{SinaiTab2}
        \begin{tabular}{|c|c|c|c|}
\hline
Number of the odd BIC & $E$ &$V_{g}$ \\
\hline
1 & 13.133 &  $-$29.6\\
2 & 14.155 &  37.1\\
3 & 20.882 &  $-$2.4\\
4 & 21.307 &  $-$34.8\\
5 & 22.927 &  25.85\\
6 & 28.844 &  26.25\\
7 & 31.099 &  48.95\\
8 & 33.063 &  $-$40.9\\
9 & 33.189 &  $-$33.5\\
\hline
\end{tabular}
\end{table}
One can see that indeed basically one eigenfunction contributes
into the BIC mode. There is also background of other
eigenfunctions which is a result of contribution of evanescent
modes into the effective Hamiltonian (see discussion in previous
section).

In what follows we will prove that if one of coupling matrix
elements $W_{b,p=1}$ vanishes then the accidental BIC occurs
embedded into the continuum of the first propagating channel $p=1$
of both waveguides $C=L,R$. Let us choose, for example, the
eigenfunction of the closed billiard, say $b=3$, whose coupling
with the first channel $p=1$ turns to zero. Then we can write the
coupling matrix (\ref{WSinai}) as follows
\begin{equation}\label{WSin}
    W_{b,1}=(W_1~~ W_2~~ 0~~ W_4~~\ldots).
\end{equation}
Because of symmetry relative  $x\rightarrow -x$ the coupling
matrix (\ref{WSin}) is invariant relative to choice of waveguides
$C=L,R$. Then there is a vector
\begin{equation}\label{null}
    \psi_3^{+}=(0~~0~~1~~0~~\ldots)
\end{equation}
which is the eigen null vector of the matrix $WW^{+}\psi_3=0$. On
the other hand, the vector (\ref{null}) is the eigenvector of the
closed billiard with the Hamiltonian
\begin{equation}
\label{HB} \widehat{H}_B=\left(\begin{array}{ccccccc} \omega_1^2 &
0 & 0 & 0 &\cdots \cr
                        0 & \omega_2^2 & 0& 0&\cdots \cr
                        0 & 0 & \omega_3^2 & 0&\cdots \cr
                        0 & 0 & 0 & \omega_4^4 &\cdots \cr
                        \vdots&\vdots&\vdots&\ddots&\cr
                        \end{array}\right).
\end{equation}
with the eigenfrequency $\omega_3$. Thus the null eigenvector
(\ref{null}) is the eigenstate of the effective non-Hermitian
Hamiltonian (\ref{HeffSinai}) with real eigenfrequency $\omega_3$,
and therefore is the BIC with this frequency. That result does not
depend on the other coupling matrix elements in (\ref{WSin}).
Following Refs. \cite{Hsu2013,Bulgakov2014} we define such BICs as
accidental. Note that conclusion is correct in neglecting of the
evanescent modes of waveguides. The contribution of evanescent
modes can be performed as it was done in Section \ref{Sect:2dres}.
However in this case the eigenstate (\ref{null}) ceases to be the
eigenstate of the effective Hamiltonian. As a result as Fig.
\ref{fig18} (a) shows the accidental BIC is blowing off the Sinai
billiard and modal expansion shows noticeable background of all
other eigenmodes of the billiard.

\section{The cylindrical resonator with non-axisymmetric waveguides.
The twisted BICs.}
\label{Sect:Twisted}
\begin{figure}[ht]
\centering{\resizebox{0.8\textwidth}{!}{\includegraphics{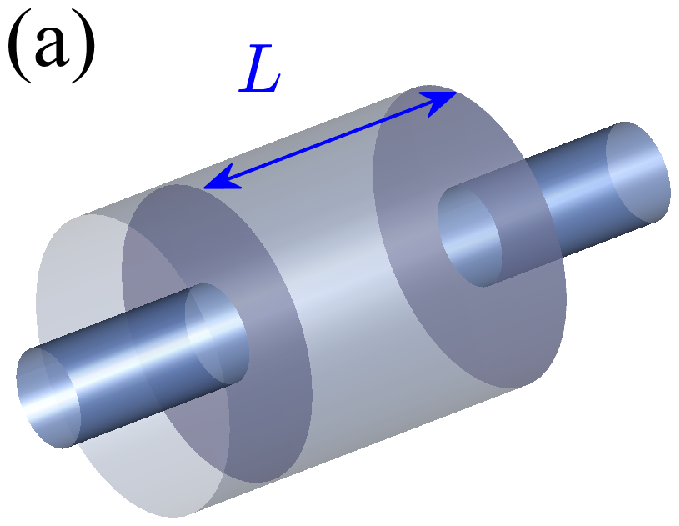},\includegraphics{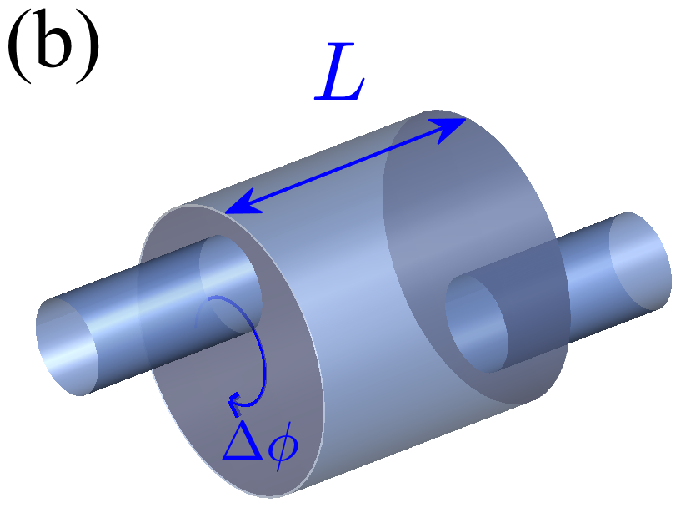}}}
\caption{Cylindrical resonator with two (a) coaxially and (b) non
coaxially attached non-axisymmetric cylindrical waveguide. The
input waveguide can freely move along the resonator axis and
rotate about the symmetry axis of resonator.} \label{fig19}
\end{figure}

The aim of this and next sections is to demonstrate nontrivial
role of the waveguides whose attachment breaks the symmetry of the
closed resonators with nontrivial BICs embedded into continua of
these waveguides. For example, the closed cylindrical resonator
with the radius $R$ and length $L$ is the typical the textbook
case \cite{Jackson} which allows separation of variables in the
cylindrical system of coordinates. If to attach cylindrical
waveguides coaxially as shown in Fig. \ref{fig19} (a) the axial
symmetry of the total open system is preserved. We skip this case
of coaxial connected waveguides where the FW BICs are accessed via
variation of the length of the resonator $L$ \cite{Lyapina2018a}
similar to the section \ref{Sect:2dres} (planar rectangular
resonators). However if one of waveguides is shifted off the
symmetry axis of the resonator as shown in Fig. \ref{fig19} (b)
the axial symmetry of the total system breaks. We consider the
case of non-axisymmetric waveguides which are identical but are
attached to the resonator by different angles so that the
waveguides are unwrapped by angular difference $\Delta\phi$ as
shown in Fig. \ref{fig19}. That does not change strength of
coupling matrix elements with continua but differ the continua by
phase. We show that nevertheless the BICs exist but have to be
twisted by the angle $\Delta\phi$.

The Helmholz equation (\ref{Helm}) can be applied for acoustic
transmission through duct-cavity structures in hard wall
approximation. The equation takes the following form in the
cylindrical system of coordinates
\begin{equation}\label{Helmcyl}
    \left[\frac{\partial^2}{\partial r^2}+\frac{1}{r}\frac{\partial }{\partial r}
    -\frac{m^2}{r^2}+\frac{\partial^2}{\partial z^2}+ \omega^2\right]\psi=0,
\end{equation}
for the non-dimensional velocity potential $\psi$ where the
non-dimensional coordinates $r$ and $z$ are normalized by the
waveguide radius $r_w$. The dimensionless frequency $\omega$ is
defined through the dimensional frequency $\widetilde{\omega}$ as
follows $\omega=\widetilde{\omega}r_w/c_0$
 and $c_0$ is the sound speed.

\begin{figure}[ht]
\centering{\resizebox{0.3\textwidth}{!}{\includegraphics{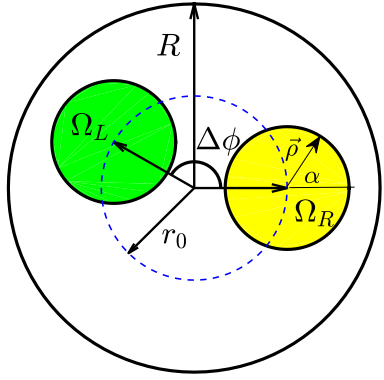}}}
\caption{Filled areas show overlapping integration area in the
coupling matrix (\ref{Wp}).} \label{overlap}
\end{figure}

The propagating
modes in the sound hard cylindrical waveguides with Neumann
boundary conditions are described by
\begin{eqnarray}\label{propag}
    \psi_{pq}(\rho,\alpha,z)=\psi_{pq}(\rho)\frac{1}{\sqrt{2\pi k_{pq}}}
    \exp({\rm i}p\alpha+{\rm i}k_{pq}z),\\
\psi_{pq}(\rho)=\left\{\begin{array}{l}
\frac{\sqrt{2}}{J_0(\mu_{0q})}J_0(\mu_{0q}\rho), p=0, \\
\sqrt{\frac{2}{\mu_{pq}^2-p^2}}
\frac{\mu_{pq}}{J_p(\mu_{pq})}J_p(\mu_{pq}\rho), p=1, 2, 3, \ldots,
\end{array}\right. \nonumber
\end{eqnarray}
where $\rho, \alpha$ are the polar coordinates shown in Fig.
\ref{overlap}, $\mu_{pq}$ is the q-th root of equation
$$\left.\frac{dJ_p(\mu_{pq}\rho)}{d\rho}\right|_{\rho=1}=0$$
imposed by the Neumann boundary condition on the walls of sound
hard cylindrical waveguide.
\begin{equation}\label{kmn}
    k_{pq}^2=\omega^2-\mu_{pq}^2
\end{equation}
The dimensional quantities $\rho,~z, ~k_{pq}$ are measured in
terms of the radius of the waveguide $\rho$ and frequency is
measured in the terms of the ratio $s/\rho$ where $s$ is the sound
velocity. The propagating bands degenerate with the respect to the
sign of azimuthal index and are classified by two indices,  the
azimuthal index $p=0, \pm 1, \pm 2, \ldots$ and radial index $q=1,
2, 3, \ldots$ . Profiles of propagating functions
$\psi_{pq}(\rho)\cos p\alpha$ are depicted in Table \ref{Tab2}.
\begin{table}
        \centering
                \caption{Cut-off frequencies and corresponding shapes of
propagating modes in the circular waveguide.} \label{Tab2}
        \begin{tabular}{|c|c|c|c|}
                \hline
                channel &cut-off frequency & indices & mode shape \\
                \hline
                \rule{0cm}{0.7cm}
                1 & 0 & $p=0, q=1$ &
                \raisebox{-.4\height}{\includegraphics[height=1cm]{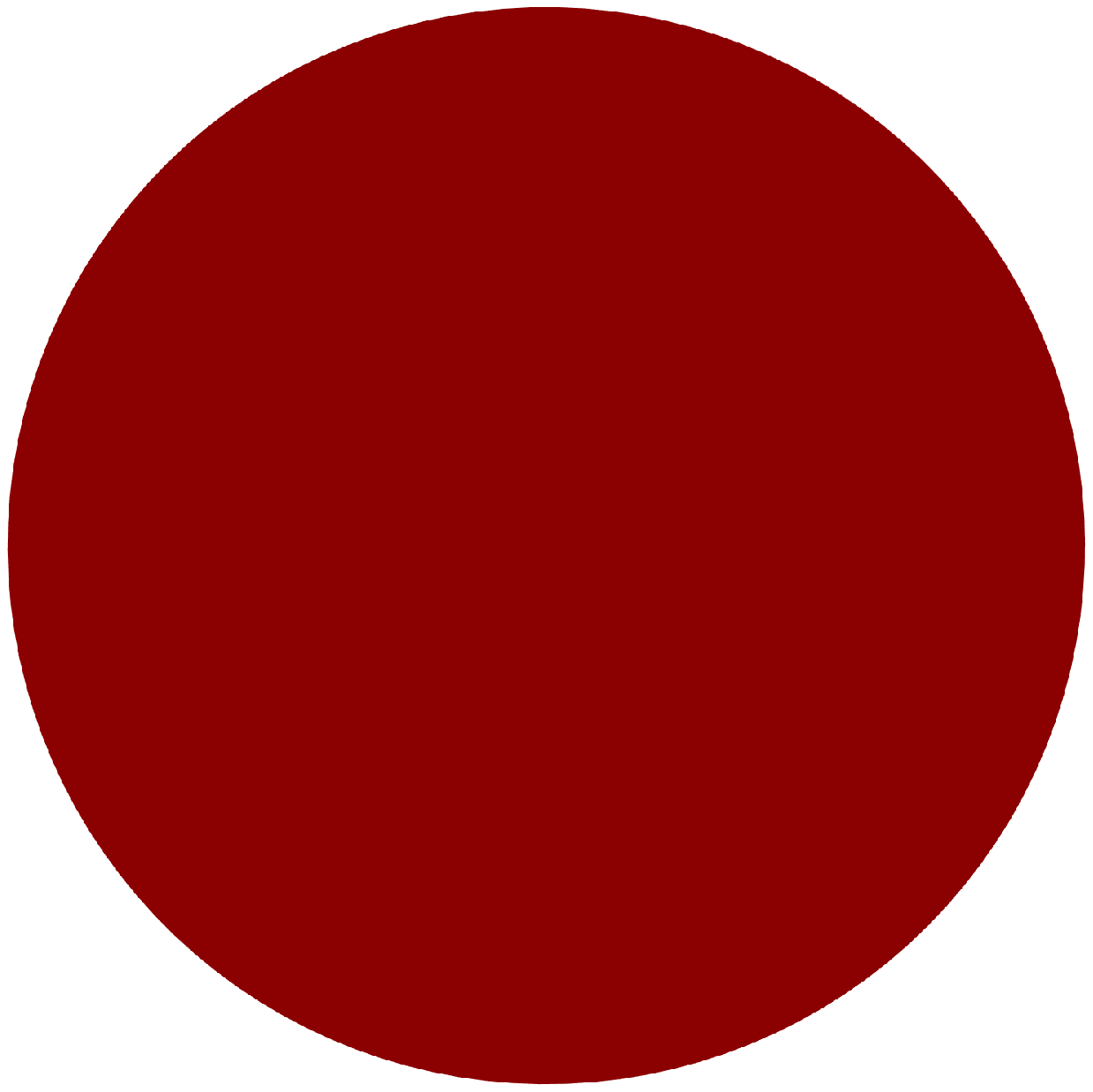}} \\
                \hline
                \rule{0cm}{0.7cm}
                2 & 1.84118 & $p=\pm 1, q=1$ &
                \raisebox{-.4\height}{\includegraphics[height=1cm]{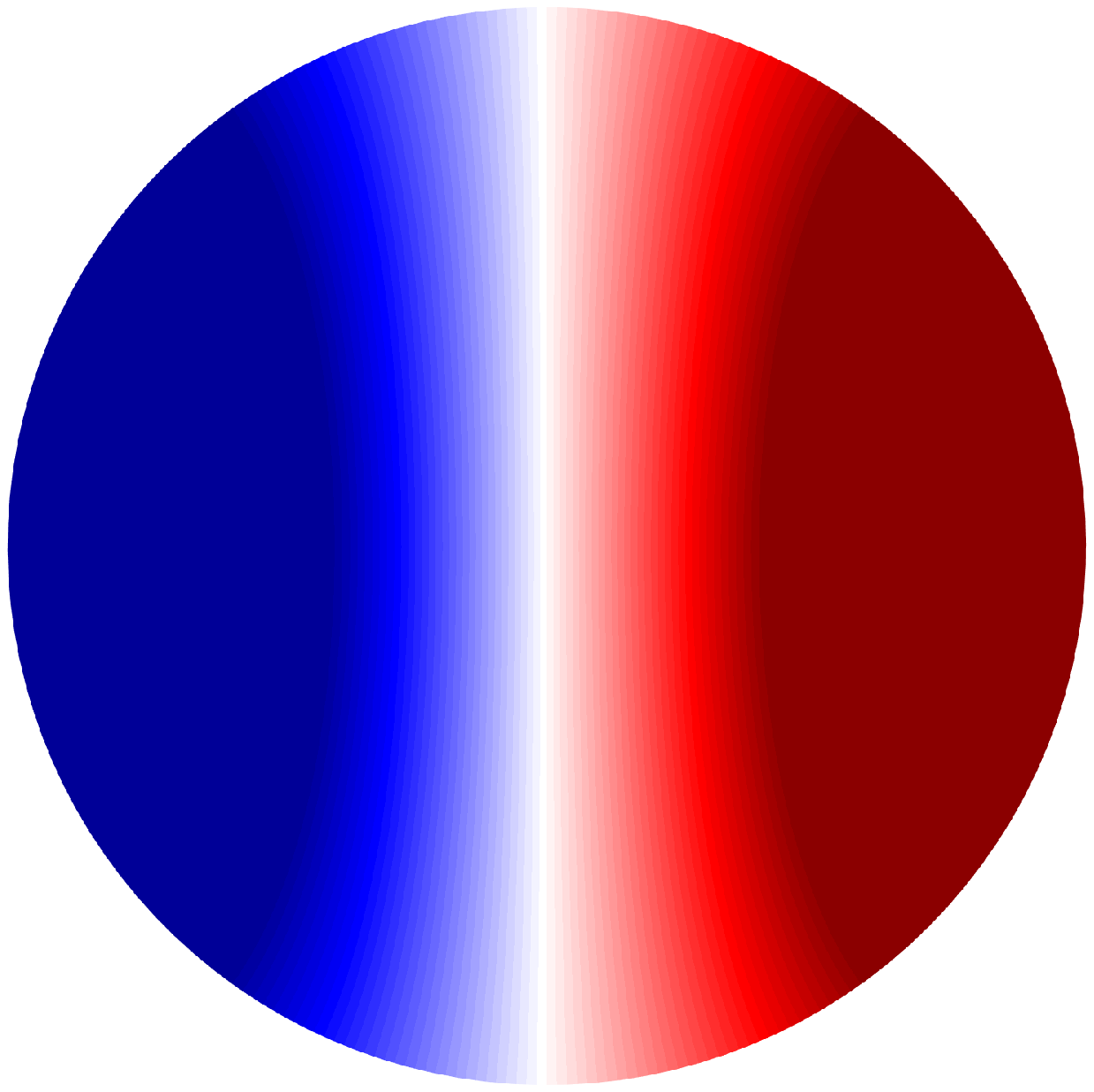}} \\
                \hline
                \rule{0cm}{0.7cm}
                3 & 3.0542 & $p=\pm 2, q=1$ &
                \raisebox{-.4\height}{\includegraphics[height=1cm]{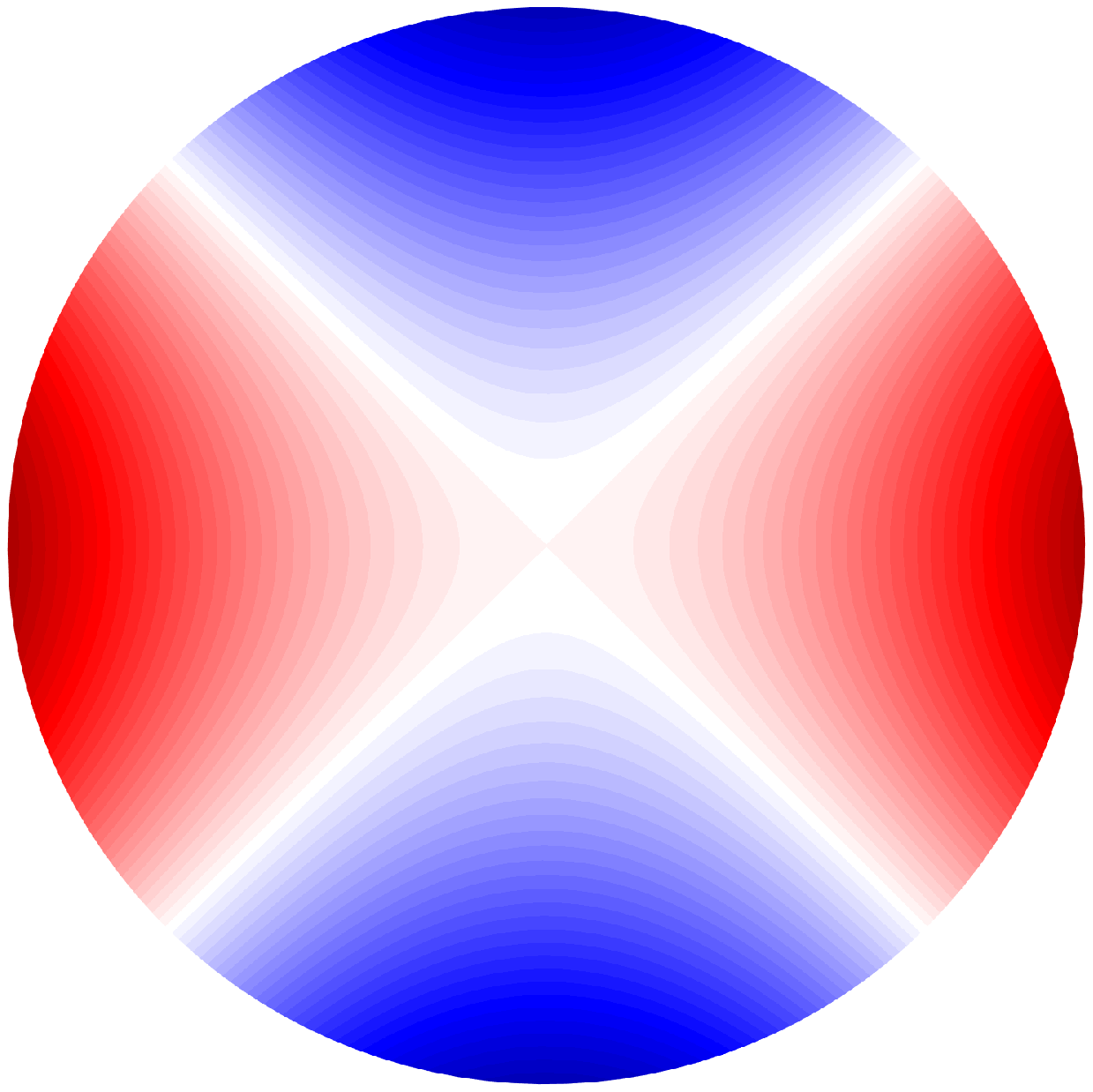}} \\
                \hline
                \rule{0cm}{0.7cm}
                4 & 3.831706 & $p=0, q=2$ &
                \raisebox{-.4\height}{\includegraphics[height=1cm]{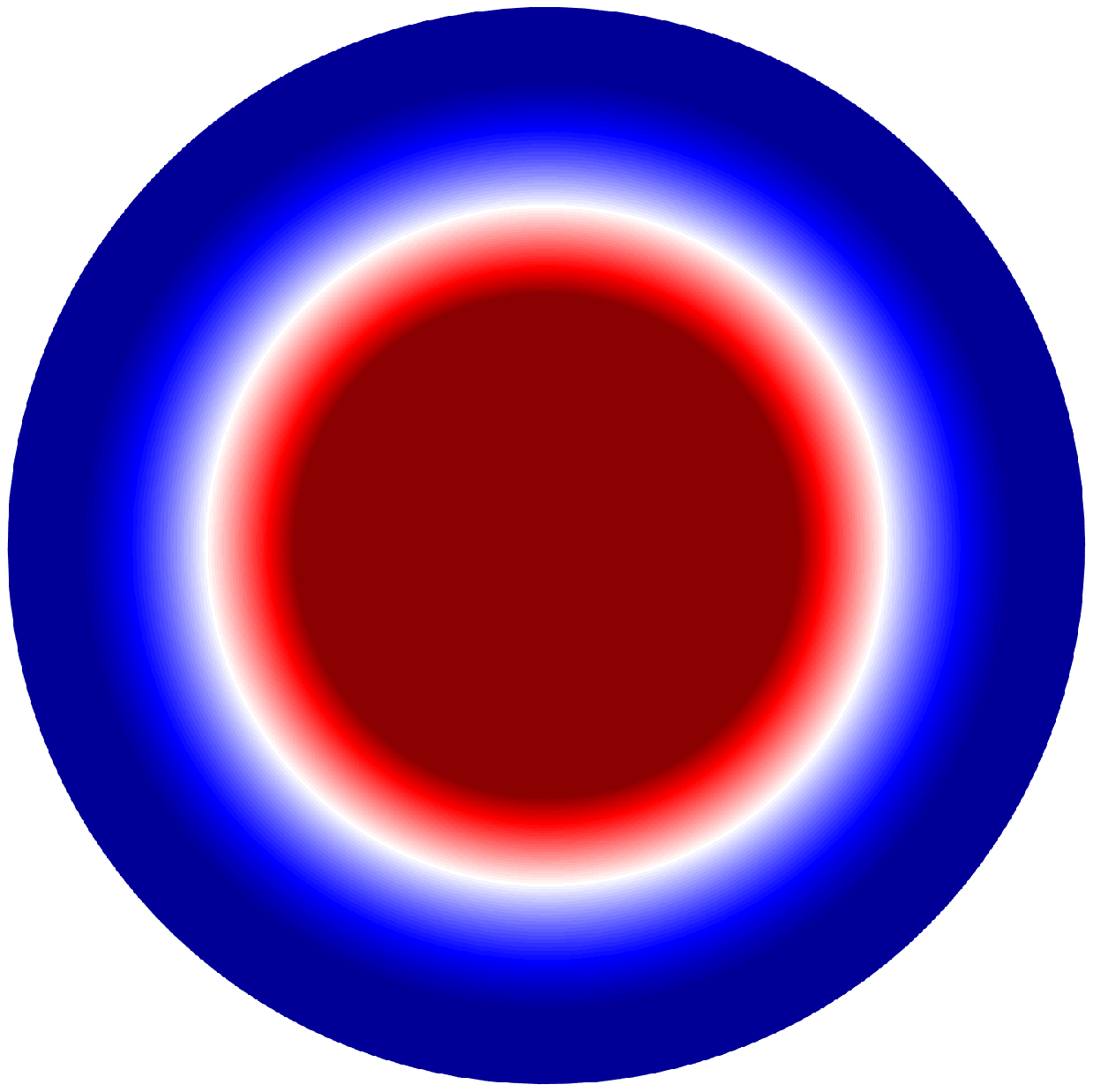}} \\
                \hline
        \end{tabular}
                \end{table}

The Hilbert space of the closed cylindrical resonator is
given by the following eigenmodes
\begin{equation}\label{eigen}
\Psi_{mnl}(r,\phi,z)=\psi_{mn}(r)\sqrt{\frac{1}{2\pi}}\exp({\rm i}m\phi)\psi_l(z),
\end{equation}
where
\begin{eqnarray}\label{eigRes}
&\psi_{mn}(r)=\left\{\begin{array}{l}
\frac{\sqrt{2}}{RJ_0(\mu_{0n}R)}J_0(\frac{\mu_{0n}r}{R}), m=0 \\
\sqrt{\frac{2}{\mu_{mn}^2-m^2}}
\frac{\mu_{mn}}{RJ_m(\mu_{mn}R)}J_m(\frac{\mu_{mn}r}{R}), m=1, 2,
3, \ldots,
\end{array}\right.&\\
   & \psi_l(z)=\sqrt{\frac{2-\delta_{l,1}}{L}}\cos[\pi(l-1)z/L],&\nonumber
\end{eqnarray}
$l=1, 2, 3, \ldots$ and $z$ is measured in terms of the waveguide
radius. The corresponding eigenfrequencies are
\begin{equation}\label{eig3D}
\omega_{mnl}^2=\left[\frac{\mu_{mn}^2}{R^2}+\frac{\pi^2
(l-1)^2}{L^2}\right]
\end{equation}
where $\mu_{mn}$ is the n-th root of the equation
$\left.\frac{dJ_m(\mu_{mn}r)}{dr}\right|_{r=R}=0$ which follows
from the Neumann BC on the walls of hard cylindrical resonator.

The matrix elements of $\hat{W}$ are given by overlapping
integrals \cite{Maksimov2015,Lyapina2018}
\begin{eqnarray}\label{Wp}
&W^{\rm C}_{mnl;pq}=\int_{\Omega_{\rm C}}\rho {\rm d}\rho {\rm
d}\alpha\psi_{pq}(\rho,\alpha)
\Psi_{mnl}^{*}(r,\phi,z=z_{\rm C})&\nonumber\\
&=\int_0^{2\pi} {\rm d}\alpha\int_0^1\rho {\rm
d}\rho\psi_{pq}(\rho,\alpha)
\Psi_{mnl}^{*}(r(\rho,\alpha),\phi(\rho,\alpha),z_{\rm C})&\nonumber\\
&=\psi_l(z_{\rm C})\int_0^{2\pi} {\rm d}\alpha\int_0^1\rho {\rm
d}\rho\psi_{pq}(\rho,\alpha)
\psi_{mn}^{*}(r(\rho,\alpha),\phi(\rho,\alpha)),&
\end{eqnarray}
where $\Omega_{\rm C, C=L,R}$ are interfaces positioned at $z_{\rm
C}=0,L$. Integration is performed over circular cross section of
the attached waveguides as shown in Fig. \ref{overlap}. One can
link the polar coordinates of the resonator with that of the
immovable waveguide $$r\sin\phi=\rho\sin\alpha,
r\cos\phi=r_0+\rho\cos\alpha$$ where $r_0$ is the distance between
the axes of the waveguide and resonator.

According to Eq. (\ref{eigen}) we have
\begin{equation}\label{psil}
\psi_l(z=0)=\sqrt{\frac{2-\delta_{l,1}}{L}},
\psi_l(z=L)=\psi_l(0)(-1)^{l-1}.
\end{equation}
Substituting (\ref{psil}) into (\ref{Wp}) we obtain the following
relation between the left and right coupling matrix elements
\begin{equation}\label{WLWR}
W^{\rm L}_{mnl;pq}=(-1)^{l-1}e^{{\rm i}(p-m)\Delta\phi}W^{\rm
R}_{mnl;pq}.
\end{equation}
Therefore the matrix of the effective Hamiltonian takes the
following form
\begin{eqnarray}\label{Heffp02}
&\langle mnl|\widehat{H}_{\rm eff}|m'n'l'\rangle=
\omega_{mnl}^2\delta_{mm'}\delta_{nn'}\delta_{ll'}&\nonumber\\
&-{\rm i}\sum_{pq}k_{pq}[1+(-1)^{l+l'}e^{{\rm
i}(m'-m)\Delta\phi}]W_{mnl;pq}W^{*}_{m'n'l';pq}.&
\end{eqnarray}
The transmittance of sound waves in the $p,q$ propagating channel
through the resonator is given by equation \cite{Maksimov2015}
\begin{equation}\label{S-matrix}
T_{pq;pq}=2{\rm i}k_{pq} \sum_{mnl}\sum_{m'n'l'}W_{mnl;pq}
e^{-{\rm i}m'\Delta\phi}G_{mnl;m'n'l'} W^{*}_{m'n'l';p'q'}
\end{equation}
where
\begin{equation}\label{Green}
    \widehat{G}=\frac{1}{\omega^2-\widehat{H}_{\rm eff}},
\end{equation}
that is propagation of waves through the resonator is described by
the Green function which is inverse of the matrix
$\omega^2-\widehat{H}_{\rm eff}$ and coupling matrices of the
resonator with the input (left) waveguide and the output (right)
waveguide. However the most remarkable feature in Eq.
(\ref{S-matrix}) is complex phases of the coupling matrix elements
between states with different azimuthal indices $m$ and $p$. As we
show below that drastically changes the transmittance.
\subsection{Variation over the length of resonator at $\Delta\phi=\pi/4$.}
\label{phi025}
The case of $\Delta\phi\neq 0$ is interesting by that we face with
problem of embedding of the BIC into two continua which differ by
phase. First, the problem of the BIC residing in a finite number
of continua was considered by Pavlov-Verevkin and coauthors
\cite{Remacle1990}. Rigorous statement about the BICs was
formulated as follows. The interference among $N$ degenerate
states which decay into K non-interacting continua generally leads
to the formation of $N-K$ BICs. The equivalent point of view
\cite{SBR} is that the linear superposition of the $N$ degenerate
eigenstates $\sum_{n=1}^N a_n\psi_n$ can be adjusted to have zero
coupling with $K$ different continua in $N-K$ ways by variation of
the $N$ superposition coefficients $a_n$. Respectively, that
involve $K$-parametric avoided crossing. The number of continua
can grow due to a number of reasons, for example, non-
symmetrically attached waveguides, multiple propagation subbands
in the waveguides, or two polarizations of the radiation continuum
in case of electromagnetic BICs. Each case puts the problem of
searching BICs embedding into  many continua on the line of art
\cite{Bulgakov2011,Hsu2013,Zhen,Yang,Bulgakov2016a,Dai2018}.

\begin{figure}
\centering{\resizebox{0.8\textwidth}{!}{\includegraphics{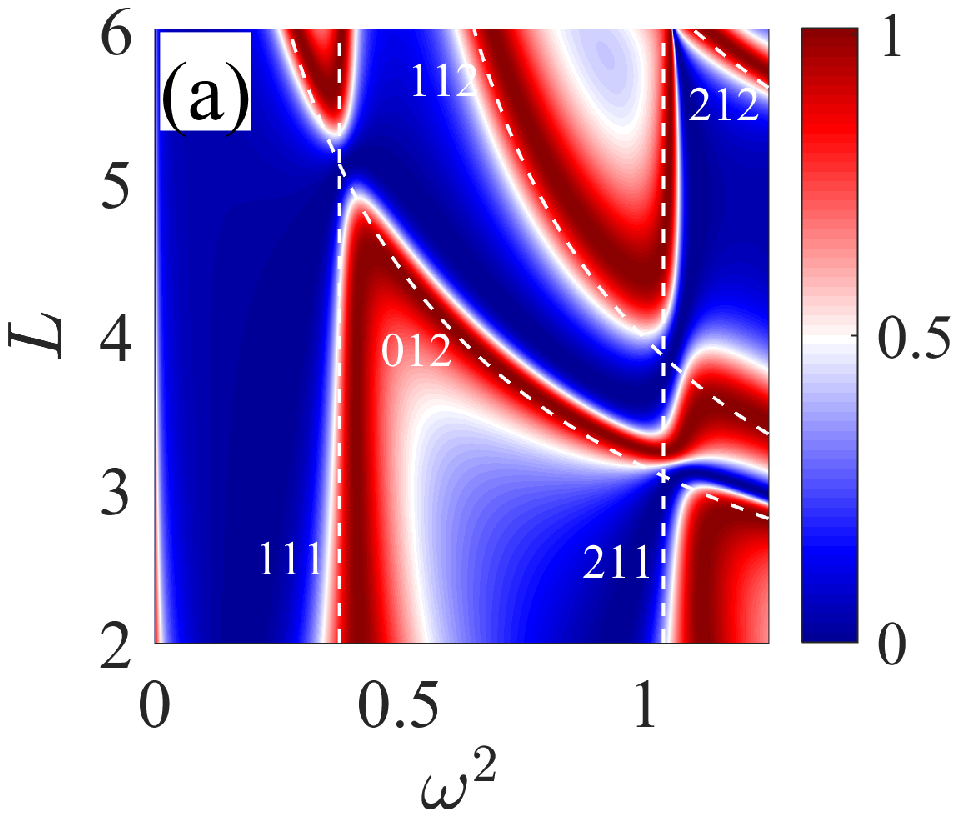},\includegraphics{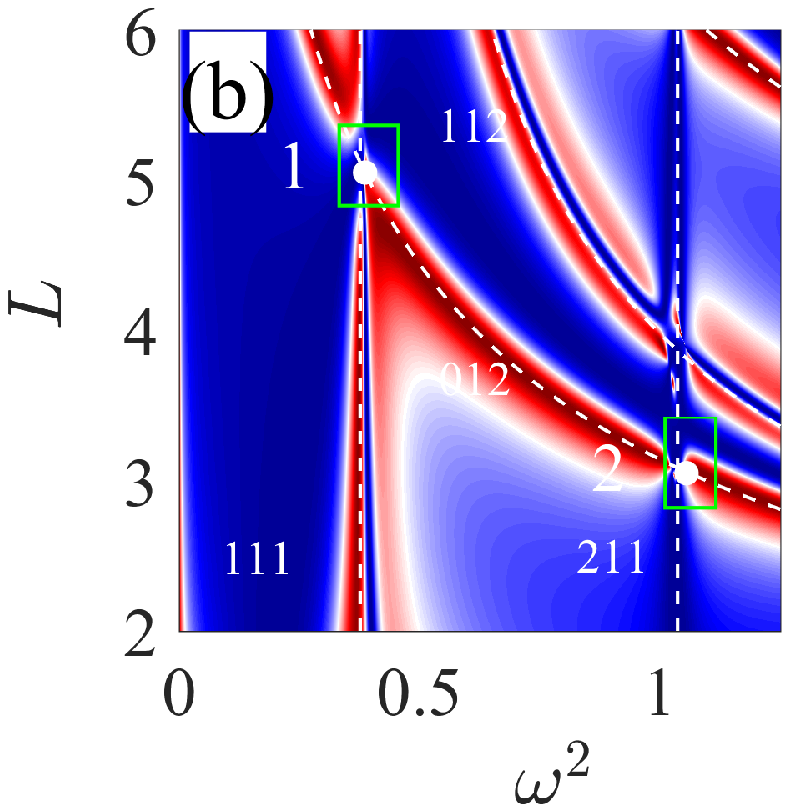}}}
\caption{Transmittance  of a  cylindrical resonator vs frequency
and length  of the  resonator $L$ at (a) $\Delta\phi=0$ and (b)
$\Delta\phi=\pi/4$. Dash lines show eigenlevels of closed
resonator with corresponding indices $m n l$. The positions of the
BICs are shown by closed circles.} \label{fig20}
\end{figure}

In what follows we take both waveguides with unit radius shifted
relative to the central axis of the resonator with radius $R=3$ by
a distance $r_0=1.5$. We consider transmission in the first
channel $p=0, q=1$ in the frequency domain $0< \omega <
\mu_{11}=1.8412$ (see Table 2). Although rotation of the waveguide
does not alter its propagating modes (continua) it provides the
complex phases in the coupling matrix elements of the resonator
eigenmodes with the continua as given by Eq. (\ref{WLWR}).  That
effects the transmittance as shown in Fig. \ref{fig20}.

\begin{figure}[ht]
\centering{\resizebox{0.45\textwidth}{!}{\includegraphics{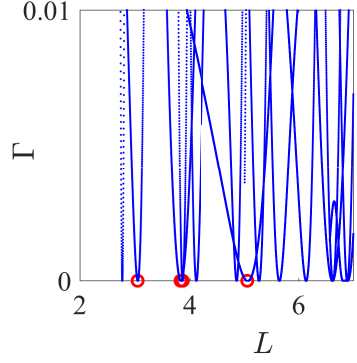}}}
\caption{Evolution of resonant widths under variation of the
resonator length at $\Delta\phi=\pi/4$. Circles mark BICs listed
in Table 3.} \label{fig21}
\end{figure}

\begin{figure}
\centering{\resizebox{0.8\textwidth}{!}{\includegraphics{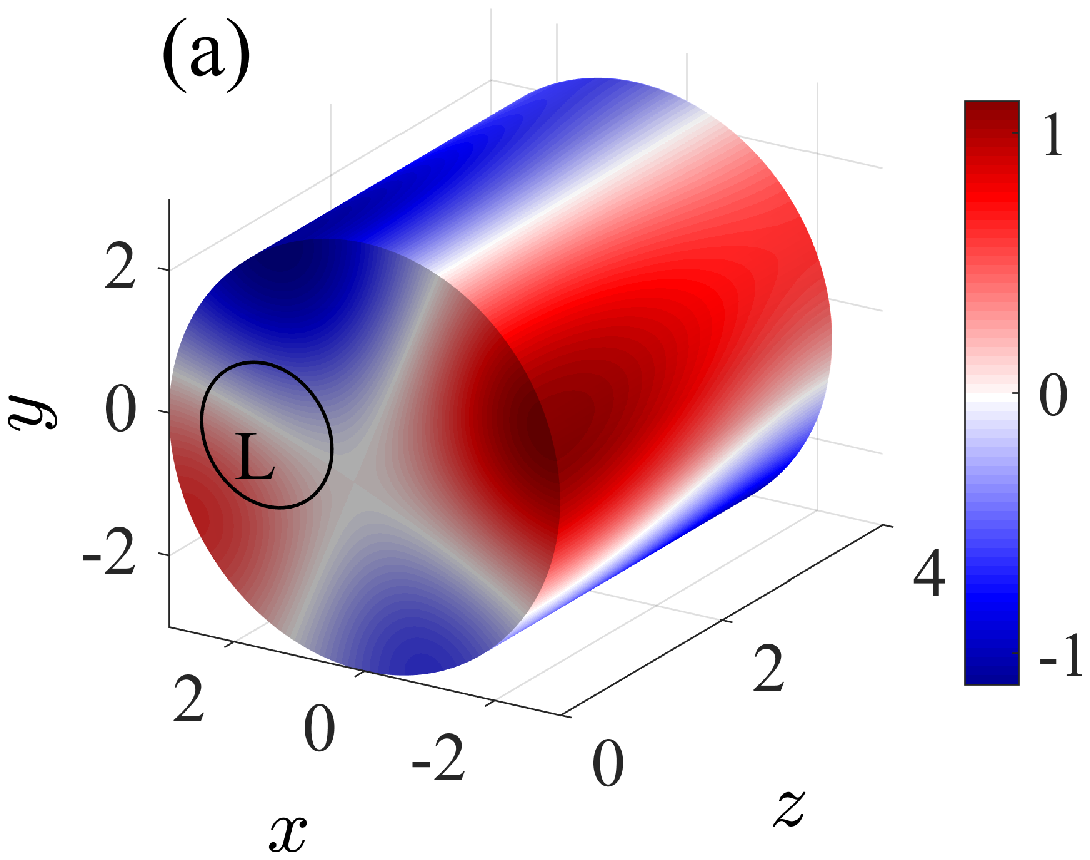},\includegraphics{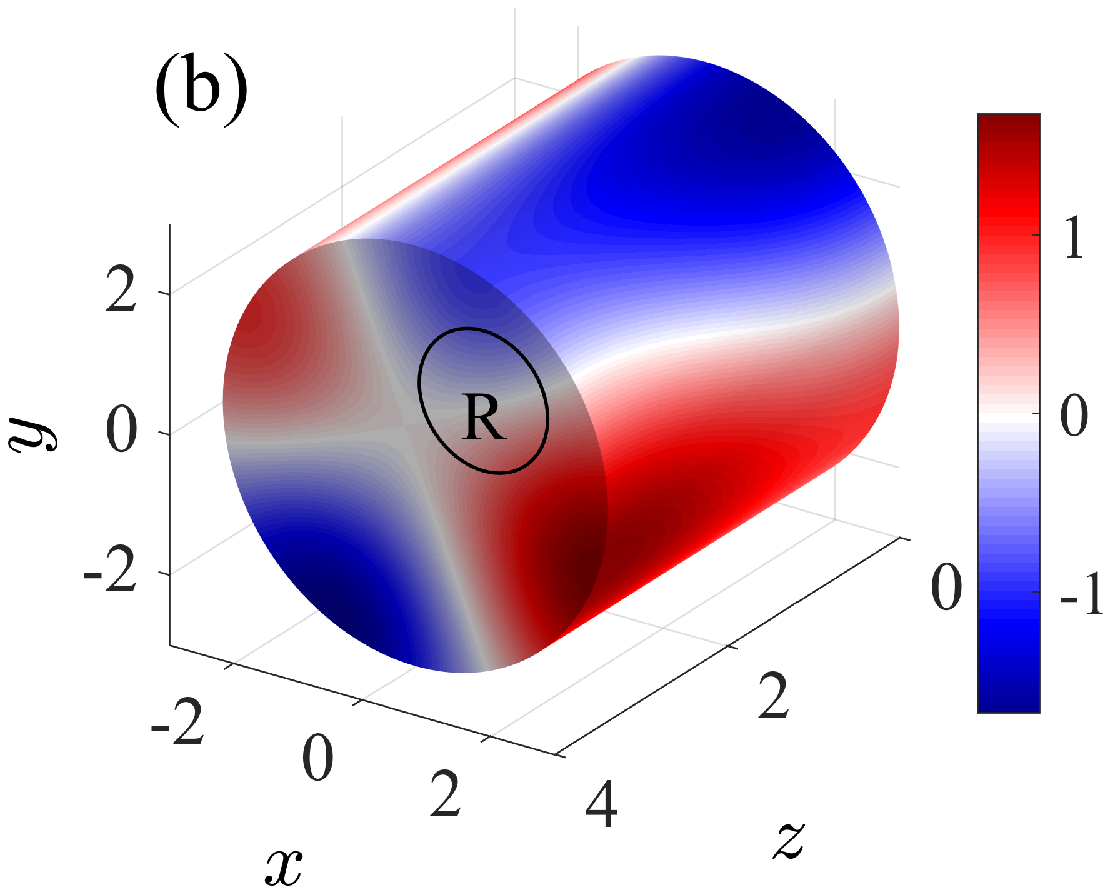}}}
\centering{\resizebox{0.8\textwidth}{!}{\includegraphics{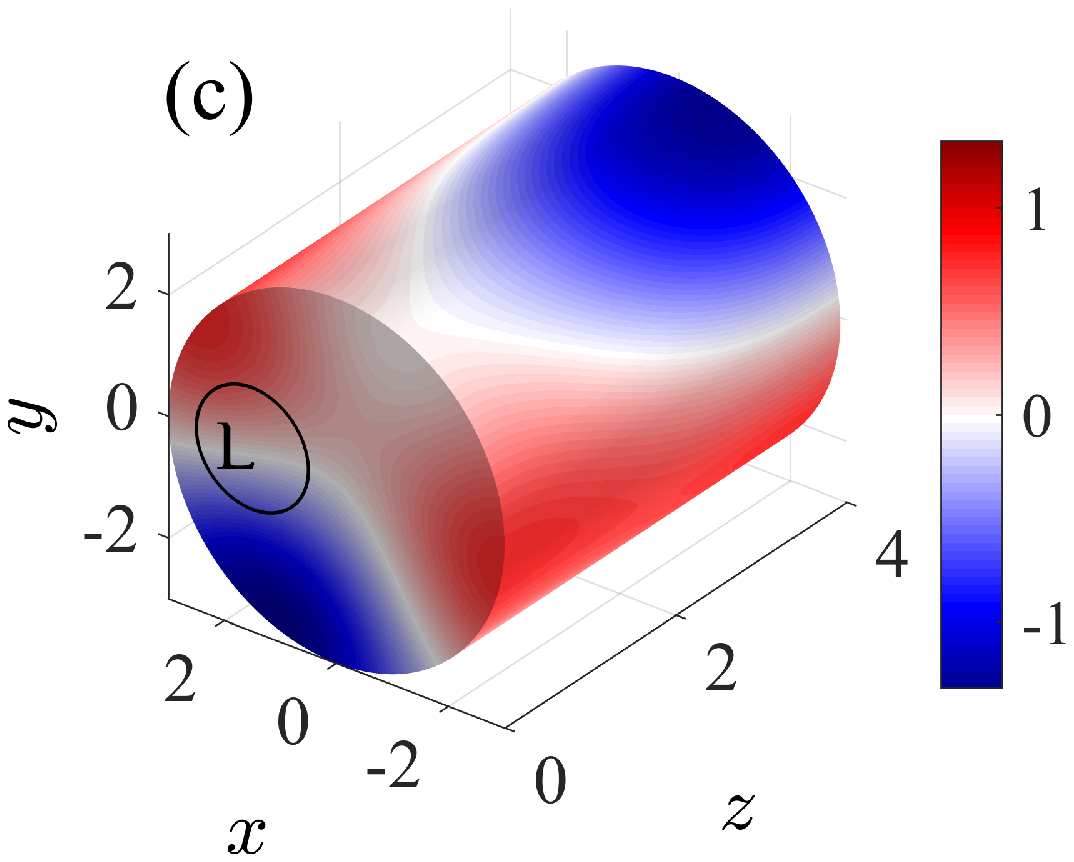},\includegraphics{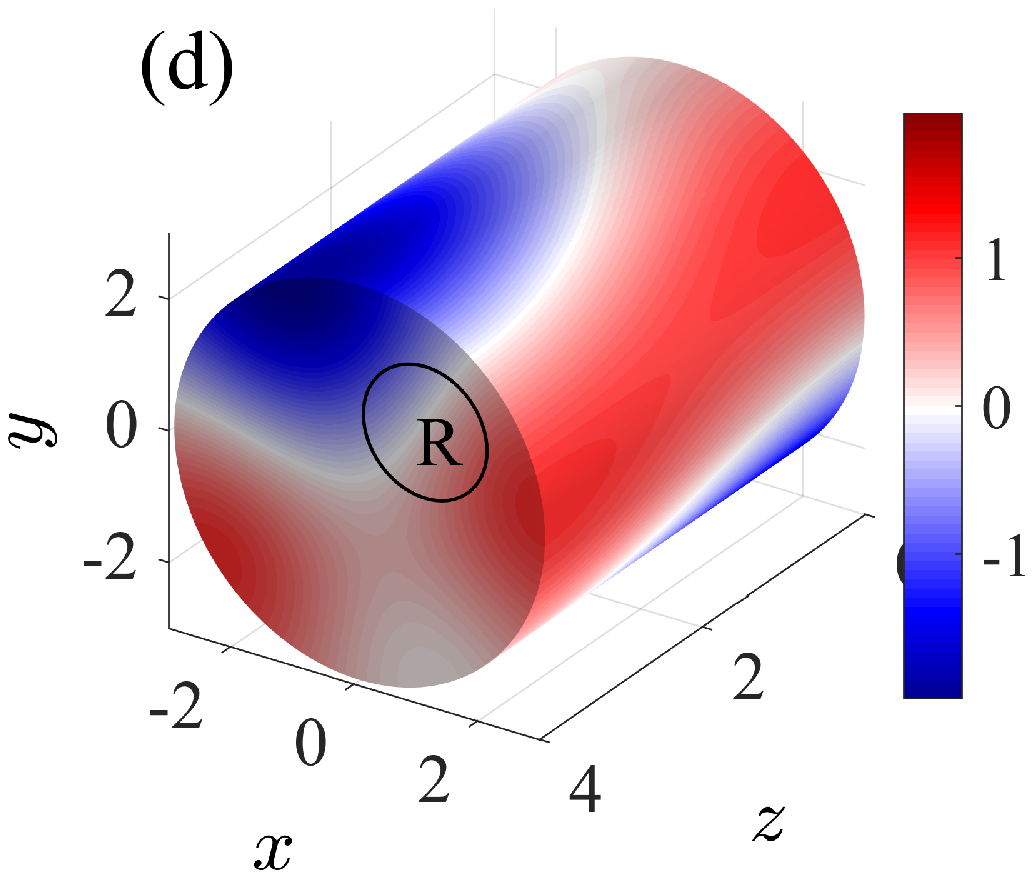}}}
\caption{Patterns of BIC 3 shown from the left (a) and right (b)
sides of the resonator and BIC 4 shown from the left (c) and right
(d) sides of resonator on the surface of the waveguide at
$\Delta\phi=\pi/4$.} \label{fig22}
\end{figure}

As before the BIC points are detected by finding zero resonant
width for variation of the resonator's length $L$ at fixed
$\Delta\phi=\pi/4$ as shown in Fig. \ref{fig21}.
We marked by circles only those BICs which are listed in Table
\ref{Tab3} and will be analyzed below.
\begin{table}
\centering
 \caption{BICs at $\Delta\phi=\pi/4$. The waveguides with the unit radius are
 shifted relative to the axis of cylindrical resonator with radius $R=3$ by
 a distance $r_0=1.5$.}
\begin{tabular}{|c|c|c|c|c|c|}
  \hline
BIC & $\omega^2$ & $L$ & $mnl$& $a_{mnl}$ &$|a_{mnl}|$ \\
  \hline
 &       &       &   012 & -0.113+0.272i & 0.294\\
1& 0.385 & 5.065 &  111 &  -0.478(1-i) &0.675\\
 &       &       &  -111& 0.675 &0.675\\
  \hline
 &       &       &  012 & -0.261(1-i)& 0.369\\
2& 1.055 & 3.051 &  211 &  0.656i &0.656\\
 &       &       & -211 &0.656 &0.656\\
\hline
 &       &       &  211 & 0.658i& 0.658\\
3& 1.0535& 3.833 &  -211 &  0.658 &0.658\\
 &       &       & 112 &-0.237-0.098i& 0.256\\
 &       &       & -112 &-0.098-0.237i& 0.256\\
\hline
 &       &       & 211 & -0.505 & 0.505\\
4& 1.065& 3.869 &  -211 &  0.505 &0.505\\
 &       &       & 112 &-0.455-0.189i&0.493\\
 &       &       & -112 &0.189+0.455i & 0.493\\
\hline
 \end{tabular}
 \label{Tab3}
\end{table}
The positions of the BICs and expansions coefficients over the
eigen modes of closed resonator (\ref{eigen})
\begin{equation}\label{BIC expansion}
    \psi_{\rm BIC}(r,\phi,z)=\sum_{mnl}a_{mnl}\Psi_{mnl}(r,\phi,z).
\end{equation}
are collected in Table \ref{Tab3}. Fig. \ref{fig22} show the 3-th
and 4-th BICs marked in Figs. \ref{fig20} (b) which are the
eigenmodes of the non hermitian effective Hamiltonian
(\ref{Heffp02}).
Fig. \ref{fig22} clearly shows that BICs at $\Delta\phi\neq 0$ are
decoupled from the first channel owing to twisting of the BIC
modes by the rotation angle $\Delta\phi$.
\subsection{Arbitrary $\Delta\phi$. Wave faucet.}
\label{Wave faucet}
\begin{figure}
\centering{\resizebox{0.8\textwidth}{!}{\includegraphics{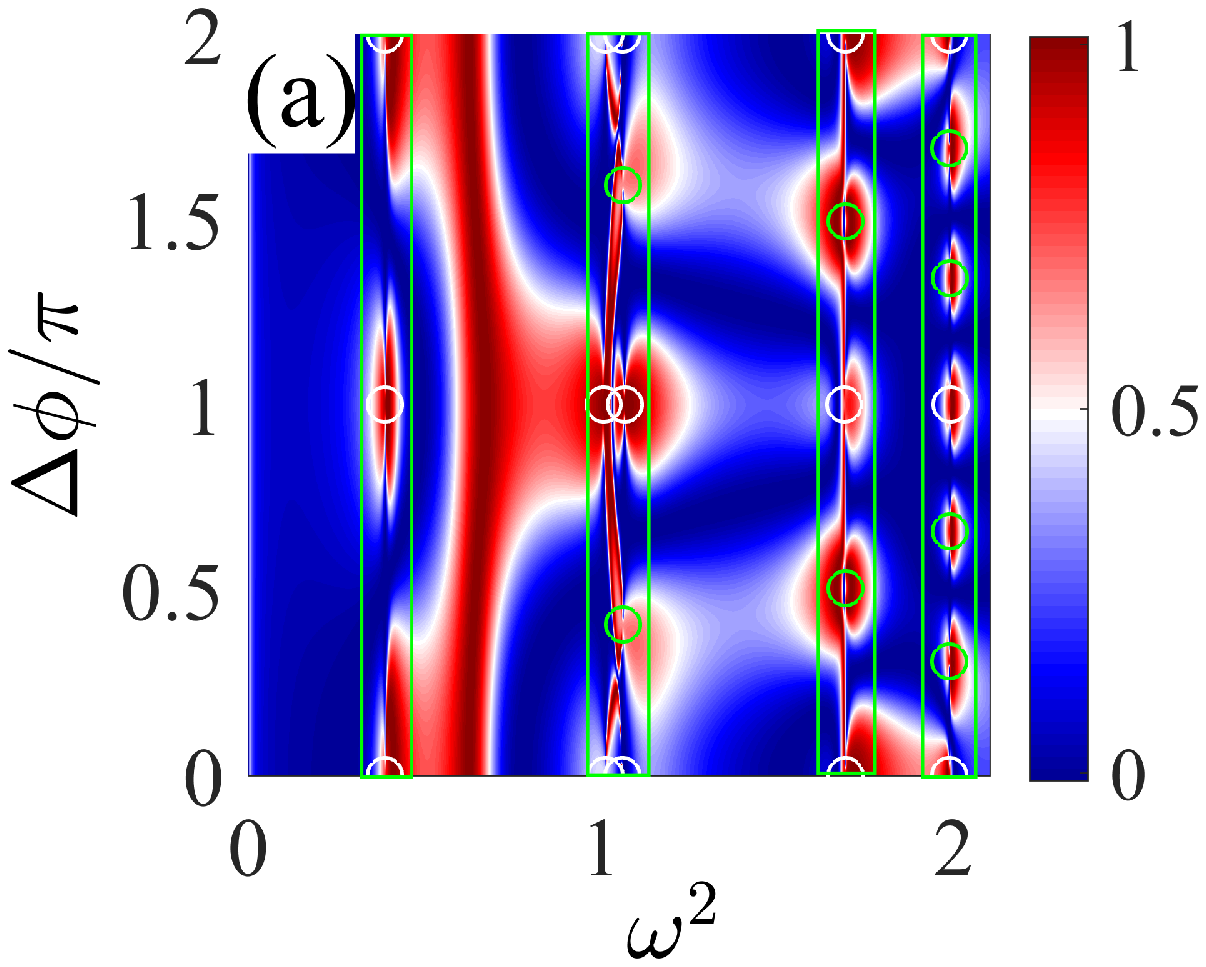},\includegraphics{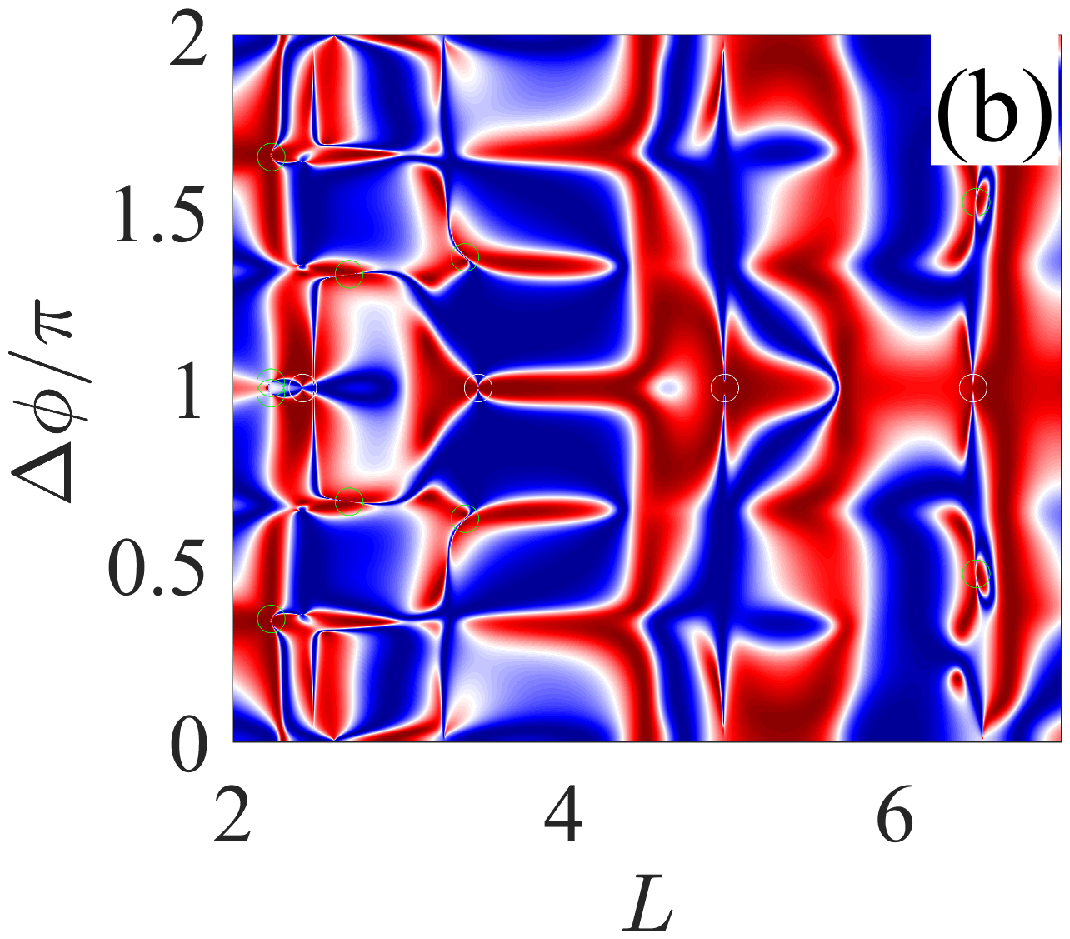}}}
\caption{(Color  online)  Transmittance  of a  cylindrical
resonator (a) vs frequency and rotation angle $\Delta\phi$ at
$L=4$ and (b) vs length and rotation angle at $\omega^2=2$. The
positions of  the BICs are shown by open circles.} \label{fig23}
\end{figure}

Eq. (\ref{S-matrix}) shows that the phase difference $\Delta\phi$
due to the rotation of the input waveguide brings an important
contribution into interference between resonances. Fig.
\ref{fig23} vividly illustrates high sensitivity of the
transmittance to the rotation angle $\Delta\phi$.
As seen from Fig. \ref{fig20}  the eigenmode $012$ crosses the
eigenmodes $\pm 111$ around $L=5$. Respectively the transmittance
is basically given by the interference of these resonances in the
vicinity of this crossing $L=5, \omega^2\approx 0.385$ (see
parameters of the 1-th BIC in Table \ref{Tab3}). According to Eq.
(\ref{WLWR}) we have $W^{\rm L}_{012;01}=-W^{\rm R}_{012;01},
W^{\rm L}_{\pm 111;01}=W^{\rm R}_{\pm 111;01}e^{\mp {\rm
i}\Delta\phi}$. Therefore for the output waves interfering
constructively we have to take $\Delta\phi=\pm \pi$, while the
full destructive interference takes place at $\Delta\phi=0$. This
simple consideration is in excellent agreement with numerics
presented in Fig. \ref{fig23} (a).
\begin{figure}[ht]
\centering{\resizebox{0.8\textwidth}{!}{\includegraphics{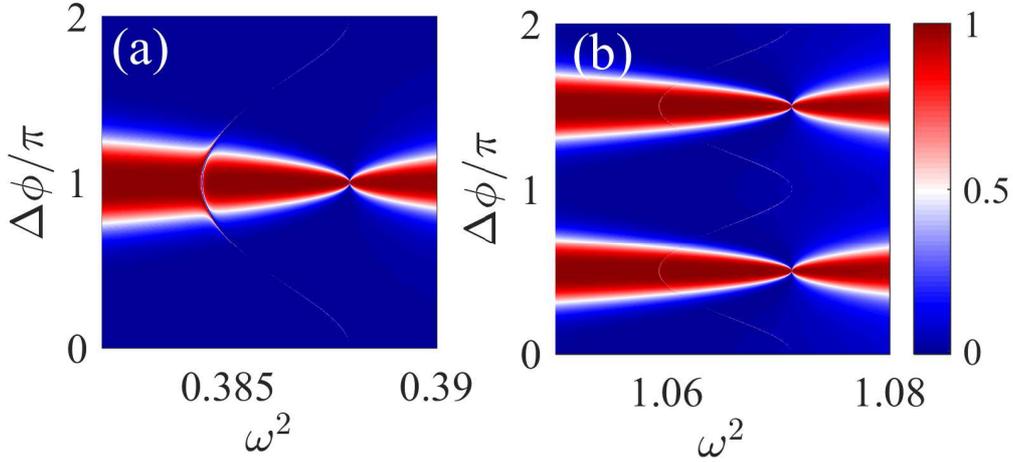}}}
\caption{Transmittance vs the frequency and rotation angle in the
vicinity of crossing of the modes (a) $012$ and $\pm 111$ at
$L=5$, (b) $012$ and $\pm 211$ at $L=3$,  and (c) $\pm 112$ and
$\pm 211$, $L=4$.} \label{fig24}
\end{figure}
Along the same line for channels $012$ and $\pm 211$ in the
vicinity of $L=3$ we have from Eq. (\ref{WLWR}) $W^{\rm
L}_{012;01}=-W^{\rm R}_{012;01}, W^{\rm L}_{\pm 211;01}=W^{\rm
R}_{\pm 111;01}e^{\mp 2{\rm i}\Delta\phi}$ to open wave flux
through the resonator at $\Delta\phi=\pi/2, 3\pi/2$. That
conclusion fully agrees with the transmittance shown in Fig.
\ref{fig23} (b). Thus, the rotation of the input waveguide
strongly tunes Fano resonance \cite{Sadreev2017}. In particular
there can be a collapse of Fano resonance when the transmission
zero approaches to the transmission maximum that is the signature
of BICs (see the section \ref{Sect:FW}).

\begin{figure}[ht]
\centering{\resizebox{0.4\textwidth}{!}{\includegraphics{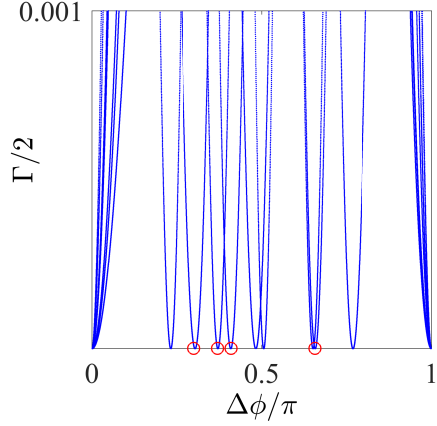}}}
\caption{Evolution of resonant widths under waveguide rotation at
$L=4$.} \label{fig25}
\end{figure}

\begin{figure}[ht]
\centering{\resizebox{0.9\textwidth}{!}{\includegraphics{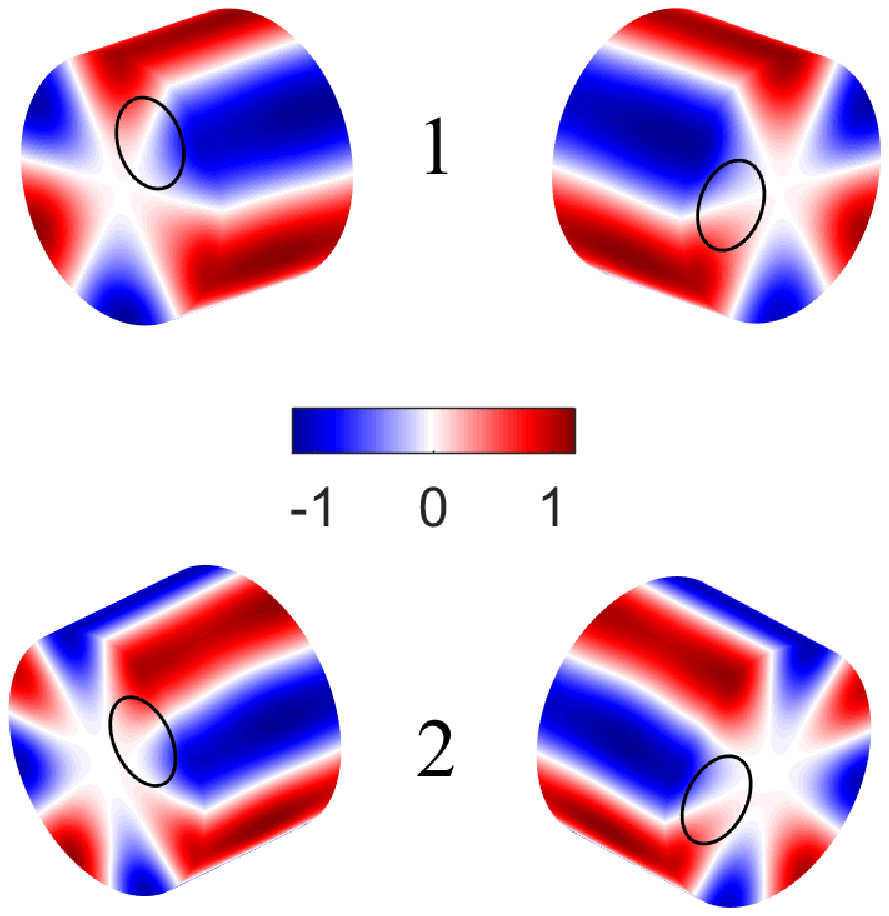},\includegraphics{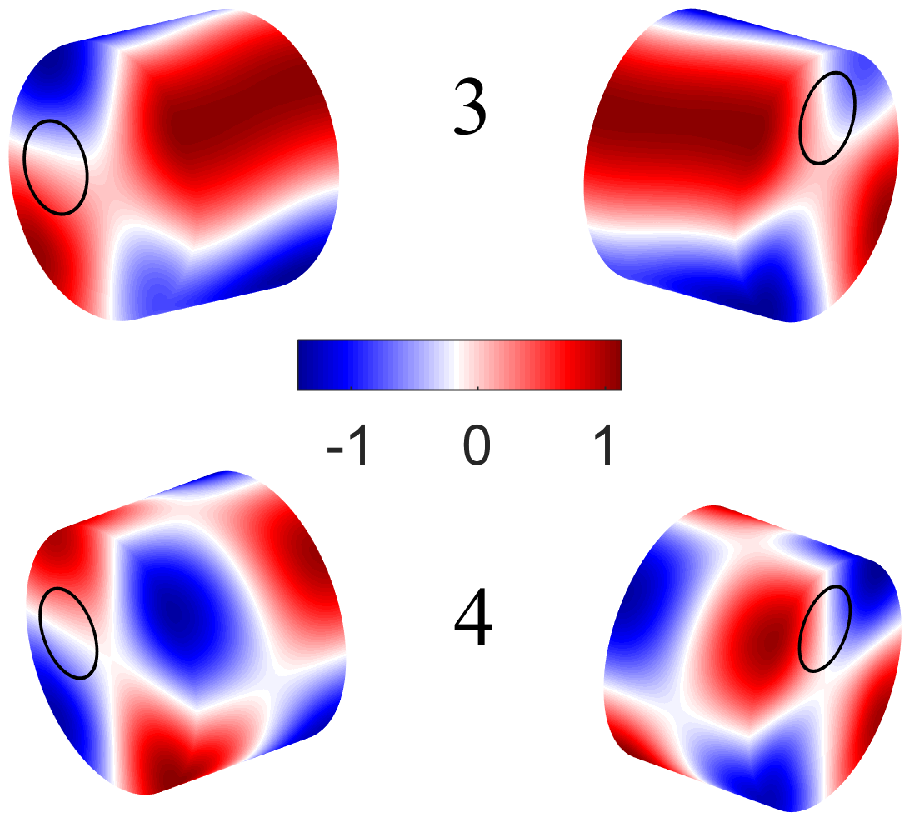}}}
\caption{Patterns of BICs marked by open circles in Fig.
\ref{fig12} and listed in Table \ref{Tab4} on the surface of the
resonator at $L=4$: 1--4. Open circles show where the left and
right waveguides are attached to the resonator. } \label{fig26}
\end{figure}

Fig. \ref{fig25} evidences that the rotation angle
$\Delta\phi=\pi/4$ is not unique for BICs to occur. In fact, we
will show below analytically that there is whole line
$L=f(\Delta\phi)$ of BICs.
Among them we select four BICs shown in Fig. \ref{fig26}.
\begin{table}
\centering
 \caption{BICs at $L=4$.}
\begin{tabular}{|c|c|c|c|c|}
  \hline
BIC &  $\Delta\phi/\pi$ & $\omega^2$ & $mnl$& $a_{mnl}$  \\
  \hline
 1& 0.308 & 1.9868 &  311 &  0.7056 \\
 &       &       &  -311& $0.7056e^{-3{\rm i}\Delta\phi}$ \\
  \hline
2& 0.2351 & 3.17304 &  411 &  0.705 \\
 &       &       & -411 & $0.705e^{4{\rm i}\Delta\phi}$ \\
\hline
 3& 0.4171& 1.05688 &  211 &  0.6898 \\
&       &       & -211 & $-0.6898e^{-2{\rm i}\Delta\phi}$ \\
 &       &       & 121 &0.0933+0.1215i\\
 &       &       & -121 &$a_{121}e^{{\rm i}\Delta\phi}$\\
\hline
 4&0.5055 & 1.68872  & 211 & 0.7043 \\
 &       &       & -211 & $0.7043e^{-2{\rm i}\Delta\phi}$\\
\hline
 \end{tabular}
 \label{Tab4}
\end{table}
Let us consider the 1-th BIC from Table \ref{Tab4} whose azimuthal
dependence is given by ${\rm cos}[3(\phi-\Delta\phi/2)]$. In order
to decouple this BIC from the right waveguide at $\Delta\phi=0$
the nodal line of the BIC mode has to be positioned at $\phi=0$
that gives us the equation $\frac{3}{2}\Delta\phi=\frac{\pi}{2}$,
i.e., $\Delta\phi=\frac{\pi}{3}$. Therefore the BIC mode is ${\rm
cos}[3(\phi-\pi/6)]$ which equals zero at $\phi=0$. The left
waveguide is rotated by the angle $\pi/3$ for which the BIC mode
is decoupled from the left waveguide too. Numerically according to
Table \ref{Tab4}  we have $\Delta\phi=0.308\pi$ which is close to
$\pi/3$. The small difference is a contribution of the evanescent
modes. Similarly, for the 2-th BIC we obtain ${\rm
cos}[4(\phi-\Delta\phi/2)]$ that gives us $\Delta\phi=\pi/4$ which
is close to numerical result $\Delta\phi=0.235\pi$ given in Table
\ref{Tab4}. For the 4-th BIC we obtain that $\Delta\phi=\pi/2$
that also well agrees with Table \ref{Tab4}. The most interesting
is the 3-th BIC which is superposed of two modes ${\rm
cos}[2(\phi-\Delta\phi/2)]$ and ${\rm cos}[(\phi+\Delta\phi/2)]$.
As the result the BIC mode is twisted as shown in Figs.
\ref{fig22} and \ref{fig26} (b) and (d).
\subsection{CMT theory of twisted BICs}
\label{CMT}
\begin{figure}[ht]
\centering{\resizebox{0.7\textwidth}{!}{\includegraphics{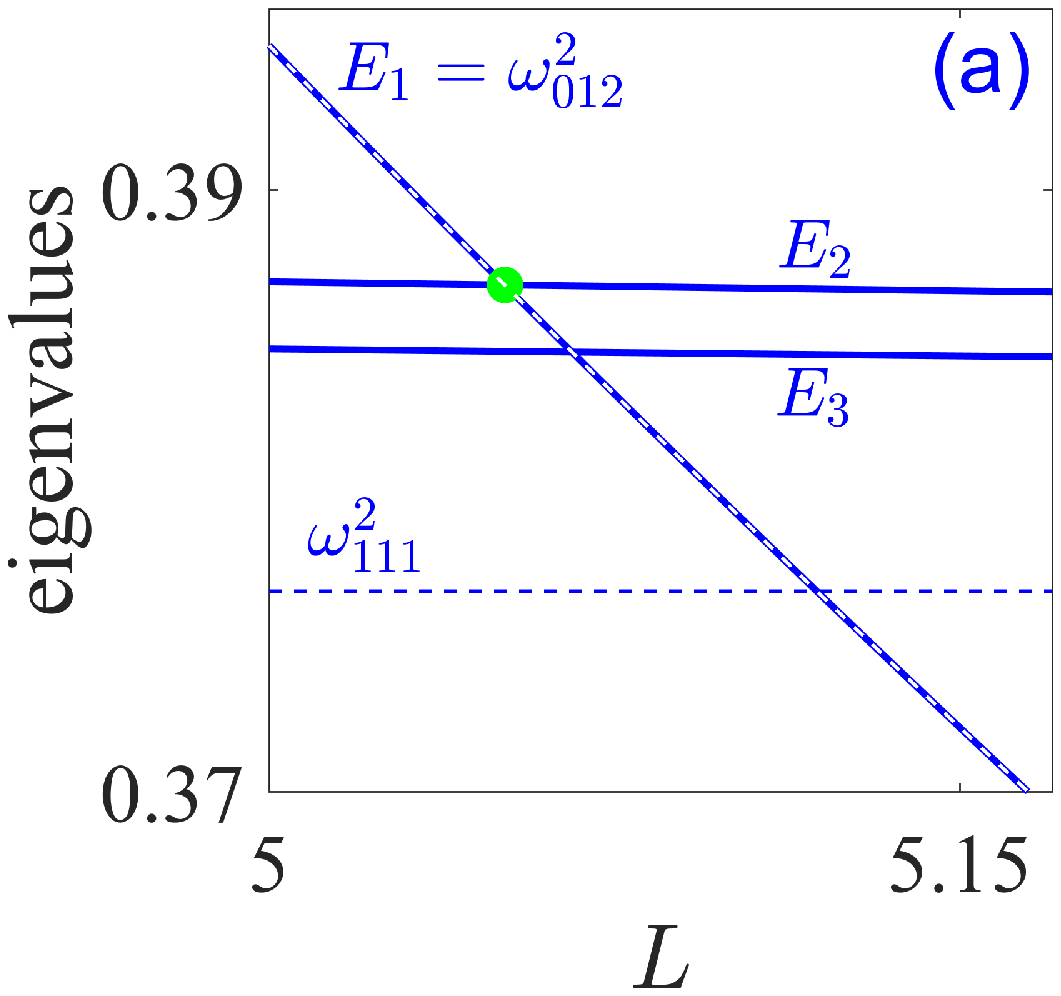},\includegraphics{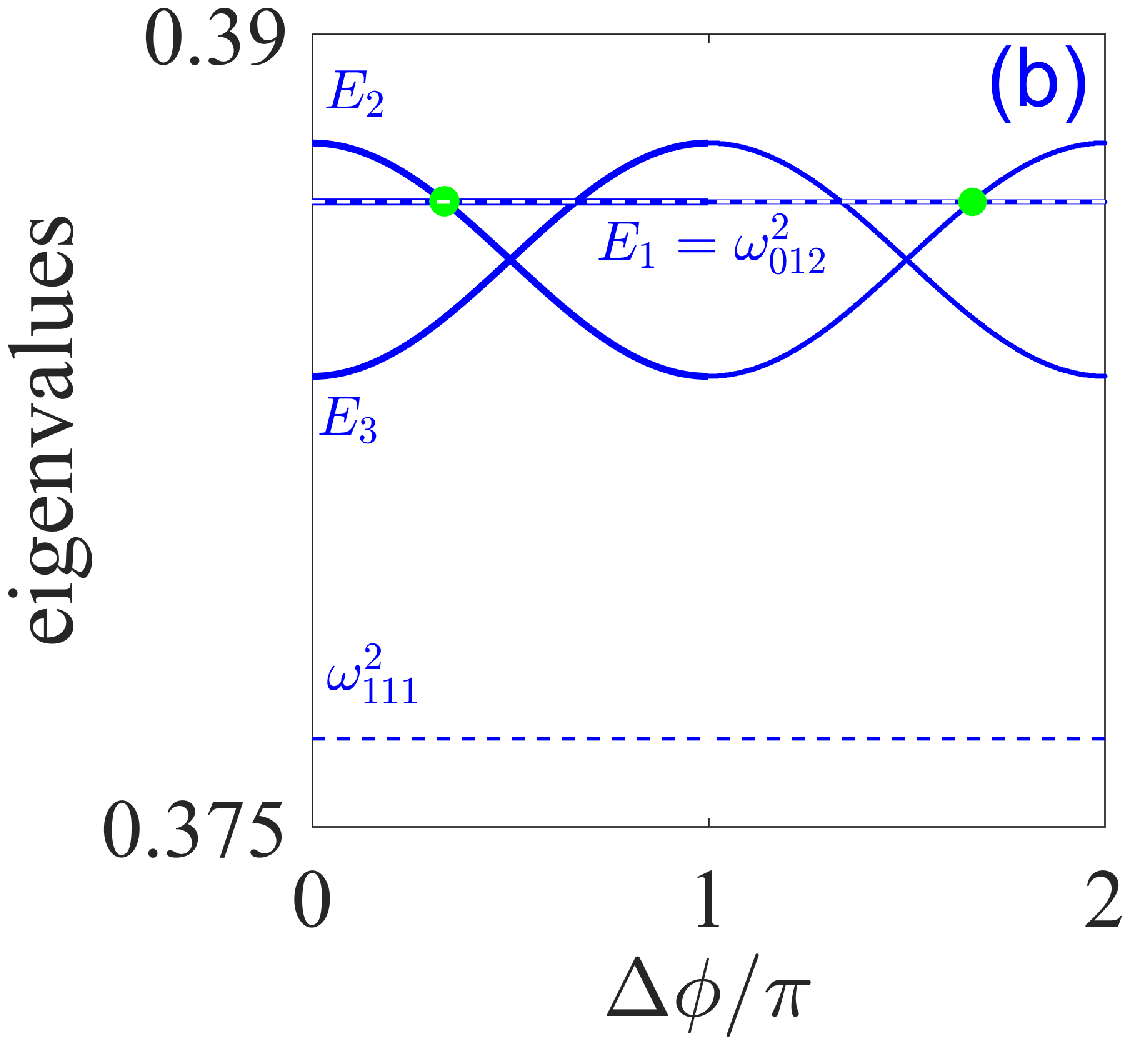}}}
\caption{The eigenvalues $\omega_{012}^2$ and $\omega_{111}^2$ of
the closed resonator are shown by dash lines while the eigenlevels
(\ref{neweig}) shifted by evanescent modes are shown by solid
lines. (a) The eigenvalues (\ref{neweig}) vs the resonator length
at $\phi=\pi/3$ and (b) vs rotation angle at $L=5.0512$.}
\label{fig27}
\end{figure}

In the vicinity of crossings of eigenlevels of closed cylindrical
resonator highlighted by green frames in Fig. \ref{fig20} (a) it
is reasonable to truncate the effective Hamiltonian
(\ref{Heffp02}) by only those modes which participate in crossing
similar to the two-level description in section 6. The only
difference is that, at least, three modes participate in
degeneracy  in the present case. For example, let consider the
case when the eigenlevel $\omega_{012}^2=\pi^2/L^2$ crosses with
the double degenerate eigenlevel $\omega_{111}^2=\mu_{11}^2/R^2$
shown in Fig. \ref{fig27} (a) by dash lines.
The coupling matrix elements of the eigenmodes with the first
propagating mode $p=0, q=1$ (see Table \ref{Tab2}) of the right
waveguide according to Eqs. (\ref{propag}), (\ref{eigRes}) and
(\ref{Wp}) equal
\begin{eqnarray}\label{W01R}
&W_{mnl;01}^L=(w_0~ w_1~ w_1), w_0=W_{012;01}^L=\frac{1}{3}\sqrt{\frac{2}{L}},&\nonumber\\
&w_1=W_{\pm 111;01}^L=0.269\sqrt{\frac{1}{L}}&
\end{eqnarray}
for the given radius of the resonator. We also take into account
the coupling with the first evanescent modes $p=\pm 1, q=1$ of the
waveguide (see Table \ref{Tab2})
\begin{eqnarray}\label{W11R}
&W_{mnl;11}^L=(0~ v_1~ v_2), ~W_{mnl;-11}^L=(0~ v_2~ v_1),&\\
&v_1=W_{012;11}^L=0.1141\sqrt{\frac{1}{L}}, ~v_2=W_{\pm
111;11}^L=-0.0141\sqrt{\frac{1}{L}}.&\nonumber
\end{eqnarray}
Because of the phase difference between the coupling matrix
elements for left and right waveguides we immediately obtain
\begin{eqnarray}\label{WL}
&W_{mnl;01}^R=(-w_0~ w_1e^{i\Delta\phi}~ w_1e^{-i\Delta\phi}),&\nonumber\\
&W_{mnl;11}^R=(0~ v_2e^{i\Delta\phi}~
v_1e^{-i\Delta\phi}),&\nonumber\\
&W_{mnl;-11}^R=(0~ v_1e^{i\Delta\phi}~v_2e^{-i\Delta\phi}).&
\end{eqnarray}
The contribution of the higher evanescent modes shown in Table
\ref{Tab2} is negligible. For open channel $p=0, q=1$ the wave
number $q_{01}=\omega$ while for the next closed channel $p=\pm 1,
q=1$ the wave number $k_{11}=iq_{11},
q_{11}=\sqrt{\mu_{11}^2-\omega^2}$ is imaginary. Then the
truncated effective Hamiltonian (\ref{Heff}) can be rewritten as
follows
\begin{equation}\label{Heff1}
\widehat{H}_{eff}=\widehat{H}_R+q_{11}\sum_{C=L,R} \sum_{p=\pm
1}\widehat{W}^C_{p=\pm 1,1}\{\widehat{W}^C_{p=\pm 1,1}\}^{\dagger}
-i\omega\sum_{C=L,R}\widehat{W}^C_{01}\{\widehat{W}^C_{01}\}^{\dagger}
=\widehat{\widetilde{H}}_R-i\omega\widehat{\Gamma},
\end{equation}
where the Hermitian term
\begin{equation}\label{HBSC1}
\widehat{\widetilde{H}}_R=\left(\begin{array}{ccc}
  \omega_{012}^2 & 0 & 0 \\
  0 & \omega_{111}^2+2q_{11}(v_1^2+v_2^2) & 2q_{11}v_1v_2(1+e^{-2i\Delta\phi}) \\
  0 & 2q_{11}v_1v_2(1+e^{2i\Delta\phi}) &  \omega_{111}^2+2q_{11}(v_1^2+v_2^2)\\
\end{array}\right)
\end{equation}
is the Hamiltonian of the resonator coupled to the evanescent
modes. The anti-Hermitian part takes the following form
\begin{equation}\label{GammaBSC1}
\widehat{\Gamma}=\left(\begin{array}{ccc}
  2w_0^2 & w_0w_1(1-e^{i\Delta\phi}) & w_0w_1(1-e^{-i\Delta\phi}) \\
  w_0w_1(1-e^{-i\Delta\phi}) & 2w_1^2 & w_1^2(1+e^{-2i\Delta\phi}) \\
  w_0w_1(1-e^{i\Delta\phi}) & w_1^2(1+e^{2i\Delta\phi}) &  2w_1^2\\
\end{array}\right).
\end{equation}
The eigenvalues of the Hermitian part of the Hamiltonian
(\ref{HBSC1}) can be easily found as
\begin{equation}\label{neweig}
E_1=\omega_{012}^2, E_{2,3}=\omega_{111}^2+2q_{11}[v_1^2+v_2^2\pm
2v_1v_2\cos\Delta\phi].
\end{equation}
Thus the evanescent modes of the waveguides non-coaxially attached
to the cylindrical resonator lift the degeneracy of eigenmodes
$\pm 1 1 1$ as shown in Fig. \ref{fig27} by solid lines. The
degeneracy is restored for $\Delta\phi=\pi/2, 3\pi/2$.  The
corresponding eigenmodes of the Hamiltonian (\ref{HBSC1}) are the
following
\begin{equation}\label{neweigmod}
 \mathbf{X}_1=   \left(\begin{array}{c}
  1 \\
  0 \\
  0 \\
\end{array}\right),
 \mathbf{X}_2=\frac{1}{\sqrt{2}}\left(\begin{array}{c}
  0 \\
  -e^{-i\Delta\phi} \\
  1 \\
\end{array}\right),
 \mathbf{X}_3=\frac{1}{\sqrt{2}}\left(\begin{array}{c}
  0 \\
  e^{-i\Delta\phi} \\
  1 \\
\end{array}\right).
\end{equation}

\begin{figure}[ht]
\centering{\resizebox{0.7\textwidth}{!}{\includegraphics{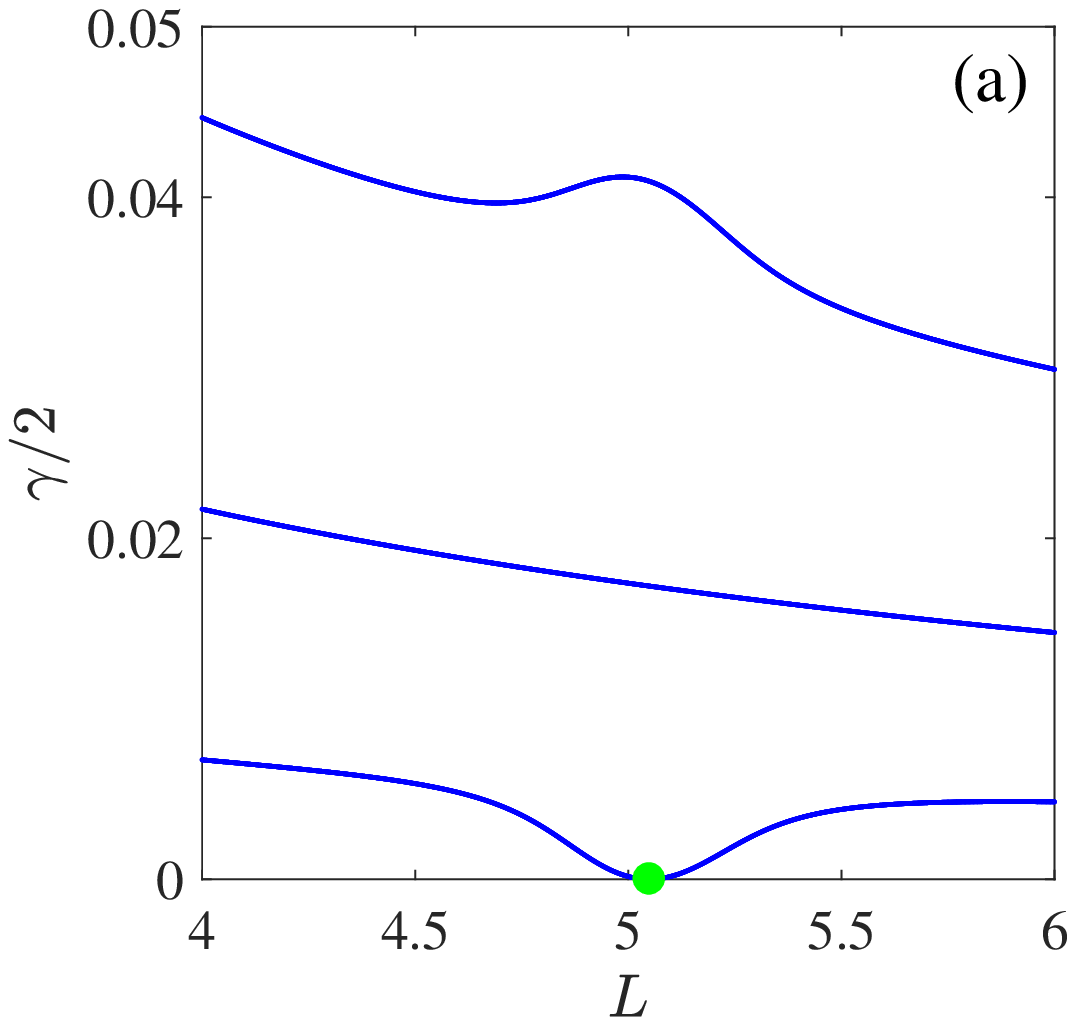},\includegraphics{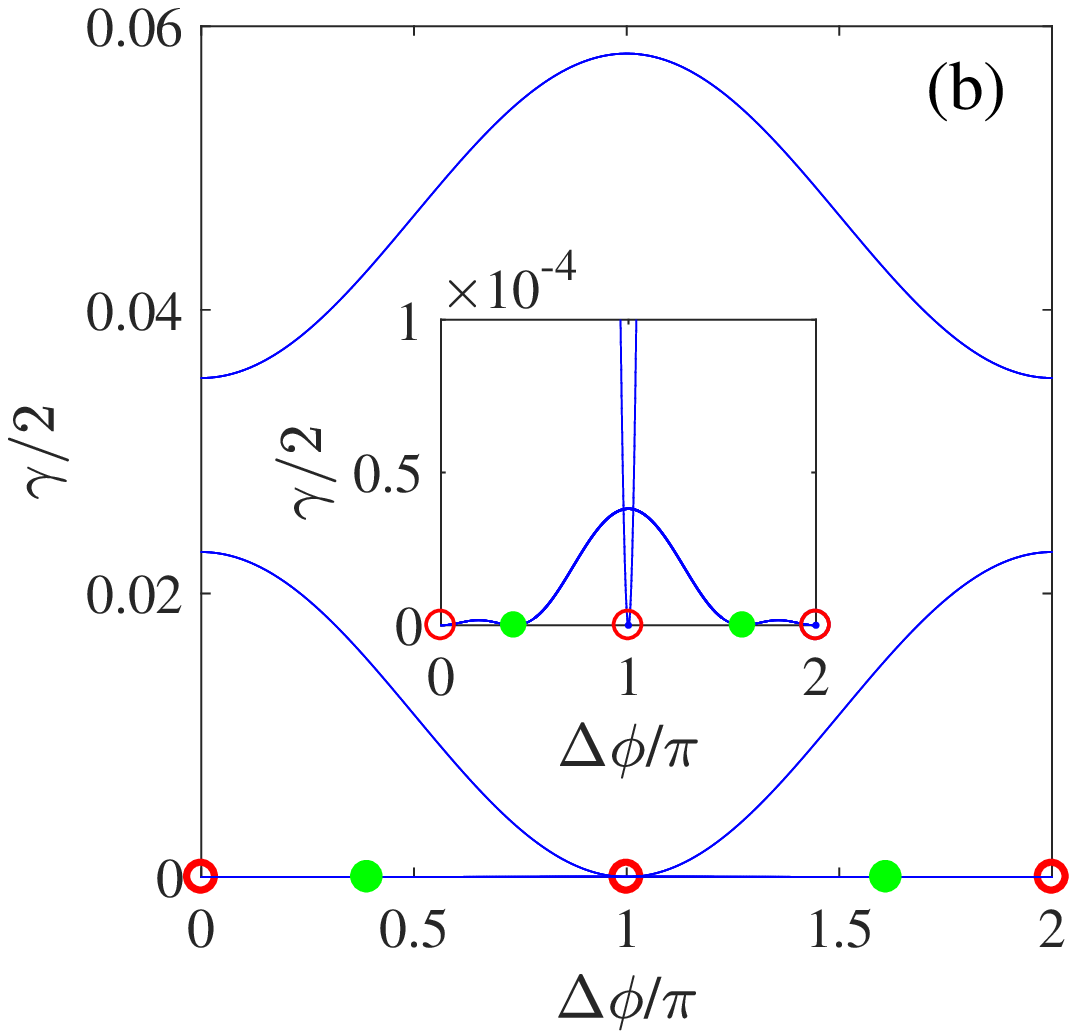}}}
\caption{The resonant width vs (a) the resonator length at
$\Delta\phi=\pi/4$ and (b) rotation angle at $L=5.0512$. Open
circles show the symmetry protected BICs, closed circles the FW
BICs .} \label{fig28}
\end{figure}

Next, let us consider the BIC in the truncated version
(\ref{Heff1}). The point of the BIC can be easily diagnosed by
zero resonant width as shown in Fig. \ref{fig28}. For
$\Delta\phi=\pi/4$ the BIC occurs at $L=L_c=5.0512$ marked by
closed green circle in Fig. \ref{fig28} (a). Respectively at
$L=L_c$ the BIC occurs at $\Delta\phi=\pi/4$ and
$\Delta\phi=2\pi-\pi/4$. These points are seen in zoomed insert in
Fig. \ref{fig28} (b).

For $\Delta\phi=0$ both continua of left and right waveguides
coincide to result in the symmetry protected BIC superposed of
degenerate eigenmodes of the closed resonator $\psi_{111}$ and
$\psi_{-111}$ to be in the following form
\begin{equation}\label{sym prot BSC}
    \psi_{\rm BSC}(r,\phi,z)=AJ_1(\mu_{11}r)\sin(\pi z/L)\sin\phi
\end{equation}
which always has zero coupling with the propagation mode
$\psi_{01}(\rho,\alpha,z)$ shown in Table \ref{Tab2}. As seen from
Eq. (\ref{sym prot BSC}) this conclusion also holds true for
$\Delta\phi=\pi$. These BICs is trivial symmetry protected ones
for arbitrary resonator length.

As soon as $\Delta\phi\neq 0$ the continua become different to
destroy the symmetry protected BICs. It could be expected that in
the case of two waveguides the point of threefold degeneracy where
the $\omega_{012}$ crosses the double degenerate $\omega_{111}$ as
shown in Fig. \ref{fig27} (a) is a BIC point in accordance with
the above consideration. However the BIC point where the resonant
width turns to zero (see Fig. \ref{fig28}) does not coincide with
this point. The computation on the basis of full basis effective
Hamiltonian gives the same result. In fact, the evanescent modes
split the eigenvalues (\ref{neweig}). Respectively the point of
threefold degeneracy $\omega_{111}^2=\omega_{012}^2(L)$ splits
into two double degenerate points $E_1(L)=E_2(L,\Delta\phi)$ and
$E_1(L)=E_3(L,\Delta\phi)$. As shown in Fig. \ref{fig27} (a) the
first case exactly corresponds to the BIC point but not the second
case.

\begin{figure}[hbt]
\centering{\resizebox{0.35\textwidth}{!}{\includegraphics{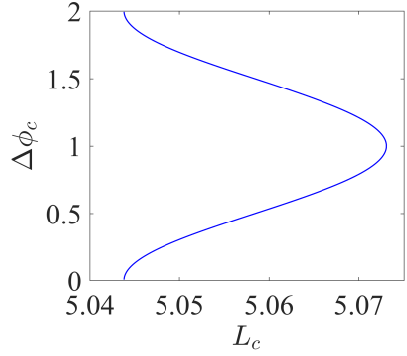}}}
\caption{Line of the BICs in the parametric space of the resonator
length and rotation angle $\Delta\phi$.} \label{fig29}
\end{figure}
In the first case we can superpose the eigenmodes
(\ref{neweigmod}) as $a\mathbf{X}_1+b\mathbf{X}_2$ and require
zero coupling of this superposed mode with the left waveguide
\begin{equation}\label{coup1}
aw_0+\frac{b}{\sqrt{2}}w_1(1-e^{-i\Delta\phi})=0
\end{equation}
according to Eqs. (\ref{W01R}) and (\ref{neweigmod}). It is easy
to show that the coupling with the phase shifted continuum of the
left waveguide takes the {\it same} form as Eq. (\ref{coup1}).
Thus, the BIC has the following form
\begin{equation}\label{BSC1}
\psi_{\rm
BSC}=w_1(1-e^{-i\Delta\phi})\psi_{012}+w_0(e^{-i\Delta\phi}\psi_{111}-\psi_{-111}).
\end{equation}
Substituting eigenmodes (\ref{eigRes}) into Eq. (\ref{BSC1}) we
obtain
\begin{equation}\label{BSC2}
\psi_{\rm
BSC}=2ie^{-i\Delta\phi/2}[w_1\sin(\Delta\phi/2)\psi_{01}(r)\psi_2(z)+
w_0\sin(\phi-\Delta\phi/2)\psi_{11}(r)\psi_1(z)].
\end{equation}
The BIC point is given by the equation $E_1(L)=E_2(L,\Delta\phi)$
which gives rise to a line of the BSC in the parametric space $L$
and $\Delta\phi$ shown in Fig. \ref{fig29}.
\begin{figure}[hbt]
\centering{\resizebox{0.65\textwidth}{!}{\includegraphics{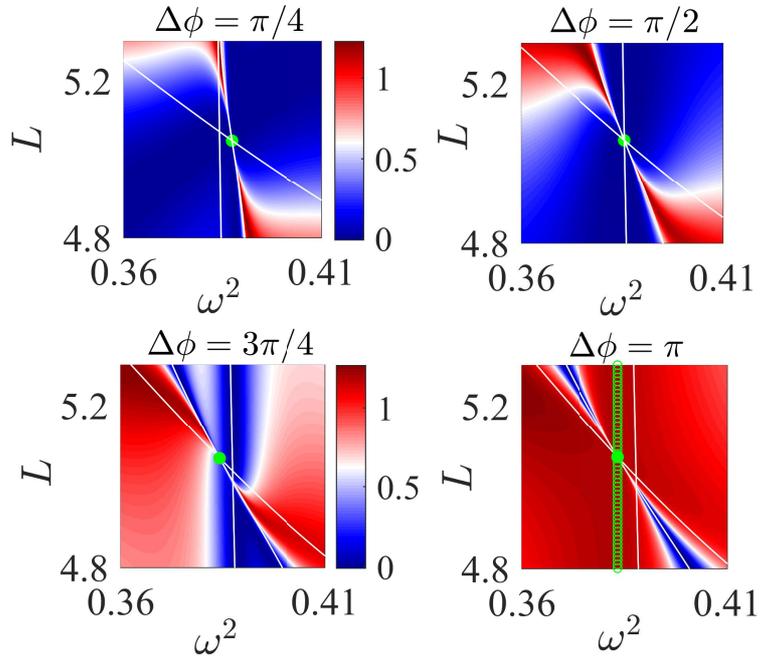}}}
\caption{(a) Transmittance vs frequency and resonator length at
four fixed rotation angles. Solid green lines show the resonances
defined by real part of the complex eigenvalues of the effective
Hamiltonian (\ref{Heff1}). Closed circles mark the BSCs which
exactly correspond to points of degeneracy of the eigenlevels
(\ref{neweig}).} \label{fig30}
\end{figure}
\begin{figure}[ht]
\centering{\resizebox{0.7\textwidth}{!}{\includegraphics{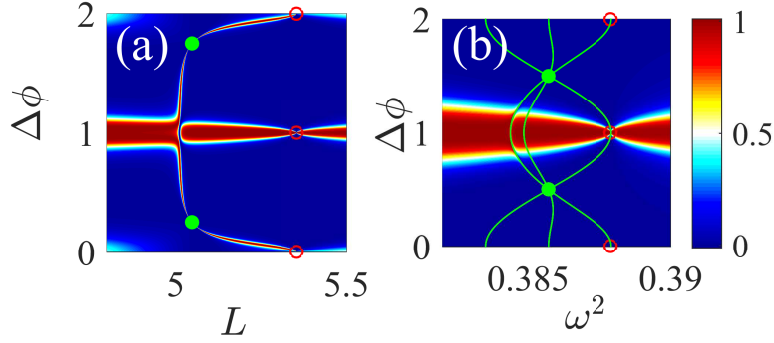}}}
\caption{(a) Transmittance vs the resonator length and rotation
angle for the frequency tuned onto the frequency of the BIC
$\omega_c^2=0.388$. (b) Transmittance vs the frequency and
rotation angle for the length tuned onto the BIC length
$L_c=5.048$. Closed circles mark BIC 1 resulted by crossing of
eigenlevels (\ref{neweig}) $\omega_{012}^2$ and $\omega_{pm
112}^2$, open circles mark the symmetry protected BICs (\ref{sym
prot BSC}).} \label{fig31}
\end{figure}

Thus, the have shown occurrence of the BICs embedded into two
continua which differ by phase in the point of twofold degeneracy.
It is important to lay stress that this degeneracy refers to the
eigenlevels of the Hamiltonian (\ref{HBSC1}) of the cylindrical
resonator modified by evanescent modes of attached waveguides.
This is necessary condition for existence of BIC but not
sufficient. Indeed let us consider the another point of degeneracy
$E_1=E_3$ (see Fig. \ref{fig27} (a)). At this point we adjust the
superposition $a\mathbf{X}_1+b\mathbf{X}_3$ for cancellation of
the coupling with both continua. The analogue of Eq. (\ref{coup1})
takes the following form
\begin{equation}\label{coup2}
\pm aw_0+\frac{b}{\sqrt{2}}w_1(1+e^{i\Delta\phi})w_1=0.
\end{equation}
These equations can not be fulfilled simultaneously to forbid this
degeneracy point as the BIC point. By the use of Eq.
(\ref{S-matrix}) and truncated effective Hamiltonian (\ref{Heff1})
we calculated the transmittance with the results presented in Fig.
\ref{fig30}. Comparison to Fig. \ref{fig20} (b) and (c) shows that
all features of the transmittance can be well reproduced in the
vicinity of the BICs by the use of truncated basis.

One can also see from Figs. \ref{fig30} and \ref{fig31} that the
resonant features follow the real parts of the complex eigenvalues
of the effective non-Hermitian Hamiltonian (\ref{Heff1}) when
$\Delta\phi\neq 0$. Fig. \ref{fig31} shows fine features of the
transmittance vs two parameters for the third parameter exactly
tuned to the BIC. Fig. \ref{fig31} (a) demonstrates a Fano
resonance collapse in the parametric space of length and rotation
angle at the BIC point $L_c=5.048$ and $\Delta\phi_c=\pi/4$ with
the frequency exactly tuned to the BIC $\omega_c=0.3873$. Fig.
\ref{fig31} (b) shows the transmittance vs the frequency and the
rotation angle for the length of the resonator tuned to the BIC
length $L_c=5.0584$. Fig. \ref{fig31} (a) and (b) shows that the
resonator is blocked when $\Delta\phi=0$ and open when
$\Delta\phi=\pi$. We skip here the case when the mode $\pm 112$
crosses the mode $\pm 211$ and refer the reader to the book
chapter \cite{Sadreev2018}. In spite of that the truncated
effective Hamiltonian includes four states still this case allows
analytical treatment of BICs.
\newpage
\section{Spherical cavity}
\label{Sect:Sphera}
\begin{figure}
\centering{\resizebox{0.8\textwidth}{!}{\includegraphics{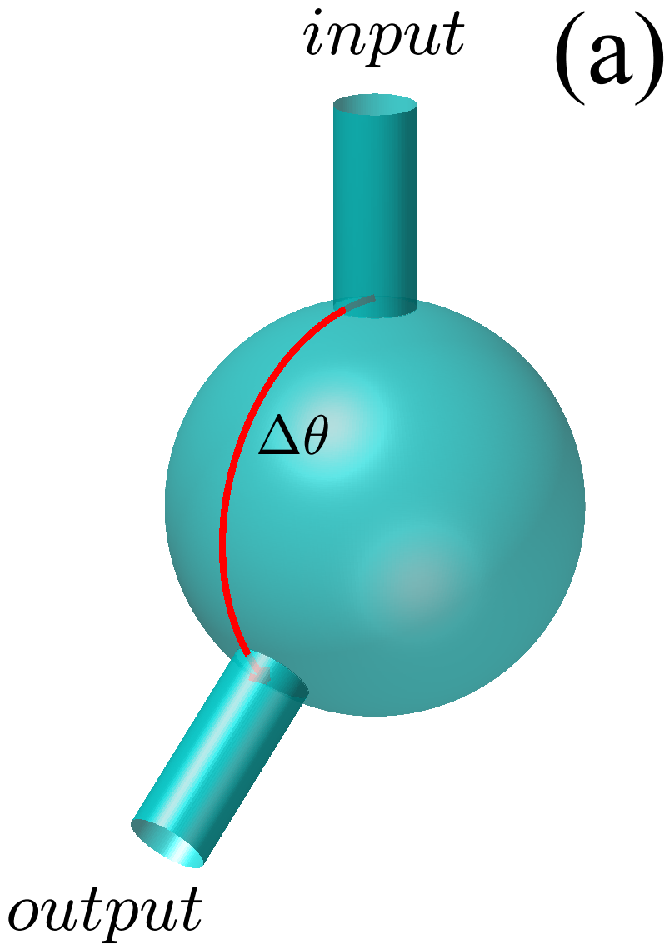},\includegraphics{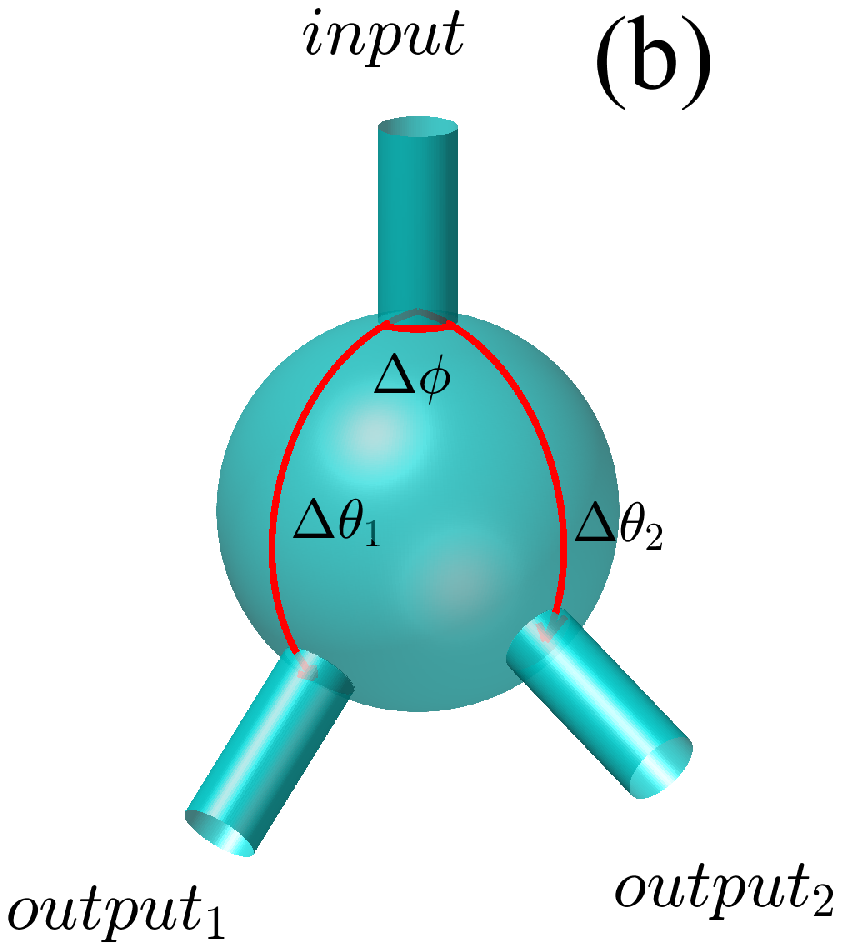}}}
\caption{Spherical cavity of radius $R$ with two (a) and (b) three
attached cylindrical waveguides of the same radii $r$.}
\label{fig32}
\end{figure}

In this section we consider the FW BICs which exist only due to a
contribution of evanescent modes of waveguides. Such an example is
open spherical cavity  shown in Fig. \ref{fig32} which presents
the system consisted of two subsystems with incompatible
symmetries. The continua obey the cylindrical symmetry while the
resonator does the spherical symmetry. Integrable spherical cavity
has the only scale to vary the sphere radius $R$, which only
scales the eigenvalues by the factor $\frac{1}{R^2}$. The
eigenmodes are spherical functions which are $2l+1$-fold
degenerated, where $l$ is the orbital index. Let us attach two
cylindrical waveguides as shown in Fig. \ref{fig32} (a) that fully
removes this degeneracy. Therefore it seems that the FW mechanism
for the BICs due to an avoided crossing can not be applied here.
The continua of the waveguides in the form of propagating Bessel
modes transform the discrete eigenfrequencies of the closed cavity
into the complex resonant frequencies whose positions depend on
overlapping of the spherical functions with the Bessel modes. In
turn, if the waveguides are angled by $\theta\neq \pi$ variation
over that angle can give rise to avoided crossings of resonant
modes with different $l$ to result in the FW BICs.

In order to demonstrate this effect we use the coupled mode theory
with the Neumann boundary conditions applicable for transmission
of acoustic or EM waves with TM polarization \cite{Maksimov2015}.
It is easy to find a solution of the Helmholtz equation in
spherical coordinates, so the eigenfunctions of a spherical cavity
are the following
\begin{eqnarray}\label{res_modes}
            & \Psi_{lmn} = \Psi_{ln}(r)Y_{lm}(\theta, \phi) \\ &
        \Psi_n(r) = \frac{1}{R_s^{3/2}}\sqrt{\frac{2}{\kappa_{l + 1/2, n}^2 - n(n + 1)}}\frac{\kappa_{l + 1/2, n}}{J_{l + 1/2}(\kappa_{l + 1/2, n})}J_{l+1/2}(\frac{\kappa_{l+1/2,n}r}{R}) \\ &
        Y_{lm}(\theta, \phi) = \sqrt{\frac{(2l + 1)(l - m)!}{4\pi(l + m)!}}
        P_{lm}(\cos{\theta})\exp{(m\phi)}
\end{eqnarray}
where $r, \theta, \phi$ are the spherical coordinates, $R$ is the
spherical cavity radius, $Y_{lm}$ are the spherical harmonics,
$P_{lm}(\cos{\theta})$ are the associated Legendre polynomials,
$J_{l + 1/2}$ are the Bessel functions, $\kappa_{l + 1/2, n}$ are
the roots of the equation $\frac{dJ_{l + 1/2, n}(\frac{\kappa_{l +
1/2, n}r}{R_s})}{dr}\big\vert_{r = R_s}$. Respective
eigenfrequencies of the closed spherical resonator are given
\begin{equation}\label{eigs}
\omega_{nl}^2=\kappa_{l+1/2,n}^2/R^2,
\end{equation}
which are $2l+1$-fold degenerate over the azimuthal index
$-l<m<l$. All the quantities are dimensionless and expressed in
terms of the cylindrical waveguides radius $a$. The dimensionless
frequency $\omega$ is expressed through the dimensional one
$\tilde{\omega}$ as follows: $\omega = \tilde{\omega}a/s$ in
acoustics or $\omega=cka$, where $s/c$ is the sound/light
velocity.

The eigenfunctions of the cylindrical waveguides are:
\begin{eqnarray}\label{wg_modes}
    &\psi_{pq}^{(C)}(\rho,\alpha,z)=\psi_{pq}^{(C)}(\rho)\frac{1}{\sqrt{2\pi k_{pq}^{(C)}}}
    \exp(ip\alpha+ik_{pq}^{(C)}z),&\\
    &\psi_{pq}^{(C)}(\rho)=\left\{\begin{array}{l}
    \frac{\sqrt{2}}{aJ_0(\mu_{0q})}J_0(\frac{\mu_{0q}\rho}{a}), p=0, \\
    \sqrt{\frac{2}{\mu_{pq}^2-p^2}}
    \frac{\mu_{pq}}{aJ_p(\mu_{pq})}J_p(\frac{\mu_{pq}\rho}{a}),
    p=1, 2, 3, \ldots,
    \end{array}\right.& \nonumber
\end{eqnarray}
where $\rho, \alpha$ are the polar coordinates in the $x0y$-plane
in the waveguides reference system, $J_p(x)$ are the cylindrical
Bessel functions of the first kind, $\mu_{pq}$ is the q-th root of
equation
$\left.\frac{dJ_p(\mu_{pq}\rho)}{d\rho}\right|_{\rho=a}=0$ imposed
by the Neumann boundary condition on the walls of sound hard
cylindrical waveguide, $C$ enumerates input and output waveguides,
$k^{(C)}_{pq}$ is the wave number:
\begin{equation}\label{kmnR}
    k_{pq}^{(C)}=\sqrt{\omega^2-\mu_{pq}^2/a^2},
\end{equation}

In order to write the non-Hermitian effective Hamiltonian it is
necessary to calculate the coupling coefficients between the modes
propagating in the waveguides and the eigenmodes of the spherical
cavity. For the waveguide connected to the pole of the resonator,
the coupling matrix elements can be calculated as follows
\cite{Maksimov2015,Lyapina2018}:
\begin{equation}\label{Wsphere}
    W_{lmn, pq} = \Psi_{ln}(r=R)\int_0^{2\pi}d\phi\int_0^1\rho d\rho
    \psi_{pq}(\rho, \phi)Y_{lm}(\theta(\rho, \phi), \phi)
\end{equation}
where $\rho$ is the radius in the cylindrical reference frame,
$\phi = \alpha$ is the azimuthal angle, $\theta$ is the polar
angle in the spherical reference frame. To perform this
integration one has to express the spherical coordinates in terms
of the cylindrical ones which could be done by a simple
mathematical transformation. We assume here that the integration
is carried out over the circular interface between the waveguides
and the cavity in the limit $R >> 1$. Then the integration
interface can be approximated by flat circle.

The calculation of the coupling matrix elements for asymmetrically
connected waveguides is a bit difficult. We assume that these
waveguides are also connected to the pole of the spherical
resonator and then rotate the cavity eigenfunctions which is
physically equivalent to rotation of  the waveguides. For that
procedure we use the Wigner $D$-matrix:
\begin{equation}\label{Wigner_matrix}
    D^l_{mk}(\alpha, \beta, \gamma) = exp(-ik\alpha)d^l_{mk}(\beta)exp(-im\gamma),
\end{equation}
where $\alpha, \beta, \gamma$ are the Euler's angles and
$d^l_{mk}(\beta)$ is the small Wigner matrix
\begin{eqnarray}\label{Wigner_small}
            & d^l_{mk} = \sqrt{\frac{(l-m)!(l+m)!}{(l-k)!(l+k)!}}\sum_{s=max(0, k-m)}^{min(l-m, l+k)}(-1)^{m-k+s} \\ &
        {l+k \choose s}{l-k \choose m-k+s}\cos^{2l-m+k-2s}\left(\frac{\beta}{2}\right)
        sin^{m-k+2s}\left(\frac{\beta}{2}\right)
    \end{eqnarray}
Then the rotated spherical harmonic can be expressed through the
non-rotated one as follows
\begin{equation}\label{Y_rot}
    \tilde{Y}^l_m(\theta, \phi) = \exp(-im\gamma)\sum_{k=-l}^l\exp(-ik\alpha)d^l_{mk}(\beta)
    Y^l_k(\theta', \alpha'),
\end{equation}
and the coupling matrix elements of the asymmetrically connected
waveguides are the following
\begin{equation}\label{W_rot}
    \tilde{W}_{lmn, pq} = \exp(-im\gamma)\sum^l_{k=-l}\exp(-ik\alpha)d^l_{mk}W_{lkn, pq}
\end{equation}

Next, we write  the effective non-Hermitian Hamiltonian of the
system, which is the result of projection of the entire Hilbert
space of the system "waveguides + cavity" onto the spherical
cavity subspace
\begin{equation}\label{Heffsphere}
    H_{eff} = H_B - i\sum_{C}\sum_{pq}k^{(C)}_{pq}W^{(C)}_{pq}W^{(C)\dagger}_{pq}
    \end{equation}
where the last term is given by the coupling matrix elements
(\ref{Wsphere}). Then the transmission coefficients from the
channel $pq$ of the waveguide $(C)$ to the channel $p'q'$ of the
waveguide $(C')$ are given by the following equations
\cite{Maksimov2015,Lyapina2018}:
\begin{equation}\label{Transmission}
    t^{(CC')}_{pq;p'q'} = 2i\sqrt{k^{(C)}_{pq}k^{(C')}_{p'q'}}\sum_{lmn}\sum_{l'm'n'}W^{(C)}_{lmn;pq}\frac{1}{\omega^2 - H_{eff}}W^{(C')*}_{l'm'n';p'q'},
\end{equation}

\subsection{Two waveguides}

\begin{figure}[ht]
\centering{\resizebox{0.4\textwidth}{!}{\includegraphics{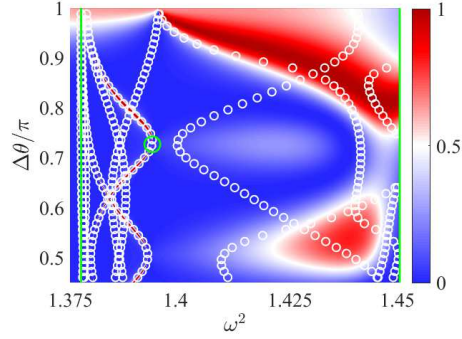}}}
\caption{Transmittance of the spherical resonator vs the frequency
of injected wave and displacement angle of the second waveguide.
Small open white circles show the real part of eigenfrequencies of
open cavity vs the second waveguide displacement angle. Large open
green circle indicates the BIC point with the collapse of Fano
resonance.}
    \label{fig33}
\end{figure}

An attachment of waveguides lifts the $2l+1$-fold degeneracy of
the eigenvalues of the closed spherical cavity as demonstrated in
the Fig. \ref{fig33}
where the real parts of the effective Hamiltonian
(\ref{Heffsphere}) complex eigenvalues are plotted by small open
circles versus the rotation angle $\Delta\theta$. One can see from
Fig. \ref{fig33} that rotation of the second waveguide relative to
the first waveguide splits resonances. The more important is,
however, that such a rotation gives rise to the avoided crossing
of resonances with different orbital indices $l$ and respectively
to the FW BIC which is marked by large open circle. Fig.
\ref{fig33} also shows the transmittance versus the injected wave
frequency and the second waveguide displacement angle
$\Delta\theta$. One can see that the narrow resonant peaks follow
to the resonant frequencies marked by open circles. The small
resonant widths are result of normalization coefficients of the
eigenmodes of the spherical cavity (\ref{wg_modes}) proportional
to $\frac{1}{R^{3/2}}$. As a result the coupling matrix elements
(\ref{Wsphere}) have the same factor and the resonant widths which
are given by squared coupling matrix elements turn out to be
proportional to $\frac{1}{R^3}$ while the distance between the
eigenfrequencies of $H_B$ are proportional to $\frac{1}{R^2}$.
Therefore for $R\ll 1$ we have the case of weak coupling of the
sphere with the waveguide continuum.

The collapse of the Fano resonance, i.e. coincidence of the unit
and zero transmittance, is the signature of the BIC
\cite{Kim1999,SBR}. Fig. \ref{fig33} shows one of these events at
which  imaginary part of the complex eigenvalues of the non
Hermitian effective Hamiltonian vanishes. A major part of the BICs
in the case of two waveguides are symmetry protected. These SP
BICs can be obtained by simple rotation of the eigenfunctions of
closed spherical resonator in order to achieve orthogonality of
the eigenfunction to the mode of waveguide. We do not show here
the symmetry protected BICs which coincide with the rotated eigen
mode of closed  cavity by use of the Wigner D-matrix.

\begin{figure}[hbt]
\centering{\resizebox{0.7\textwidth}{!}{\includegraphics{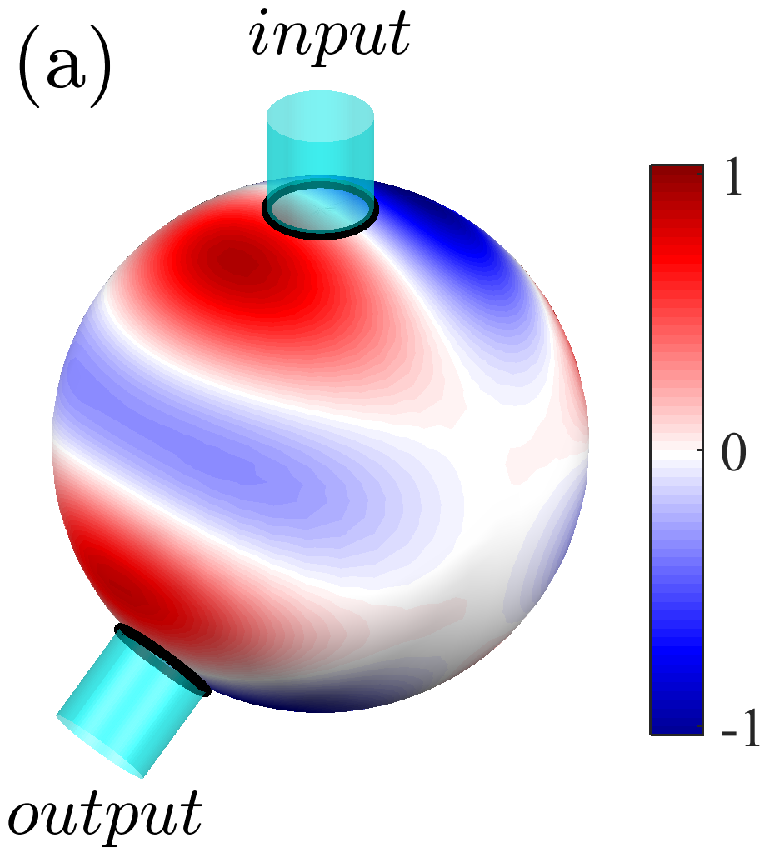},\includegraphics{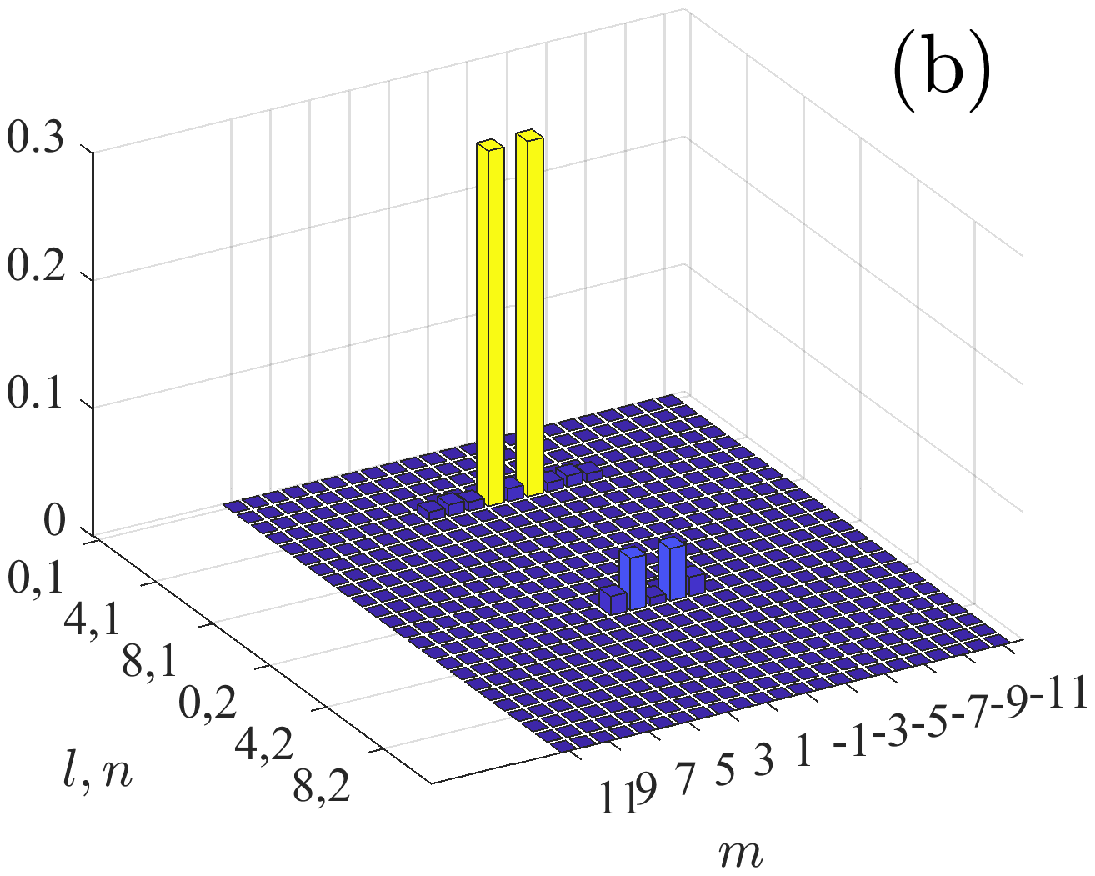}}}
\caption{(a) The pressure field of the FW  BIC at $\omega =
1.3937$ and $\Delta\theta = 0.727\pi$. (b) The modal decomposition
of the BIC.} \label{fig34}
\end{figure}

However Fig.~\ref{fig33} marks the FW BIC at point
$\Delta\theta=0.7\pi, \omega=1.378$ by open green circle. Fig.
\ref{fig34} (a) shows the FW BIC wave function (the pressure
field/magnetic field) on the resonator surface. One can from nodal
lines on the surface on sphere the BIC mode is decoupled from the
first continuum of the waveguide with indices $p = 0, q =1$. The
modal expansion of this FW BIC over the eigen modes of the closed
spherical cavity $\Psi_nl(r)Y_{lm}(\theta,\phi)$ is shown in
Fig.~\ref{fig34} (b). The eigenmodes with quantum numbers  $l=4,
m=\pm 1, n=1$ and $l=1, m=\pm 1, n=2$ contribute into the FW BIC.
Thus, the FW BIC is the result of full destructive interference of
resonant modes with different orbital indices, despite that the
eigenmodes of the closed spherical cavity with different orbital
momentum $l$ have different frequencies (\ref{kmnR}).
\subsection{Three waveguides}
\begin{figure}[hbt]
\centering{\resizebox{0.7\textwidth}{!}{\includegraphics{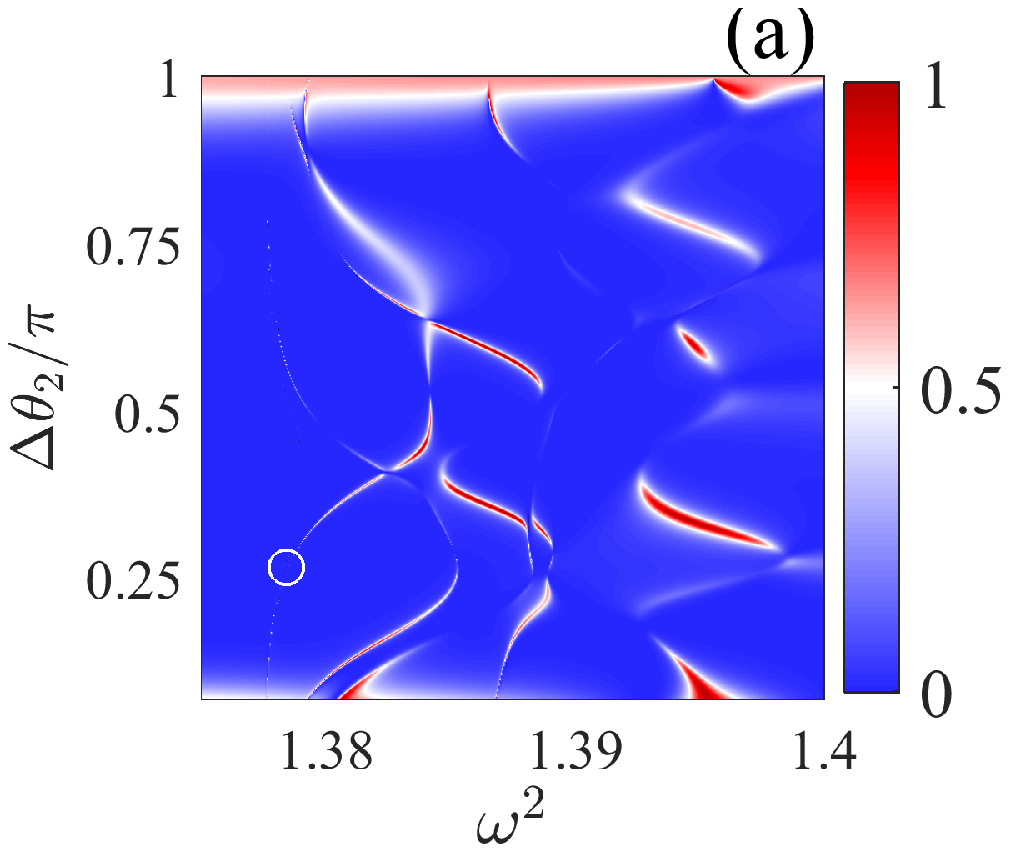},\includegraphics{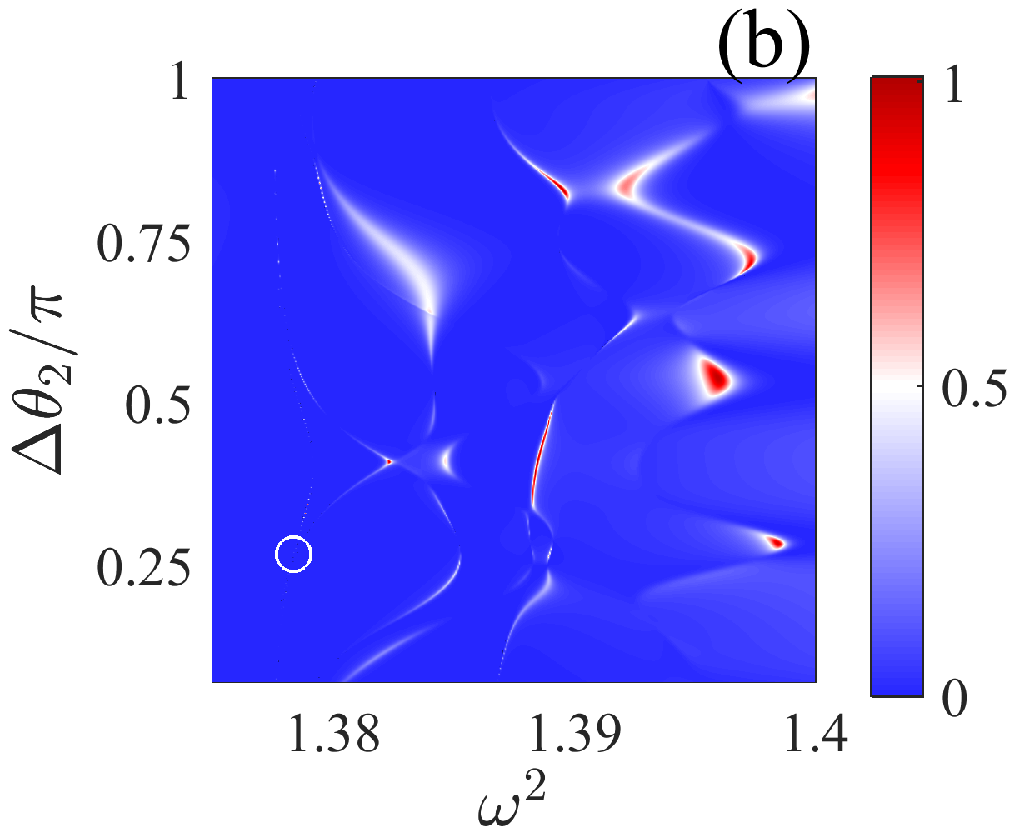}}}
\caption{Transmittance between "input" and "output 1" (a) and
"input" and "output 2" vs frequency and the displacement angle
$\Delta\theta_2$ of the third waveguide for $\Delta\theta_1 =
\pi/4$. The displacement angle of the second waveguide is
$\Delta\theta_1 = 3\pi/4$. Crosses mark the FW BICs.}
\label{fig35}
\end{figure}

\begin{figure}[hbt]
\centering{\resizebox{0.7\textwidth}{!}{\includegraphics{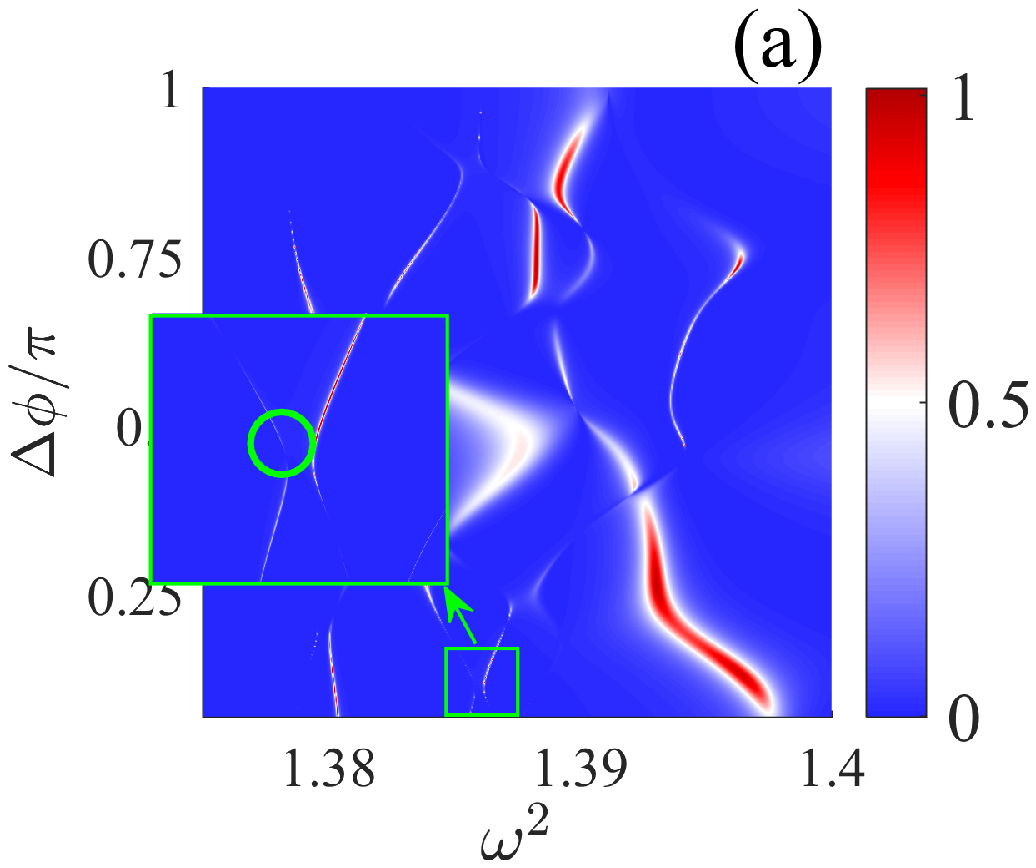},\includegraphics{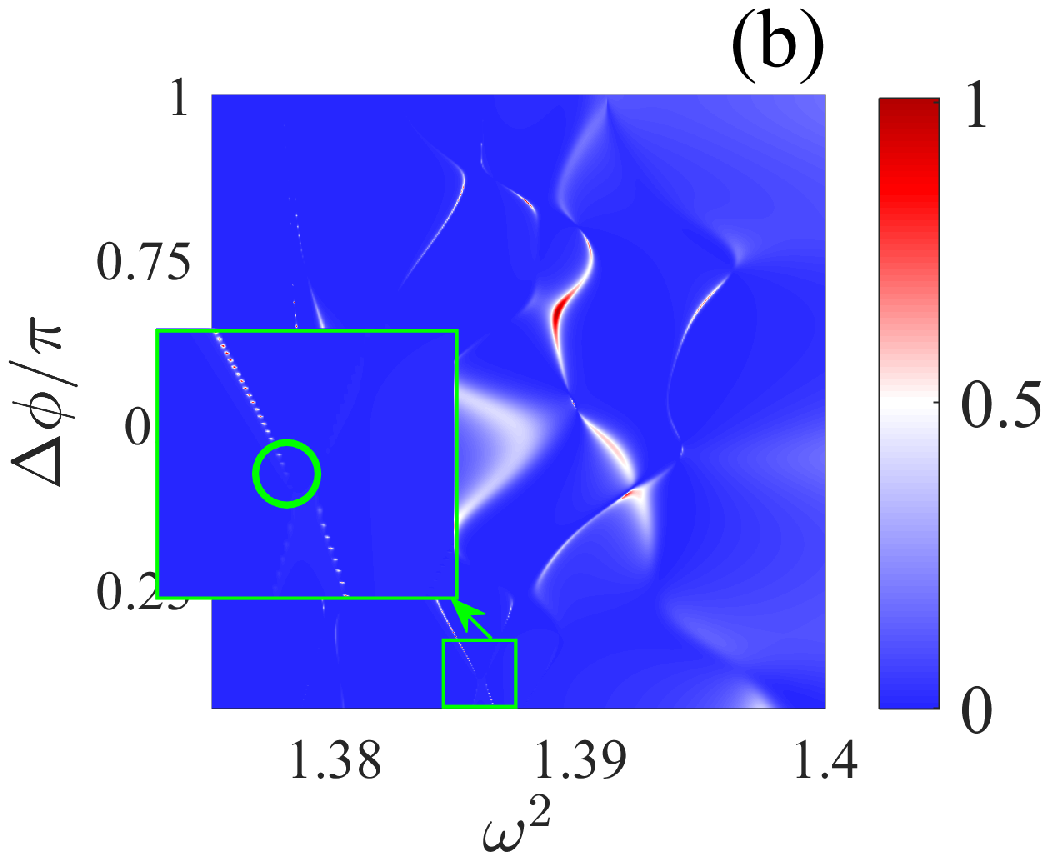}}}
\caption{Transmittance between "input" and "output 1" (a) and
"input" and "output 2" vs frequency and the displacement angle
$\Delta\theta_2$ of the third waveguide for $\Delta\theta_1 =
\sqrt{5}$. } \label{fig36}
\end{figure}

Although the position of the second waveguide relative to the
first one at the pole of sphere is given by two angles in general,
only the polar angle $\Delta\theta_1$ is physically relevant for
resonances and, in particular, for the BICs. Introduction of the
third waveguide as shown in Fig. \ref{fig32} (b) substantially
changes effects of the continua onto the resonances because of
three relevant angles, two polar angles $\Delta\theta_1$ and
$\Delta\theta_2$ and one azimuthal angle $\Delta\phi$. Figs.
\ref{fig35} and \ref{fig36} show the transmittance versus the
frequency of injected wave and rotation angles $\Delta\theta_2$
and $\Delta\phi$ of the third waveguide which evident importance
of mutual orientations of the all three waveguides. The regions in
which avoid crossing phenomenon occurs, as well as the collapse of
the Fano resonance, are highlighted by frames in Fig. \ref{fig36}.
One can see that these phenomena take place  irrespective to which
waveguide goes wave.

\begin{figure}[ht]
\centering{\resizebox{0.7\textwidth}{!}{\includegraphics{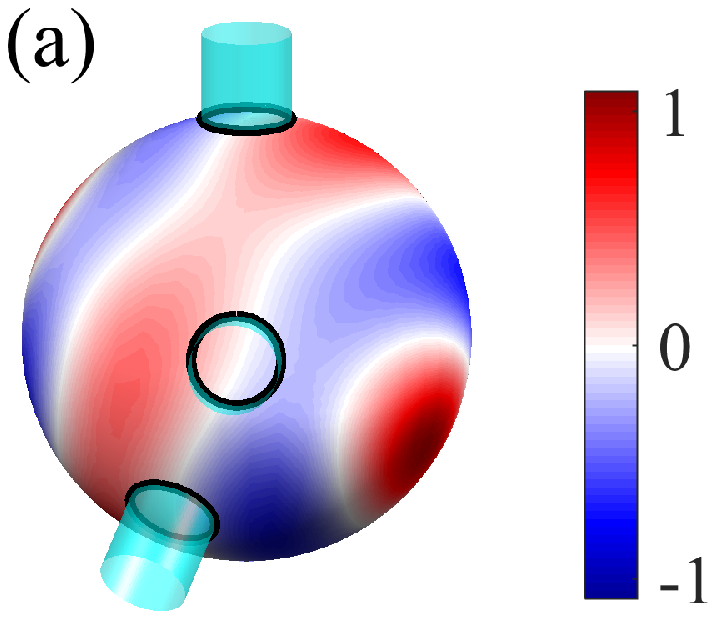},\includegraphics{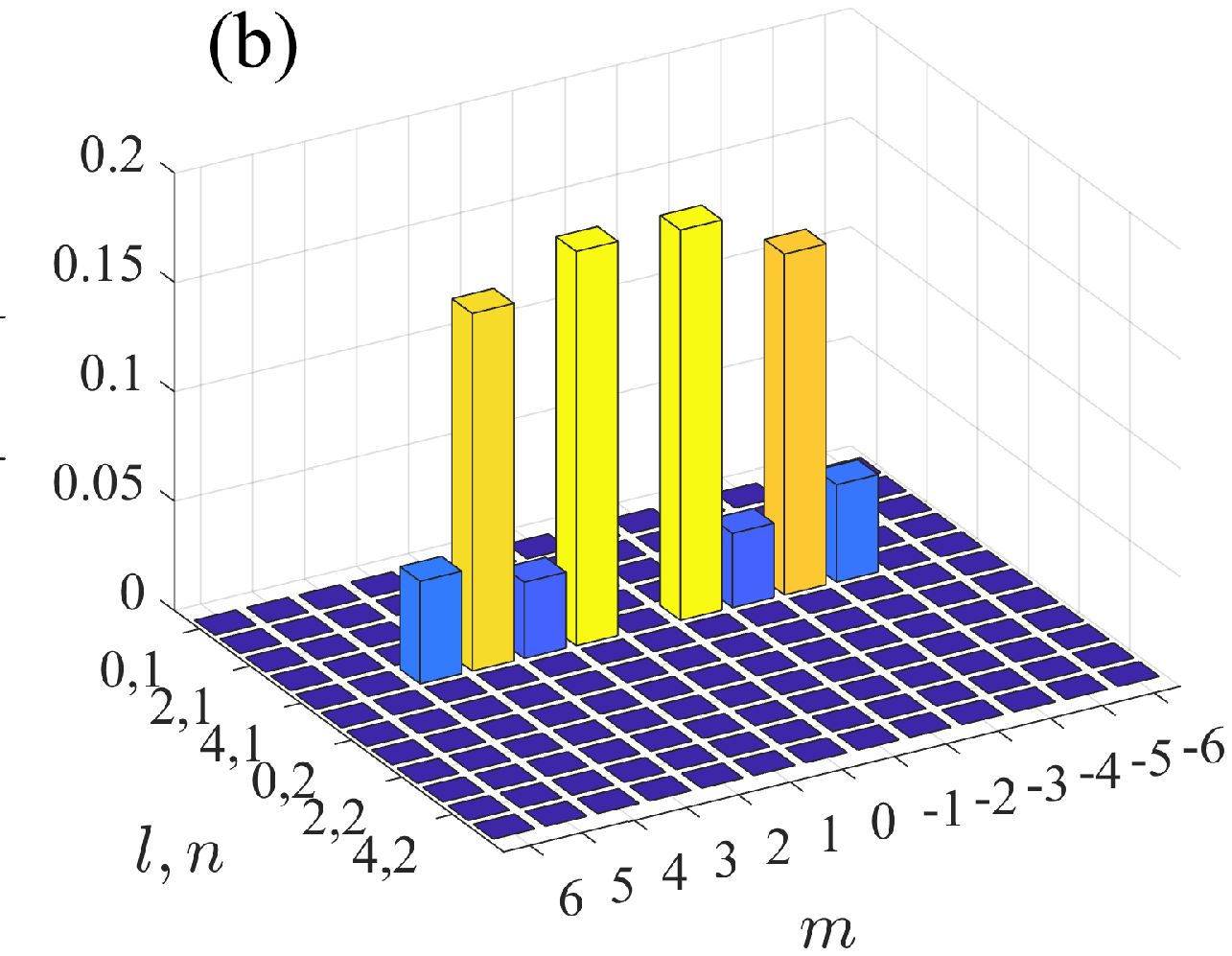}}}
\caption{(a) The BIC pattern (pressure field) on surface of
spherical cavity at the BIC point with $\omega = 1.38575$,
$\Delta\theta_1 = \sqrt{5}$, $\Delta\theta_2 = \sqrt{2}$ and
$\Delta\phi = 0.1222\pi$.  (b) The modal decomposition of the BIC.
} \label{fig37}
\end{figure}

The circle in the Figs. \ref{fig35} and \ref{fig36} marks the
position of the FW BIC whose pattern in the form of surface
pressure/magnetic field on surface of the resonator is shown in
the Fig. \ref{fig37} (a). The amplitudes $a_{nlm}$ of
superposition of spherical harmonics are chosen in so way that the
nodal lines shown by white pass through the overlapping areas of
waveguides with the spherical cavity. As a result the coupling
constants of the FW BIC with the first propagating channel or
continuum vanish.

\newpage
\section{The Fabry-Perot mechanism of BICs in the system of two
coupled resonators} \label{FPR}
\begin{figure}[ht]
\centering{\resizebox{0.8\textwidth}{!}{\includegraphics{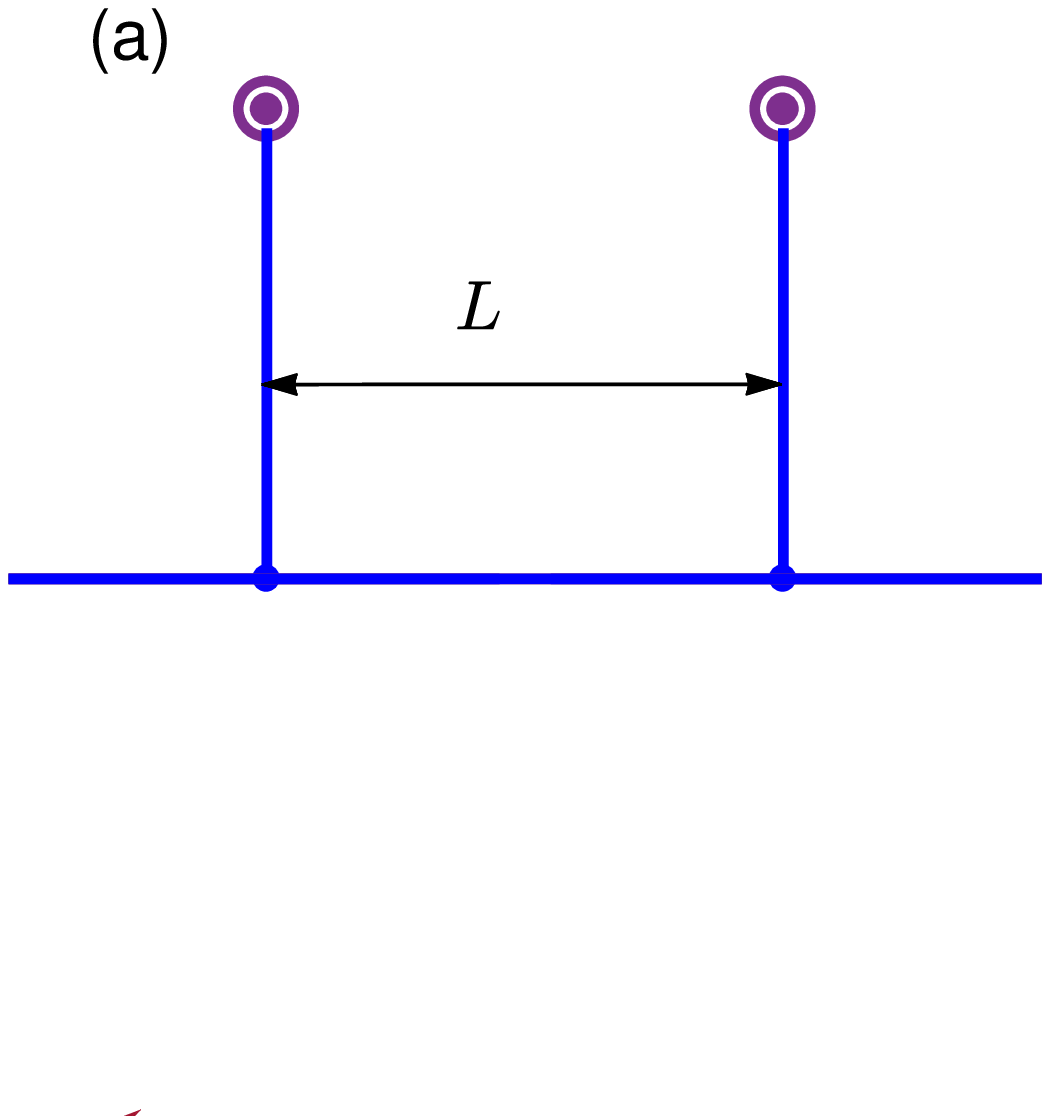},\includegraphics{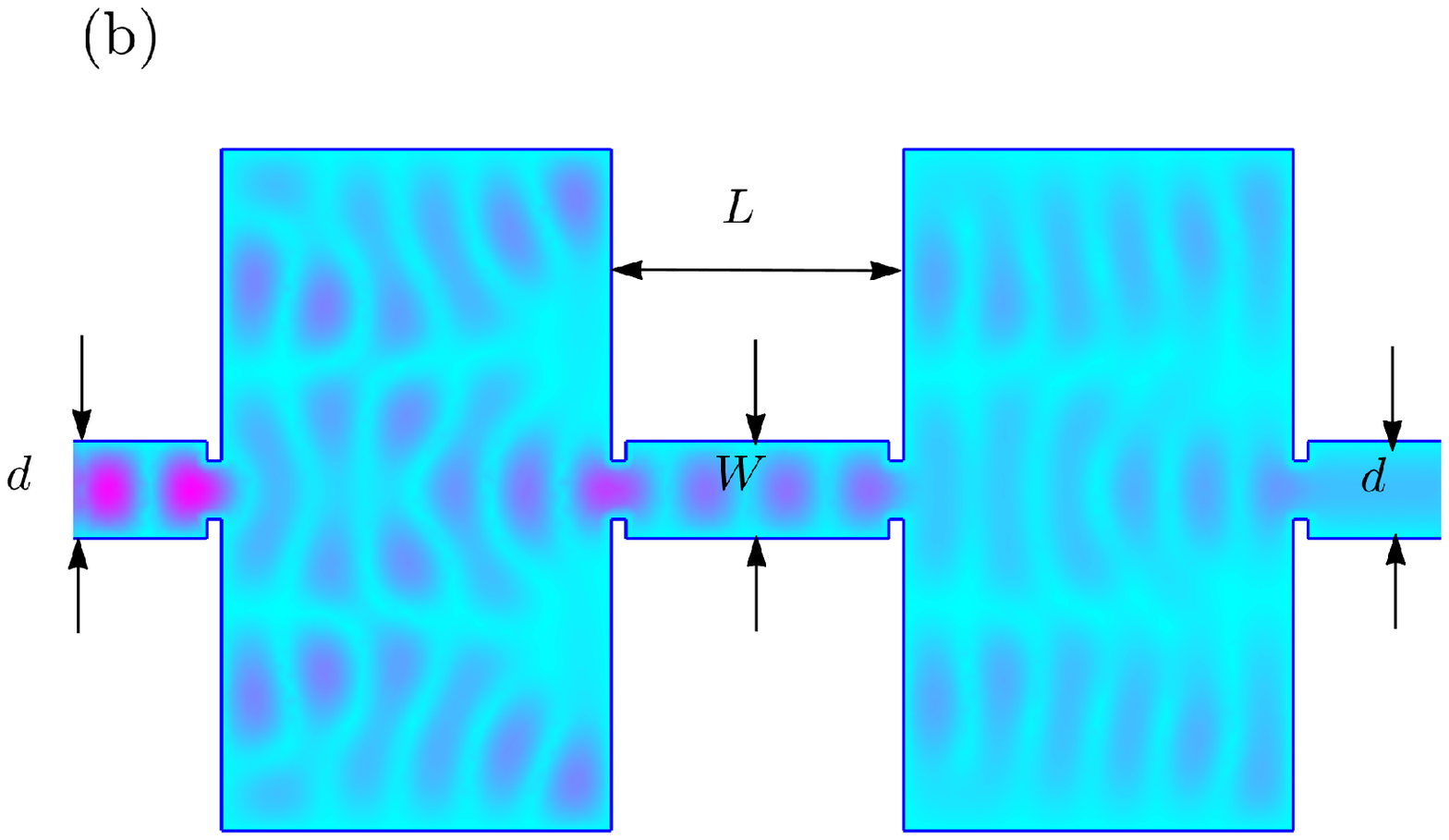}}}
\caption{(a) 1d wire with two off-channel cavities. (b) 2d wire
with two inserted identical resonators. } \label{fig38}
\end{figure}

If the double-barrier resonant structure had the infinitely high
barriers the eigenmodes were localized between the barriers. For
finite height of barriers these eigenmodes transform to the
resonant modes with finite resonant widths defined by probability
of tunnelling. Such a one-dimensional QM structure has one by one
equivalency to the Fabry-Perot resonator (FPR) \cite{Born_Wolf}
and has no BICs as was discussed in section 4. Let us substitute
the two-dimensional resonators instead of the barriers or mirrors
in the FPR as presented in Fig. \ref{fig38}.

We start with the simplest case of 1d wire to which two off-side
or off-channel cavities are attached as illustrated in Fig.
\ref{fig38} (a). The case of a single off-channel defect realizes
the simplest way for Fano resonance due to interference of two
wave paths, direct path over the wire and second path through the
off-channel defect. As a result that gives rise to $N$
transmission zeros at $\omega=\omega_n, n=1, 2, \ldots, N$ where
$N$ is the number of eigenfrequencies of the defect
\cite{SR2003,Miros2005}. Thus, the off-channel defects can serve
as ideal Fano mirrors and support BICs provided that an integer of
half-waves is placed between mirrors, i.e.,
\begin{equation}\label{FPRcond}
\pi cn=\omega_n L
\end{equation}
where $c$ is the light velocity. Therefore, the underlying
mechanism of the bound states in the FPR is (i) perfect
reflections at mirrors and (ii) the integer number of the half
waves between mirrors. This mechanism, exclusively transparent,
for the bound states which we call as the Fabry-Perot (FP) BIC was
applied to photonic crystal structure with one and two waveguides
coupled with two single-mode cavities
\cite{Fan1999,BS2008,Wang2003,Lin2005,Marinica}. Technologically a
tuning of eigenfrequencies of the off-channel defects can be
performed by variation of their refractive index or size. However
in Ref. \cite{Pichugin2016}  stable light trapping in  the
nonlinear Fabry-Perot  resonator without necessity to tune the the
distance between the off-channel defects was reported by
implementation of an auxiliary nonlinear  resonator.

A different way is to implement two-dimensional cavities into
waveguide as shown in Fig. \ref{fig38} (b). Each resonator has
transmission zeroes \cite{LeePRL} at some frequencies to serve as
Fabry-Perot mirrors. Therefore the total system consists of two
cavities and a wire between them. In the simplest form the
Hamiltonian of closed system has the following matrix structure
\begin{equation}\label{HB5}
H_B=\left(\begin{array}{ccccc}\epsilon_1 & 0 & u & 0 & 0   \cr
                               0 & \epsilon_2 &  u & 0 & 0 \cr
                               u &  u & \epsilon_w & u & u \cr
0 & 0 &  u & \epsilon_2 & 0 \cr
            0  &  0 & u & 0 & \epsilon_1  \end{array}\right).
\end{equation}
We can consider the eigenlevel of the wire $\epsilon_w$ is the
parameter by which the system can be controlled.

The minimal rank of matrix (\ref{HB5}) is five, so that let $E_n$
and $|n\rangle $ with $n=1, ..., 5$ denote  the five eigenlevels
and eigenstates of (\ref{HB5}).  The amplitudes $\langle
j=1,2|n\rangle$ describe the left resonator, $\langle
j=3|n\rangle$ the waveguide, and $\langle j=4,5|n\rangle$ the
right resonator. Two semi-infinite waveguides attached to the
resonators provide continua and therefore transform the states of
closed system into resonances which are described by the effective
non-Hermitian Hamiltonian \cite{SR2003}
\begin{equation}\label{FPRHeff}
\langle  m|H_{\rm eff}|n\rangle = E_m\delta_{mn}-2\pi
i(V_L(m)V_L(n)+V_R(m)V_R(n))
\end{equation}
with the coupling matrix elements
\begin{eqnarray}\label{VLR}
&V_L(m)  = v(k)\sum_{j=1, 2}\langle
j|m\rangle,&\nonumber\\
&V_R(m)  = v(k)\sum_{j=4, 5}\langle j|m\rangle,&
\end{eqnarray}
where the factors $\sqrt{\frac{k}{2\pi}}$ originated from the
normalization of propagating states of 1d waveguides are absorbed
by $v(k)$.
\begin{figure}[ht]
\centering{\resizebox{0.8\textwidth}{!}{\includegraphics{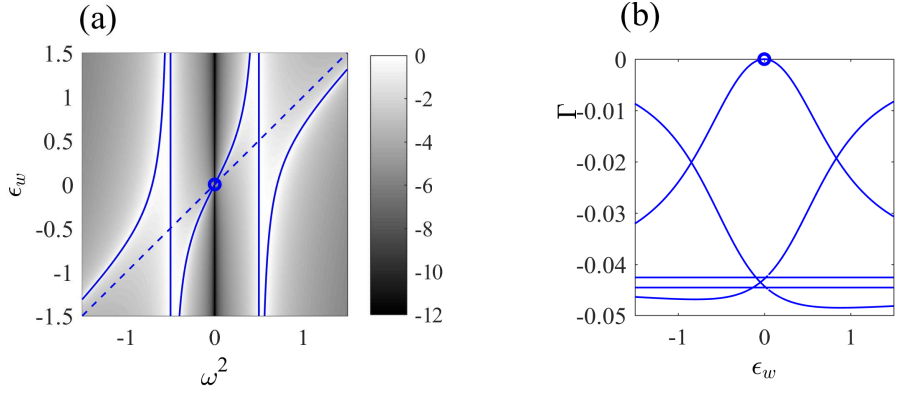}}}
\caption{(a) The transmittance in Log scale through double
resonator vs the frequency $\epsilon_W$ of wire and incident
frequency $E=\omega^2$. (b) The resonant widths as dependent on
$\epsilon_W$ at $E=0.5$ at the following parameters of the system:
$\epsilon_{1,2}=\pm 1/2, v=0.5, u=1/4$.} \label{fig39}
\end{figure}

The transmittance through the system given by Eq.
(\ref{Transmission}) is shown in Fig. \ref{fig39} (a) in Log scale
in order to follow transmission zeros and resonances. Because of
small coupling constant $v(k)=0.5\sqrt{\frac{k}{2\pi}}$ the
transmittance demonstrates resonant behavior which follows the
eigenlevels of the Hamiltonian (\ref{HB5}) of closed system
\begin{equation}\label{EB5}
  E_{1,5}=\pm\eta, ~~E_2=\epsilon_1, ~~E_3=0,
  ~~E_4=\epsilon_2.
\end{equation}
The eigenvalues 2 and 4 of the effective Hamiltonian are
independent of the wire's eigenvalue $\varepsilon_w$, while those
of the other states depend on it. The eigenvalue 3, lying in the
middle of the spectrum, crosses the transmission zero at
\begin{equation}\label{epsc}
 \epsilon_w=\epsilon_b=\frac{\epsilon_1+\epsilon_2}{2}=0.
  \end{equation}
At this eigenvalue we observe the collapse of the Fano resonance
that witnesses the BIC that fully agrees with turning to zero of
resonant width as seen from Fig. \ref{fig39} (b).

However still it is remaining a question where the BIC is
localized, in the wire between the resonators or entirely in whole
structure including resonators. It might be seemed that the latter
answer taking into account that resonator provide large rooms for
localization. Below by use of exact analytic equations we show
that the first answer is correct, at least, in the present model
case of 1d wire. The eigenstates of the Hamiltonian (\ref{HB5})
are the following
\begin{eqnarray}\label{psiB5}
\langle 1| &=&\frac{\sqrt{2}u}{\eta}\left(
\frac{u}{\eta-\Delta\epsilon},~ \frac{u}{\eta+\Delta\epsilon},
~-1,~ \frac{u}{\eta+\Delta\epsilon},~
\frac{u}{\eta-\Delta\epsilon}\right) \nonumber\\ \langle
2|&=&\frac{1}{\sqrt{2}}\left(1,~ 0,~ 0,~ 0,~ -1\right) \nonumber\\
\langle 3|&=&\frac{u}{\eta}\left(1,~ -1,~
\frac{\Delta\epsilon}{u},~ -1,~ 1\right) \nonumber\\ \langle 4|
&=&\frac{1}{\sqrt{2}}\left(0,~ 1,~ 0,~ -1,~ 0\right)\nonumber\\
\langle 5|
&=&\frac{\sqrt{2}u}{\eta}\left(\frac{u}{\eta+\Delta\epsilon},~
\frac{u}{\eta-\Delta\epsilon},~1,~ \frac{u}{\eta-\Delta\epsilon},~
            \frac{u}{\eta+\Delta\epsilon} \right)
\end{eqnarray}
where $\eta^2=\Delta\epsilon^2+4u^2,
~\Delta\varepsilon=(\epsilon_2-\epsilon_1)/2$. Substituting
(\ref{psiB5}) into (\ref{VLR}) we obtain
\begin{equation}\label{V5}
\langle m|V|E,C=L,R\rangle=v\sqrt{\frac{k}{8\pi}}(1~ \pm 1~~
\frac{\Delta\epsilon}{u}~ \pm 1~~ 1)
\end{equation}
for the elements of the coupling matrix. One can see that, under
the condition (\ref{epsc}), the wire decouples from the rest of
the system with zero imaginary part of the third eigenvalue of
$H_{\rm eff}$, i.e. the width of the third eigenstate vanishes at
$\varepsilon_w=\varepsilon_b$.

\begin{figure}[ht]
\centering{\resizebox{0.5\textwidth}{!}{\includegraphics{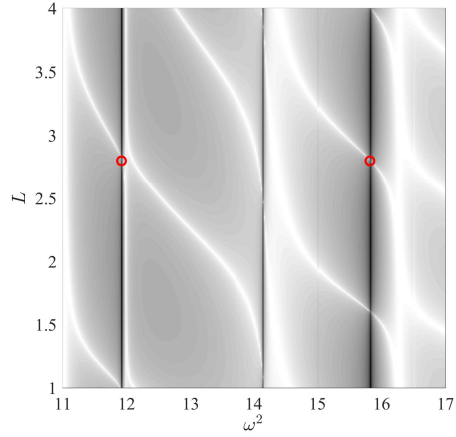}}}
\caption{The transmittance in Log scale of the double resonator
shown in Fig. \ref{fig38} (b) versus frequency and length of the
waveguide between the resonators. The bold open circles mark two
points of BICs shown in the next Fig. \ref{fig41}.} \label{fig40}
\end{figure}

In conclusion we present numerically computed transmittance
through planar metallic double resonator connected by planar
two-dimensional waveguide in Fig. \ref{fig40}. Total view of the
double resonator connected to semi-infinite waveguide through
diaphragms is shown in Fig. \ref{fig38} (b). This figure also
shows the scattering wave function.
\begin{figure}[ht]
\centering{\resizebox{0.9\textwidth}{!}{\includegraphics{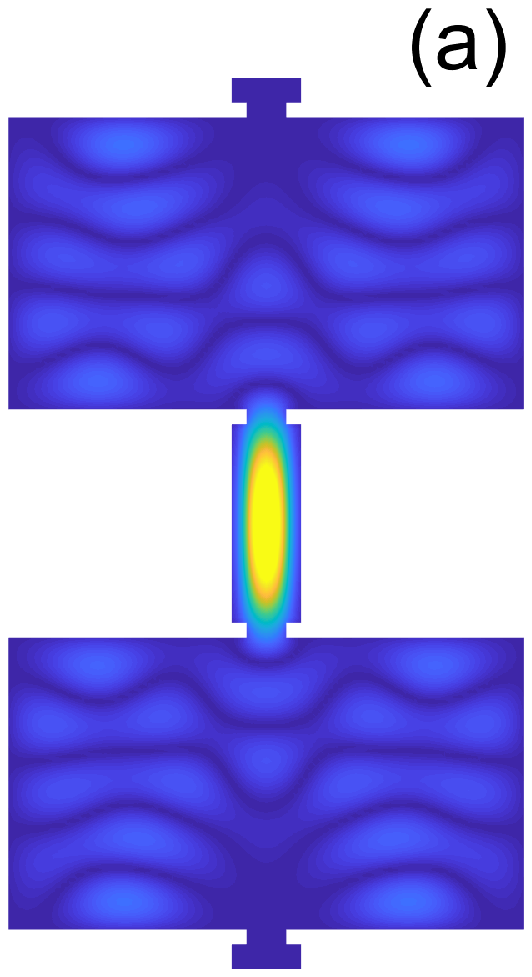},\includegraphics{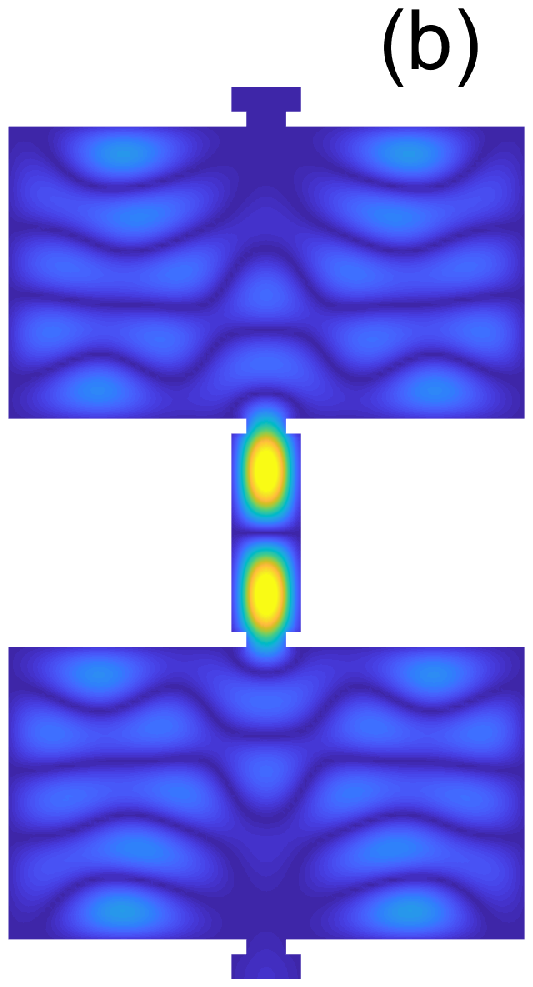}}}
\caption{BICs at points marked by open circles in Fig.
\ref{fig40}: (a) at $\omega^2=11.92$ and $L=2.8$ and (b)
$\omega^2=15.83$ and $L=2.83$.} \label{fig41}
\end{figure}
Fig. \ref{fig41} presents two patterns of the BICs which
correspond to the Fabry-Perot resonances $n=1$ (a) and $n=2$ (b)
in Eq. (\ref{FPRcond}). Details of the Fabry-Perot BICs the reader
can find in Refs. \cite{Sadreev2005a,Sadreev2005b}. Note that
different scheme for the Fabry-Perot BICs was presented by
Marinica {\it et al} in which arrays of dielectric cylinders play
role of mirrors at the frequency of full reflection
\cite{Marinica,Ndangali2010}.

\section{Conclusions}
\label{Conclusions}
The present review is addressed first of all to the
Friedrich-Wintgen mechanism of localization of waves in the open
electromagnetic (metallic) and acoustic cavities. The mechanism is
based on full destructive interference of two resonant modes
outgoing from the cavity. Although we presented the
three-dimensional symmetrical cavities, cylindrical and spherical,
in which BICs are result of destructive interference of more
resonances. The cavities are open by attachment of directional
waveguides which provide well separated continua of propagating
modes. Therefore such waveguide systems have the advantage of
controlling the number of continua by crossing the cutoff
frequencies. Throughout the review we almost entirely used two
identical waveguides to have identical the continua of waveguides.
Importantly, the open resonators are one of the best systems where
the effective non-Hermitian Hamiltonian can be derived
analytically with exact expressions for the coupling matrix.
Moreover the identical waveguides can be attached to the resonant
cavities of cylindrical or spherical shapes in such a way that the
coupling matrices for the two waveguides differ by phase. That
simple way to distinguish the continua gives us an additional
parameter to control the wave transmission (wave faucet) and
realize twisted BICs.

The evanescent modes with cutoffs above the  BIC frequencies have
also principal importance for the BICs: first due to the boundary
conditions between localized BIC mode and evanescent modes the
BICs exist and slightly stand out from the cavity. Because of
absence of evanescent modes in one-dimensional wires there is no
BICs in the cavity opened by attachment half-infinite wires. That
is only true for the one-dimensional quantum wires or layered
structures where TE and TM polarizations are separated. For the
case of spinor fields transmission like one-dimensional electron
transmission through the quantum dot we show that the FW BICs can
occur due to the full destructive interference of resonances with
opposite spins. The same idea can be applied to defect anisotropic
layer where EM waves with TE and TM polarization can destructively
interfere \cite{Pankin2020}.

Second, the evanescent modes contribute into the Hamiltonian of
the closed cavity similar to the Lamb shift in atomic physics. The
coupling to evanescent modes shifts the BIC point from the point
of degeneracy of the closed cavity. However, the most striking
effect is that the FW BICs exist only owing to the evanescent
modes as it was demonstrated in the open spherical cavity.

Although the FW mechanism of the BICs is the most generic and
interesting we have reviewed also another mechanisms of the BICs.
The second mechanism is vanishing of the coupling of some
eigenmode of the cavity with the continuum that results in
accidental BICs. As an example we considered the open chaotic
Sinai billiard (rectangular resonator with hole inside)  in which
variation of the hole's diameter gives changes the coupling
constants and finally to the accidental BICs.

There are no BICs in one-dimensional system except specially
chosen long-range oscillating potentials by von Neumann and Wigner
\cite{Neumann}. However that is truth only for scalar waves. For
vectorial waves again the FW mechanism of BICs can be applied
however as a result of full destructive interference of resonances
corresponding to different components of the vectorial field
\cite{Pankin2020}.

Finally we have reviewed the Fabry-Perot mechanism for the BICs
when the off-channel defects or two-dimensional cavities
integrated into waveguide can serve as ideal mirrors due to
transmission zeroes of the cavities.

 One of the most remarkable results for BICs is their existence in
 photonic crystal systems embedded into the radiation continuum
 which has infinite number of continua because of dispersion
 equation $\omega=ck$. It may be seemed that there are not
 possible BICs embedded into the radiation continuum. Indeed
 rigorous theorem forbids BICs in finite dielectric structure \cite{Colton}.
 However, if we take the infinite periodic PhC structures like 2d
 PhC surface, one-dimensional array of dielectric particles we
 obtain an analogue of diffraction lattices which are coupled with
 only discretized continua defined as the diffraction orders. That
 is physical expalantion  for BIC  in such infinite PhC
 structures \cite{Hsu2013,Bulgakov2015}.

Here we skipped  the majority of results on  BICs in photonics for
two reasons. First, this research direction is so rapidly
developing and huge that it could hardly be put into a single
review. We only included one example of BICs in one-dimensional
photonic crystal holding the defect anisotropic layer in which the
BICs are realized because of full destructive interference of
resonance with TE and TM polarizations. The second reason is that
the recent reviews  have filled this gap
\cite{Hsu16,Krasnok2019,Koshelev2019,Peng2020} already.

 {\bf Acknowledgments}
 First of all I am grateful to Ingrid Rotter who introduced me to
 concept and machinery of non Hermitian Hamiltonian and with whom we first revealed
 the phenomenon of vanishing of resonant width.
I would like express gratitude to my colleagues with whom I worked
long time in the field of bound states in the continuum, Evgeny N.
Bulgakov, Dmittrii N. Maksimov, Konstantin N. Pichugin, Artem A.
Pilipchuk, and Alina Pilipchuk. I also had a lot of discussions
with researches over all the world: Andrey A. Bogdanov, Yi Xu,
Dezhuan Han, Egor Muliarov, Ivan Timofeev, Pavel Pankin, Evgeny
Kamenetskii, Andrey Miroshnichenko, Yurii Kivshar, Kirill
Koshelev, Ya Yan Lu, Evgeny Sherman. My special thanks to Monti
Segev who initiated me to write this review.

The work was partially supported by Russian Foundation for Basic
Research projects No. 19-02-00055.



\bibliographystyle{elsarticle-num}
\bibliography{RMP}
\end{document}